\documentclass[twocolumn]{aastex631}

\usepackage[T1]{fontenc}
\usepackage{ae,aecompl}

\usepackage{graphicx}	
\usepackage{amsmath}	
\usepackage{amssymb}	

\usepackage{natbib}

\bibpunct{(}{)}{;}{a}{}{,}

\usepackage{url}

\usepackage{longtable}
\usepackage{booktabs}

\usepackage[flushleft, referable]{threeparttablex}

\usepackage{subfloat}

\DeclareGraphicsExtensions{.jpg,.pdf,.png}

\usepackage{parskip}

\usepackage{lipsum}

\usepackage{dsfont}

\usepackage{comment}

\usepackage{tablefootnote}

\newcommand\degrees{^{\circ}}

\newcommand\Msun{\rm M_{\odot}}

\newcommand\WISE{\textit{WISE}}

\newcommand\HST{\textit{HST}}
\newcommand\JWST{\textit{JWST}}

\newcommand\lenstool{\textsc{lenstool}}

\newcommand\ElA{{\it El Anzuelo}}

\newcommand\ElG{{\it El Gordo}}

\newcommand\GALFIT{{\sc galfit}}

\newcommand\EAZY{{\sc eazy}}

\graphicspath{{./}{figures/}}

\shorttitle{Color gradients in a $z=2.3$ DSFG lensed by the El Gordo cluster}
\shortauthors{Kamieneski et al.}

\begin{document}


\title{Are JWST/NIRCam color gradients in the lensed $z=2.3$ dusty star-forming galaxy {\it El Anzuelo} due to central dust attenuation or inside-out galaxy growth?}

\correspondingauthor{Patrick S. Kamieneski}
\email{pkamiene@asu.edu}

\author[0000-0001-9394-6732]{Patrick S. Kamieneski}
\affiliation{School of Earth and Space Exploration, Arizona State University, 
PO Box 871404,
Tempe, AZ 85287-1404, USA}

\author[0000-0003-1625-8009]{Brenda L. Frye}
\affiliation{Steward Observatory, University of Arizona, 933 N Cherry Ave, Tucson, AZ, 85721-0009}

\author[0000-0002-2282-8795]{Massimo Pascale}
\affiliation{Department of Astronomy, University of California, 501 Campbell Hall \#3411, Berkeley, CA 94720, USA}

\author[0000-0003-3329-1337]{Seth H. Cohen}
\affiliation{School of Earth and Space Exploration, Arizona State University, 
PO Box 871404,
Tempe, AZ 85287-1404, USA}

\author[0000-0001-8156-6281]{Rogier A. Windhorst}
\affiliation{School of Earth and Space Exploration, Arizona State University, 
PO Box 871404,
Tempe, AZ 85287-1404, USA}

\author[0000-0003-1268-5230]{Rolf A. Jansen}
\affiliation{School of Earth and Space Exploration, Arizona State University, 
PO Box 871404,
Tempe, AZ 85287-1404, USA}

\author[0000-0001-7095-7543]{Min S. Yun}
\affiliation{Department of Astronomy, University of Massachusetts, Amherst, MA 01003, USA}

\author[0000-0003-0202-0534]{Cheng Cheng}
\affiliation{Chinese Academy of Sciences South America Center for Astronomy, National Astronomical Observatories, CAS, Beijing 100101, China}

\author[0000-0002-7265-7920]{Jake S. Summers}
\affiliation{School of Earth and Space Exploration, Arizona State University, 
PO Box 871404,
Tempe, AZ 85287-1404, USA}

\author[0000-0001-6650-2853]{Timothy Carleton}
\affiliation{School of Earth and Space Exploration, Arizona State University, 
PO Box 871404,
Tempe, AZ 85287-1404, USA}

\author[0000-0001-5429-5762]{Kevin C. Harrington}
\affiliation{European Southern Observatory, Alonso de C{\'o}rdova 3107, Vitacura, Casilla 19001, Santiago de Chile, Chile}

\author[0000-0001-9065-3926]{Jose M. Diego}
\affiliation{Instituto de F\'isica de Cantabria (CSIC-UC). Avda. Los Castros s/n. 39005 Santander, Spain}

\author[0000-0001-7592-7714]{Haojing Yan}
\affiliation{Department of Physics and Astronomy, University of Missouri, Columbia, MO 65211, USA}

\author[0000-0002-6610-2048]{Anton M. Koekemoer}
\affiliation{Space Telescope Science Institute, 3700 San Martin Dr., Baltimore, MD 21218, USA}

\author[0000-0001-9262-9997]{Christopher N. A. Willmer}
\affiliation{Steward Observatory, University of Arizona, 933 N Cherry Ave, Tucson, AZ, 85721-0009}

\author[0000-0003-4030-3455]{Andreea Petric}
\affiliation{Space Telescope Science Institute, 3700 San Martin Dr., Baltimore, MD 21218, USA}

\author[0000-0001-6278-032X]{Lukas J. Furtak}
\affiliation{Physics Department, Ben-Gurion University of the Negev, P.O. Box 653, Be'er-Sheva 84105, Israel}

\author[0000-0002-7460-8460]{Nicholas Foo}
\affiliation{University of Arizona, Department of Astronomy/Steward Observatory, 933 N Cherry Ave, Tucson, AZ 85721, USA}

\author[0000-0003-1949-7638]{Christopher J. Conselice}
\affiliation{Jodrell Bank Centre for Astrophysics, University of Manchester, Oxford Road, Manchester M13 9PL, UK}

\author[0000-0001-7410-7669]{Dan Coe} 
\affiliation{AURA for the European Space Agency (ESA)} 
\affiliation{Space Telescope Science Institute, 3700 San Martin Dr., Baltimore, MD 21218, USA}

\author[0000-0001-9491-7327]{Simon P. Driver}
\affiliation{International Centre for Radio Astronomy Research (ICRAR) and the International Space Centre (ISC), The University of Western Australia, M468, 35 Stirling Highway, Crawley, WA 6009, Australia}

\author[0000-0001-9440-8872]{Norman A. Grogin}
\affiliation{Space Telescope Science Institute, 3700 San Martin Dr., Baltimore, MD 21218, USA}

\author[0000-0001-6434-7845]{Madeline A. Marshall}
\affiliation{National Research Council of Canada, Herzberg Astronomy \& Astrophysics Research Centre, 5071 West Saanich Road, Victoria, BC V9E 2E7, Canada}
\affiliation{ARC Centre of Excellence for All Sky Astrophysics in 3 Dimensions (ASTRO 3D), Australia}

\author[0000-0003-3382-5941]{Nor Pirzkal}
\affiliation{Space Telescope Science Institute, 3700 San Martin Dr., Baltimore, MD 21218, USA}

\author[0000-0003-0429-3579]{Aaron S. G. Robotham}
\affiliation{International Centre for Radio Astronomy Research (ICRAR) and the International Space Centre (ISC), The University of Western Australia, M468, 35 Stirling Highway, Crawley, WA 6009, Australia}

\author[0000-0003-0894-1588]{Russell E. Ryan, Jr.}
\affiliation{Space Telescope Science Institute, 3700 San Martin Dr., Baltimore, MD 21218, USA}

\author[0000-0001-9052-9837]{Scott Tompkins}
\affiliation{School of Earth and Space Exploration, Arizona State University, 
PO Box 871404,
Tempe, AZ 85287-1404, USA}

\begin{abstract}
Gradients in the mass-to-light ratio of distant galaxies 
%
impede our ability to
characterize their size and compactness.
The long-wavelength filters of \JWST's NIRCam 
%
offer a significant step forward.
For galaxies at Cosmic Noon ($z\sim2$), 
this 
regime 
corresponds to the
rest-frame near-infrared,
which is less biased towards young stars and 
captures emission from the bulk of a galaxy's stellar population.
We present an initial analysis of an extraordinary lensed dusty star-forming galaxy (DSFG) at $z=2.3$ behind the {\it El Gordo} cluster ($z=0.87$), 
named {\it El Anzuelo} (``The Fishhook") after its 
partial Einstein-ring morphology.
%
The FUV-NIR 
SED
suggests an intrinsic star formation rate of $81^{+7}_{-2}~M_\odot~{\rm yr}^{-1}$ and dust attenuation $A_V\approx 1.6$, in line with other DSFGs 
on the star-forming main sequence.
We develop a parametric lens model 
to reconstruct the source-plane structure of 
dust 
imaged by the Atacama Large Millimeter/submillimeter Array, far-UV to optical light from {\it Hubble}, and 
%
near-IR imaging with 8 filters of {\it JWST}/NIRCam, as part of 
the 
Prime Extragalactic Areas for Reionization and Lensing Science (PEARLS) program. 
%
The source-plane half-light radius is remarkably consistent from $\sim 1-4.5~\mu$m, despite a clear color gradient where the inferred galaxy center is redder than the outskirts. 
We interpret this to be the result of both a radially-decreasing gradient in 
attenuation
and substantial spatial offsets between UV- and IR-emitting components.
A 
spatial decomposition of the 
SED
reveals modestly suppressed star formation in the inner kiloparsec,
%
which suggests that we are witnessing the early stages of inside-out quenching. 
\end{abstract}



\section{Introduction} 
\label{sec:intro}

The recent advent of 
\JWST\
has already offered substantial new insight into the Universe at the epoch of Cosmic Noon, $1 \lesssim z \lesssim 3$, during which the cosmic star-formation history was at its zenith \citep{Madau:2014aa}. Dusty star-forming galaxies (DSFGs)\textemdash also referred to historically as submillimeter galaxies, or SMGs (see reviews by \citealt{Blain:2002aa,Casey:2014aa})\textemdash have been studied extensively in the past few decades 
as tracers of 
the rapid stellar mass assembly 
of the Universe
that peaked 10 billion years ago.
Their substantial shrouds of dust are efficient at reprocessing the intense ultraviolet (UV) and visible light from sites of rapid, active star formation into infrared (IR) radiation, resulting in large observed-frame (sub)-millimeter fluxes. 
Yet, their dust content also has the effect of 
strongly 
attenuating and reddening their rest-frame UV/optical emission, such that it has been a challenge to study them with even the most sensitive optical telescopes, including the {\it Hubble Space Telescope} (\HST). 
When detected, the peak of UV emission in DSFGs is often significantly offset from that of the far-IR (e.g., \citealt{Goldader:2002aa, Chapman:2004aa,  Hodge:2015aa, Chen:2017aa, Calistro-Rivera:2018aa, Cochrane:2021aa}).
%
Yet, a subset of these objects have gone undetected in deep $H$-band imaging (to limiting $5\sigma$ depths of $H > 27$ mag), leading to their classification as ``HST-dark" or ``OIR-dark" galaxies (e.g., \citealt{Huang:2011aa, Simpson:2014aa, Franco:2018aa, Wang:2019ab}, and references therein). \JWST\ has proven capable of shedding new light on these elusive galaxy populations, which will offer substantial refinements to theories of galaxy evolution (e.g., \citealt{Barrufet:2023ab, Ferreira:2022aa}). 

Galaxy size is a key observable property that has been used to advance our understanding of the assembly history of galaxies throughout cosmic time \citep{Mo:1998aa}. In particular, the size-mass relation ($M_\star-R_e$; e.g., \citealt{Shen:2003aa, Trujillo:2004aa, van-der-Wel:2014aa, Mowla:2019aa})
indicates that more massive galaxies are typically larger in size, with markedly steeper slopes for quiescent (late-type) galaxies than for star-forming (early-type) galaxies. 
\citet{Ryan:2012aa} found that the evolution of the size of passive galaxies with redshift is itself stellar mass-dependent (such that the most massive galaxies experienced the greatest evolution from $z$$\sim$2 to the present).
However, it is not trivial to assign the effective radius of a galaxy, as the distribution of light is known to vary strongly with wavelength, such that $K$-band sizes can be as small as half those at $r$-band
(e.g., \citealt{Evans:1994aa, Cunow:2001aa, La-Barbera:2002aa, La-Barbera:2010aa, Kelvin:2012aa, Haussler:2013aa, Vulcani:2014aa, Kennedy:2015aa, Kennedy:2016aa}).
This often results in 
radial color gradients, or variations in the mass-to-light ratios, which are usually explained by some combination of\footnote{Observation-specific effects, such as unmatched PSFs at different wavelengths, can also be responsible.}:
\begin{itemize}
    \item centrally-concentrated dust, leading to greater attenuation in the inner regions of a galaxy (e.g., \citealt{Jansen:1994aa, Peletier:1996aa, Mollenhoff:2006aa, Graham:2008aa, Pastrav:2013aa, Pantoni:2021ab})
    \item inside-out galaxy growth and quenching \citep{Carrasco:2010aa, van-Dokkum:2010aa, Hopkins:2010ab}, by which stellar populations in the center are predominantly older, especially when a spheroidal/bulge component is present (e.g., \citealt{Driver:2007aa, Driver:2007ab, La-Barbera:2010aa, Marian:2018aa})
    \item gradients in metallicity, often driven by inflow of primordial gas to galaxies' gravitational potential (e.g., \citealt{Cresci:2010aa, Jones:2010ab, Yuan:2011aa, Pilkington:2012aa})
\end{itemize}
Typically, galaxies with color gradients are redder in the center, but this is especially true for more luminous objects \citep{Jansen:2000aa} and for late-type disk-dominated galaxies \citep{de-Jong:1996aa, Gadotti:2001aa}.

In effect, the 
implication
of these color gradients is that optical filters are not always effective proxies for the underlying stellar mass distribution, and the mass-to-light ratio can vary with radius in even a single galaxy (e.g., \citealt{Suess:2019aa}).
More worrisome is the finding by \citeauthor{Suess:2019aa} that the amplitude of color gradients can correlate with properties like stellar mass or surface mass density, such that our interpretation of scaling relations (like $M_\star-R_e$) that depend on light-weighted sizes might be impacted significantly.  
Rest-frame near-infrared light, on the other hand, arises primarily from 
stellar populations of cooler, low-mass stars 
that make up most of a galaxy's stellar mass. 
Additionally, near-IR is comparatively much less affected by the dust attenuation that plagues the rest-frame UV and bluer optical regime (by approximately an order of magnitude; e.g., \citealt{Li:2001aa}).
Until recently, however, observations of the high-$z$ Universe at this wavelength have been limited to unresolved, galaxy-integrated studies.

Now, \JWST\ has enabled resolved measurements of the light distribution at wavelengths beyond $\sim$2 $\mu$m in $z>1$ galaxies for the very first time. 
In an examination of $\sim 1000$ galaxies from the Cosmic Evolution Early Release Science program (CEERS; see \citealt{Finkelstein:2023aa}), \citet{Suess:2022aa} found that galaxies were systematically smaller at 4.4 $\mu$m ($\lambda_{\rm rest}\sim1.6 \mu$m) than at 1.5 $\mu$m ($\lambda_{\rm rest}\sim550$ nm) by about 9\% on average (but reaching $\sim 30\%$ for high-mass galaxies, $M_\star \sim 10^{11}~M_\odot$). 
It is now feasible to directly probe the distribution of stellar mass in the extragalactic Universe with the long-wavelength filters of \JWST's NIRCam instrument, without needing to model the galaxy spectral energy distribution (SED) or infer any gradients in mass-to-light ratio (e.g., \citealt{Suess:2019aa}).

\begin{figure}[ht!]
\plotone{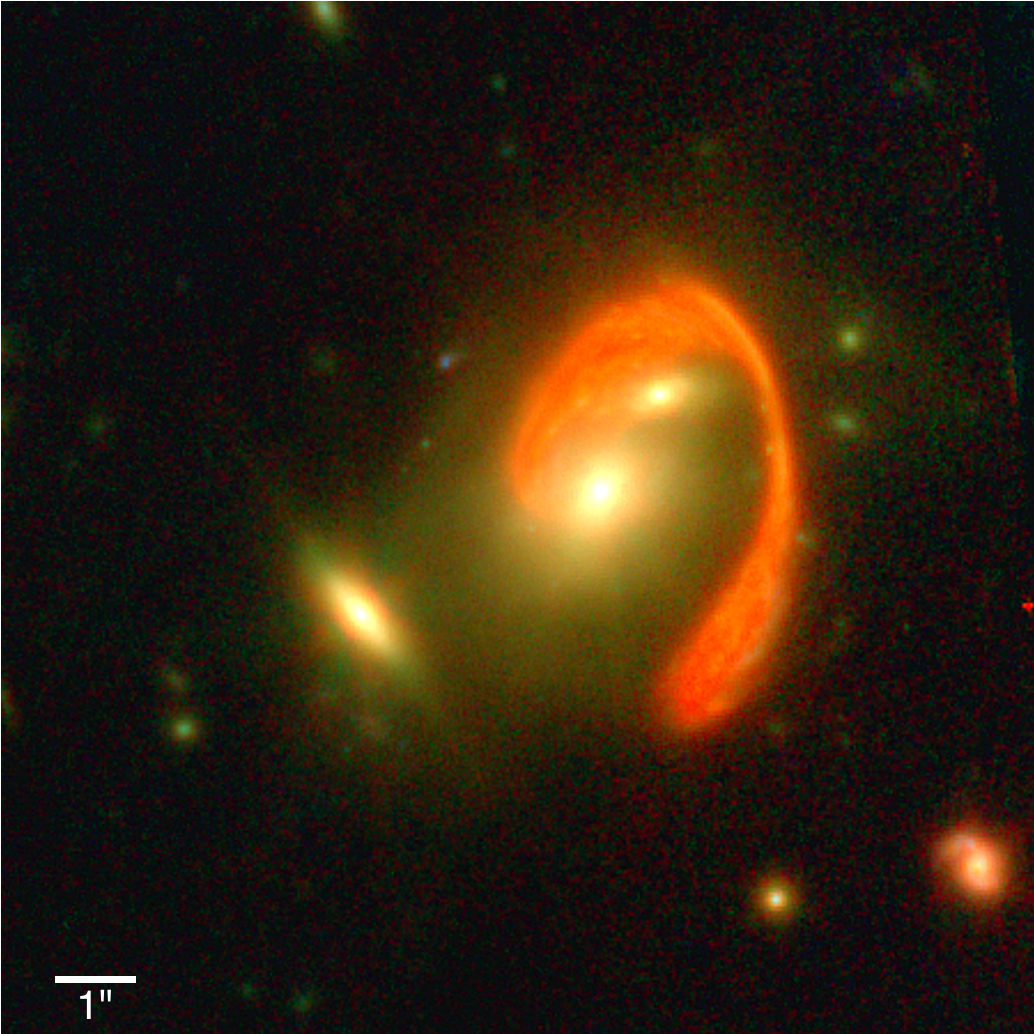}
\caption{
    RGB image of \ElA, which is aligned with north, $13.5\arcsec$ on each side, and centered on $(\alpha, \delta)=$ (01$^h$02$^m$49\fs 452,$-49^d$15$^m$07\fs 05). 
    For this \JWST-only image, the red, green, and blue channels are constructed by weighting filters as follows: $R = [(0.3 \times {\rm F356W}) + (0.6 \times {\rm F410M}) + (1.0 \times {\rm F444W})]$; $G = [(0.8 \times {\rm F200W}) + (1.0 \times {\rm F277W}) + (0.5 \times {\rm F356W})]$; $B = [(1.0\times {\rm F090W}) + (1.0 \times {\rm F115W}) + (1.0 \times {\rm F150W}) + (1.0 \times {\rm F200W}) + (1.0 \times {\rm F277W}) + (0.8 \times {\rm F356W}) + (0.5 \times {\rm F410M}) + (0.5 \times {\rm F444W})]$.
    \label{fig:cutout}
}
\end{figure}

To fully take advantage of this new capability, the best approach is to use \JWST\ to observe objects affected by strong gravitational lensing. The amplification of flux helps to facilitate the mapping of the stellar continuum out to even larger radii, where the light profile would otherwise fall below the detection limit. This amplification also allows one to probe less-luminous galaxies that are more representative of the luminosity function at $z>1$. Of paramount importance to this work
is the areal magnification provided by lensing, which helps push angular resolution beyond even the \JWST\ diffraction limit, albeit with a complex pattern of distortion that must be accounted for.
A prime opportunity for using \JWST\ to map the assembly of stellar mass at high-redshift is with the Prime Extragalactic Areas for Reionization and Lensing Science (PEARLS) program, which targets seven massive lensing clusters \citep{Windhorst:2023aa}. One such cluster, ACT-CL J0102-4915, or ``{\it El Gordo}" \citep{Menanteau:2012aa}, has been identified as a foreground for a rich galaxy group at $z$=4.3 \citep{Caputi:2021aa, Frye:2023aa}, including DSFGs (see also \citealt{Cheng:2023aa}). 
In this work, 
we focus on an extraordinary, recently-discovered lensed DSFG J010249-491507 (shown in Fig.~\ref{fig:cutout}), which we find to be near the peak of Cosmic Noon, at $z$=2.291.
The two primary lensed images were denoted as EG-ALMA2a and EG-ALMA2b by \citet{Cheng:2023aa}. \citet{Diego:2023ab} first gave this object the moniker \ElA\ (Spanish for ``Fishhook," given the shape of the lensing morphology), which we will use for the remainder of this work. 
This lensed system has been revealed in much greater detail than was possible in previous optical imaging with {\it Hubble}. Thus, we consider it an excellent prospect for an in-depth analysis of mapping the light vs. mass profile in an extreme star-forming object\textemdash all with superb, sub-diffraction-limit angular resolution thanks to lensing.

This paper is organized as follows: in Section \ref{sec:data}, we describe the \JWST, {\it Hubble} Space Telescope (\HST), and Atacama Large Millimeter/submillimeter (ALMA) data used in this work.
In Section \ref{sec:analysis}, we detail our analysis, including lens modeling, subtraction of foreground light, photometric redshift estimation, characterization of source-plane size and light profile, and determination of the (wavelength-dependent and location-dependent) magnification.
In Section \ref{sec:discussion}, we discuss this differential magnification, and weigh the evidence of various effects that might be responsible for the observed radial color gradients. We discuss a decomposition of the spectral energy distribution for the inner- versus outer-disk, and what this reveals about possible inside-out galaxy quenching.
Our conclusions are summarized in Section \ref{sec:conclusions}.

Throughout this work, we use a concordance $\Lambda$CDM cosmological model, with $\Omega_m=0.3$, $\Omega_\Lambda=0.7$, and $H_0 = 70~{\rm km~s}^{-1}~{\rm Mpc}^{-1}$. At the redshift of \ElA\ ($z=2.291$), the angular-to-physical size conversion with this cosmology is $1\arcsec=8.209$ kpc, and at the redshift of the \ElG\ cluster ($z=0.87$), the conversion is $1\arcsec=7.713$ kpc \citep{Wright:2006aa}. The magnitudes in this paper are given according to the AB system. All images are aligned with north up and east to the left, unless otherwise noted.

\section{Data} 
\label{sec:data}

\subsection{Object selection}
\label{sec:object}

\ElA\ was identified as a candidate for an in-depth analysis because of the relative simplicity of the foreground lensing environment, in contrast with the large-scale cluster mass profile.
As there are fewer cluster galaxies contributing significantly to the lensing deflection, we can derive a robust model with few parameters. Moreover, we can fine-tune the model to minimize the separations between observed and reconstructed multiple image locations for this lensed object alone, rather than a large collection of image systems (e.g. \citealt{Diego:2023ab, Frye:2023aa}).
As we discuss in \S \ref{sec:EAZY}, we adopt a source redshift of $z=2.291$ based on SED-fitting and a CO line detection.

\subsection{JWST NIRCam} 
\label{sec:jwst}

The {\it El Gordo} cluster was selected to be included in \JWST's %
PEARLS program \citep{Windhorst:2023aa} because of its extremely large halo mass, on the order of $M_h$ $\sim$ 2$\times10^{15}~\Msun$ \citep{Menanteau:2012aa}. 
A more detailed description of this data, collected on 2022 July 29, and its subsequent reduction are provided in \citet{Windhorst:2023aa}, \citet{Diego:2023ab}, and \citet{Frye:2023aa}, but we summarize relevant details here.
\ElG\ was observed
in the wide-band filters F090W, F115W, F150W, F200W, F277W, F356W, F444W, and the medium-band filter F410M. Exposure times were 2491 seconds for F090W, F115W, F410M, and F444W; 2104 seconds for F200W and F277W; and 1890 seconds for F150W and F356W. 
The data were processed with pipeline version 1.6.2 and the context file jwst\_0942.pmap.
The $5\sigma$ limiting magnitudes for all filters are in the range $m_{\rm AB} 
\simeq 28 - 29$ mag.

\subsection{HST} 
\label{sec:hst}

Existing ancillary imaging with the {\it Hubble} Space Telescope (\HST) provides helpful constraints for the rest-frame UV regime in \ElA.
The filters used in this analysis include 
Advanced Camera for Surveys (ACS) Wide Field Channel (WFC) 
F606W, F625W, F775W, F814W, F850LP; and 
Wide Field Camera 3 (WFC3) IR channel F105W, F125W, F140W, and F160W. 
These data were collected as part of the Reionization Lensing Cluster Survey\footnote{\url{https://relics.stsci.edu/}} (RELICS), through \HST\ programs GO 12755 (PI: J. Hughes), GO 12477 (PI: F. High), and GO 14096 (PI: D. Coe). The observations were taken in 2012 September, 2012 October, and 2016 July/August, respectively. 
Data were retrieved from the Mikulski Archive for Space Telescopes\footnote{\url{https://archive.stsci.edu/}} (MAST).
We refer the reader to \citet{Coe:2019aa} for a more complete description of these observations.

\subsection{ALMA Bands 3, 6, and 7} 
\label{sec:ALMA}

\ElA\ in the \ElG\ field was observed with the Atacama Large Millimeter/submillimeter Array (ALMA) in programs 2015.1.01187.S (PI: K. Basu) and 2017.1.01621.S (PI: K. Basu) in Band 3 (3 mm or 100 GHz), 2013.1.01358.S (PI: A. Baker) and 2018.1.00035.L (PI: K. Kohno) in Band 6 (1 mm or 270 GHz), and 2013.1.01051.S (PI: P. Aguirre) in Band 7 (870 $\mu$m or 340 GHz). All data are public and were retrieved from the ALMA archive\footnote{\url{https://almascience.nrao.edu/aq/}}.

The reduced 3 mm data are presented by \citet{Basu:2016aa}. Coarse-resolution imaging with a synthesized beam size of $3.6\arcsec \times 2.7\arcsec$ (PA $= 88\degrees$) was taken in 2015 December with the most compact antenna configuration, with a noise rms of 6 $\mu$Jy beam$^{-1}$. While designed as a continuum experiment, these data were used for spectroscopic redshift confirmation of the background DSFG (see \S \ref{sec:EAZY}). The angular resolution is not sufficient to perform a reasonable source-plane reconstruction.

The two 1 mm measurement sets were combined in the complex visibility (i.e., Fourier) space and imaged together by \citet{Cheng:2023aa}. The resulting mosaic has a beam size of $1.28\arcsec \times 0.93\arcsec$ (PA $=88\degrees$) and a rms sensitivity of 
$\sim$0.065 mJy beam$^{-1}$. Likewise, the 870 $\mu$m map has a beam size of $0.33\arcsec \times 0.40\arcsec$ (PA $=-35\degrees$), with noise rms 0.25 mJy beam$^{-1}$. For both the 870 $\mu$m and 1mm data, images were created in \textsc{casa} v.6.2.1 \citep{McMullin:2007aa} with Briggs weighting (\textsc{robust}$=2$) and cleaned with \textsc{tclean} down to the $3\sigma$ level.

\section{Analysis} 
\label{sec:analysis}

For this case study of \ElA, we develop a gravitational lens model, measure filter-dependent magnifications, reconstruct the observed data in the source plane, and characterize the spatial extent as a function of wavelength.

\startlongtable
\begin{deluxetable*}{c|ccccc}
\tablecaption{Lens modeling results for \ElA. All potentials are held fixed at the cluster redshift of $z=0.87$. \label{tab:lensmodel}}
\tablehead{
\colhead{Potential ID \& type} & \colhead{RA} & \colhead{Dec} & \colhead{$e$} & \colhead{PA} & \colhead{$\sigma$ (or $\gamma$)}\\
\colhead{} & \colhead{[h:m:s]} & \colhead{[d:m:s]} & \colhead{} & \colhead{[$^\circ$]} & \colhead{[km s$^{-1}$]}
}
\startdata
L1 (SIE) &  1:02:49.347$^\dagger$   &   -49:15:06.60$^\dagger$	&	0.16 $\pm$ 0.13		& 27 $\pm$ 39	 & 302.6 $\pm$ 4.6	\\
L2 (SIE) &  1:02:49.269$^\dagger$  &  -49:15:05.33$^\dagger$	&	0.44 $\pm$ 0.19$^\ddagger$		& 71 $\pm$ 29	 & 219.7 $\pm$ 7.4	\\
L3 (SIE) &  1:02:49.665$^\dagger$  &  -49:15:08.22$^\dagger$	&	0.75$^\dagger$ 	& -50$^\dagger$	 & 27 $\pm$	68$^\ddagger$\\
L4 (ex. shear) &  \textemdash  &  \textemdash & \textemdash 	& 99 $\pm$ 6	 & 0.36 $\pm$ 0.08	\\
\enddata
\tablenotetext{^\dagger}{Parameter value held fixed. For lens L3, the ellipticity and orientation are set at those derived with \GALFIT, as the lens model appears to be quite insensitive to the mass at this projected distance.}
\tablenotetext{^\ddagger}{Parameters for which there is a sizable disagreement between the highest-likelihood and median solution include the ellipticity of L2 (which has a median $e = 0.23$) and the velocity dispersion of L3 (which has a median $\sigma = 98$ km s$^{-1}$).} 
\tablecomments{
Here we report the parameter values for the highest-likelihood MCMC iteration. These are consistent within 1$\sigma$ uncertainties with the median of the posterior distribution except for the two cases mentioned above. Where possible in this work, we account for the uncertainty inherent to our lens model by randomly selecting from the full posterior, as discussed in \S \ref{sec:mag_by_filter}.}
\end{deluxetable*}

\subsection{Lens modeling} 
\label{sec:lens_modeling}

\begin{figure*}[ht!]
\centering
\includegraphics[width=0.3\textwidth]{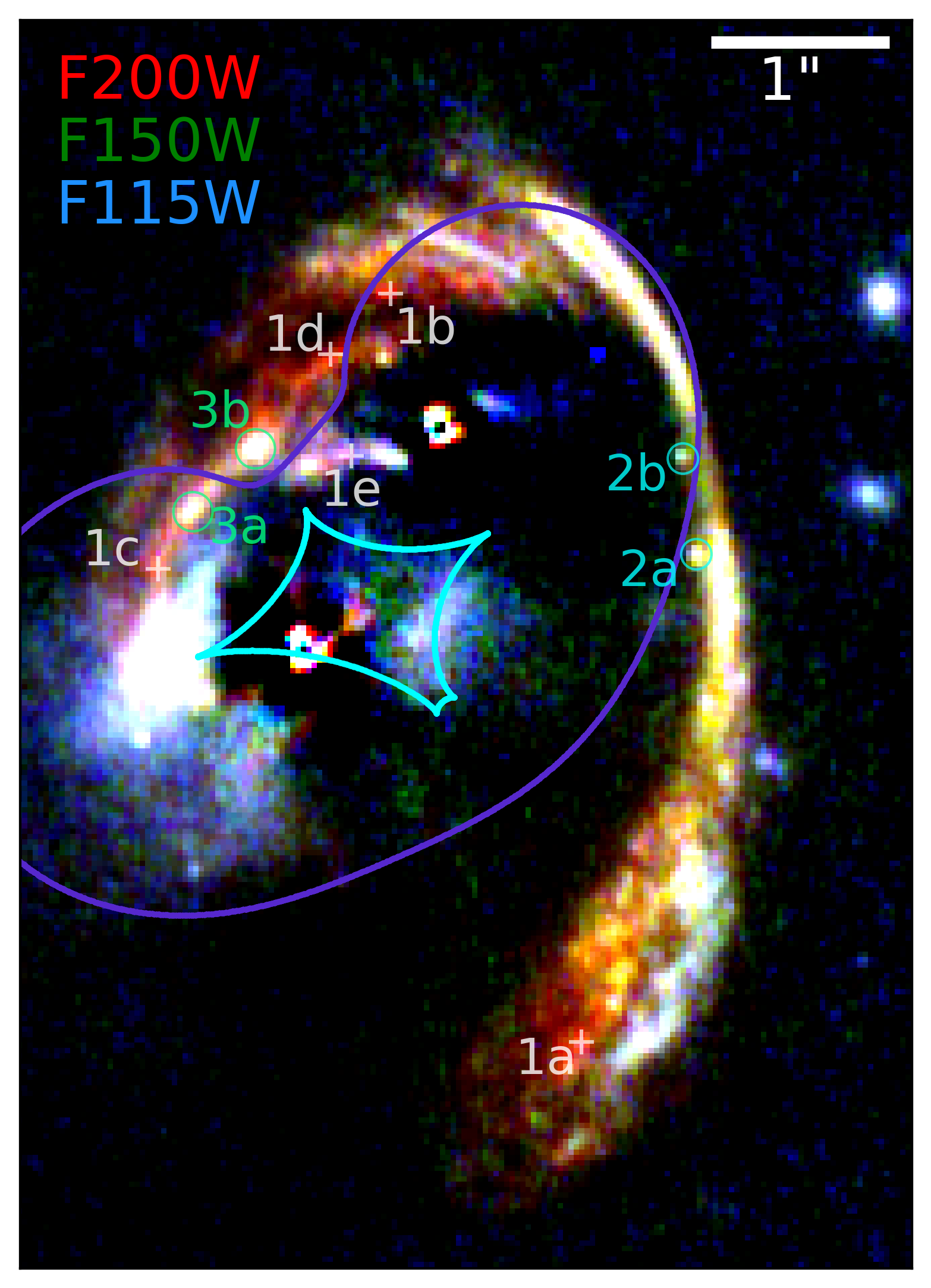}
\includegraphics[width=0.3\textwidth]{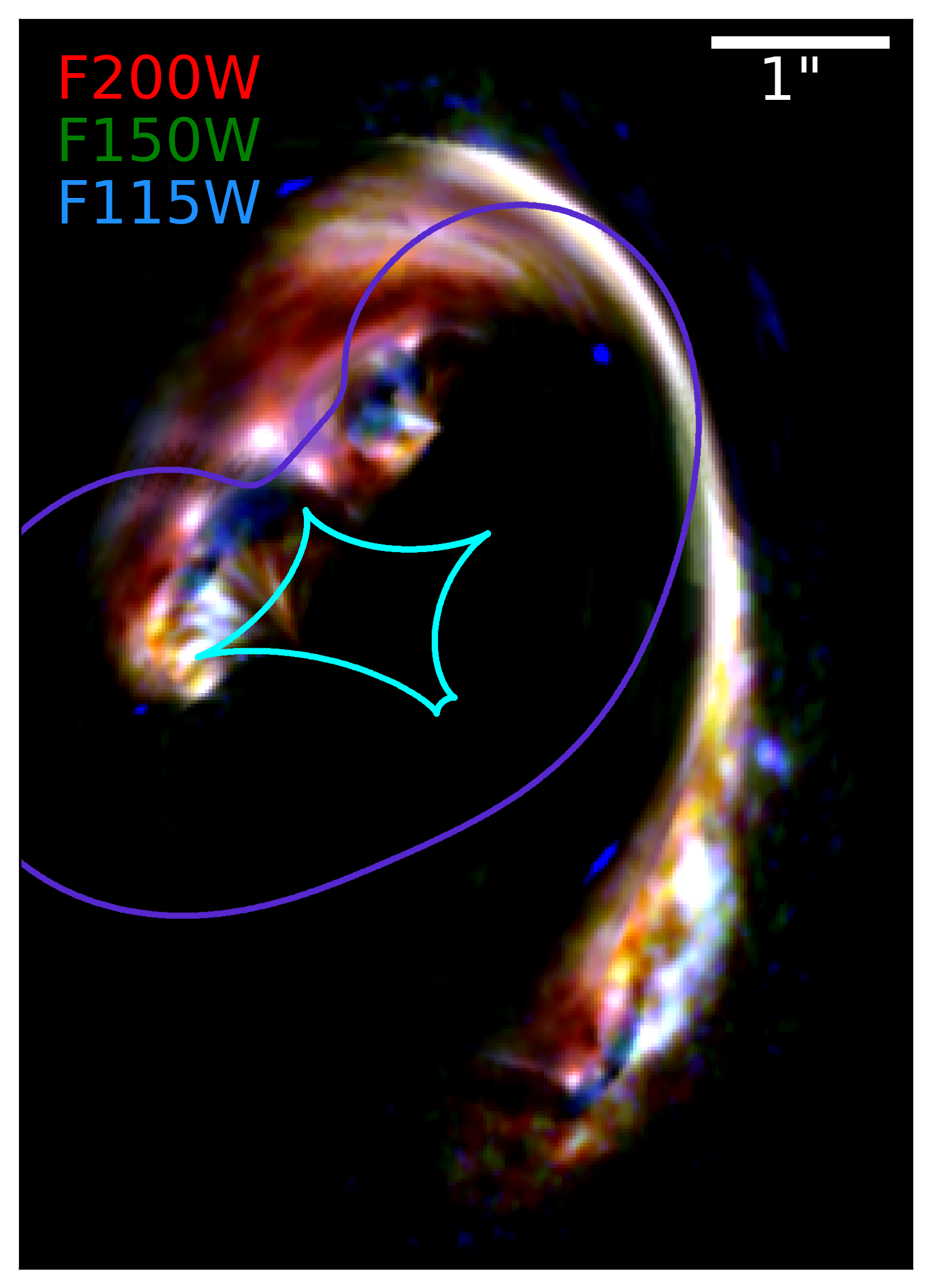}
\raisebox{0.1\height}{\includegraphics[width=0.35\textwidth]{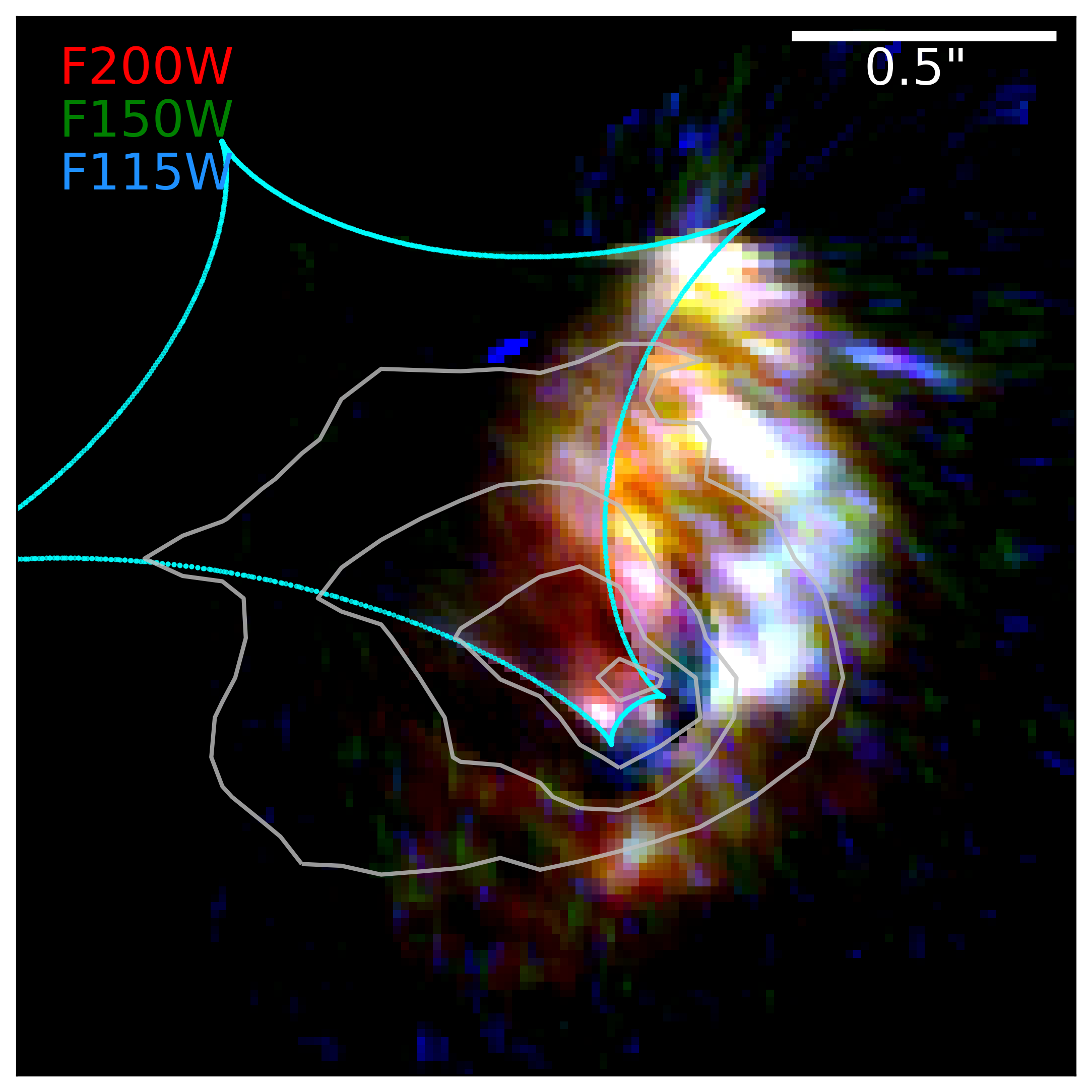}}
\caption{Observed and modeled image plane light distribution for three short-wavelength NIRCam filters in an RGB image (red: F200W, green: F150W, blue: F115W). The caustic curve is shown in cyan, and the corresponding image-plane critical curve is shown in violet.
    {\it Left:} Observed image, after removing foreground lens galaxy light with \GALFIT. Any residuals nearest the foreground galaxies (i.e., for F115W) are not of great concern, as they are masked out during the ray-tracing. Identified image families used as constraints in the lens model are labeled, including $1abcde$ (identified primarily with the long-wavelength NIRCam filters) $2ab$, and $3ab$.
    {\it Center:} Model-reconstructed image-plane distribution of the same filters, made by ray-tracing the observed data to the source-plane with the best-fit model, then ray-tracing back to the image plane (and finally, convolving with a smoothing kernel to remove ray-tracing discontinuity artifacts). 
    {\it Right:} Model-reconstructed source-plane distribution, zoomed-in to show more spatial detail. Gray contours show ALMA 1 mm continuum, ray-traced to the source-plane.
    \label{fig:IP_SW_reconstruction}
}
\end{figure*}

\begin{figure*}[ht!]
\centering
\includegraphics[width=0.3\textwidth]{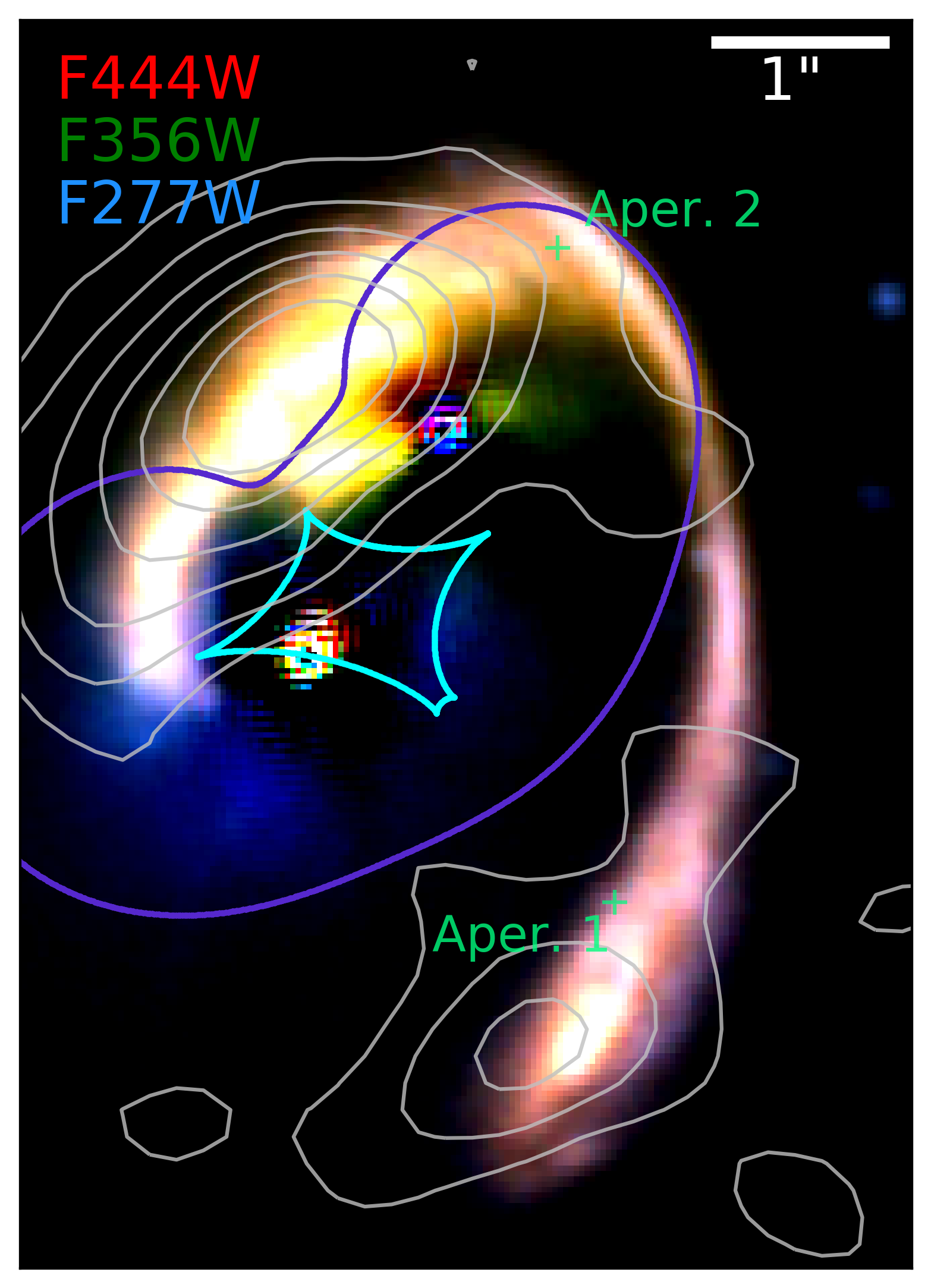}
\includegraphics[width=0.3\textwidth]{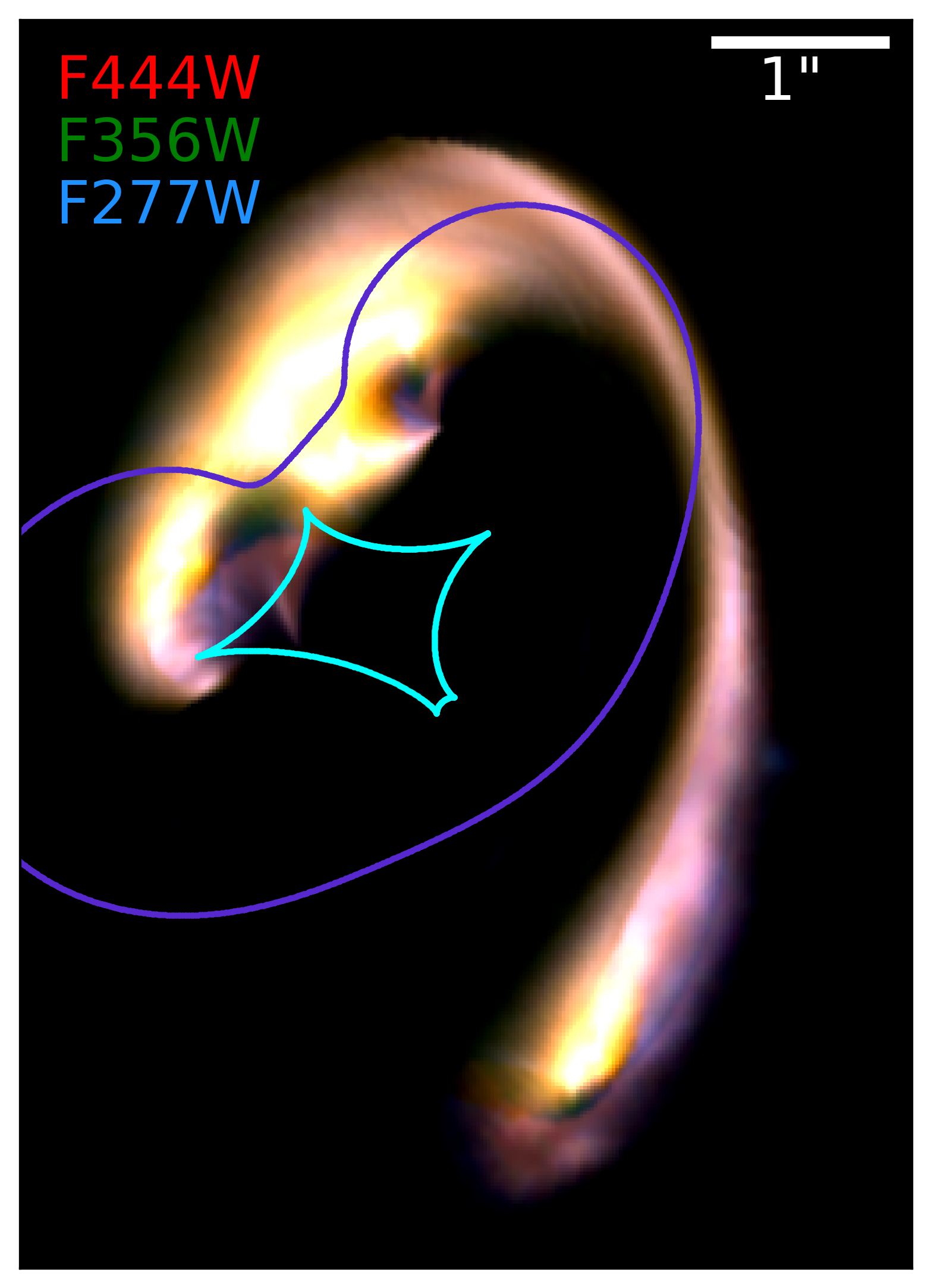}
\raisebox{0.1\height}{\includegraphics[width=0.35\textwidth]{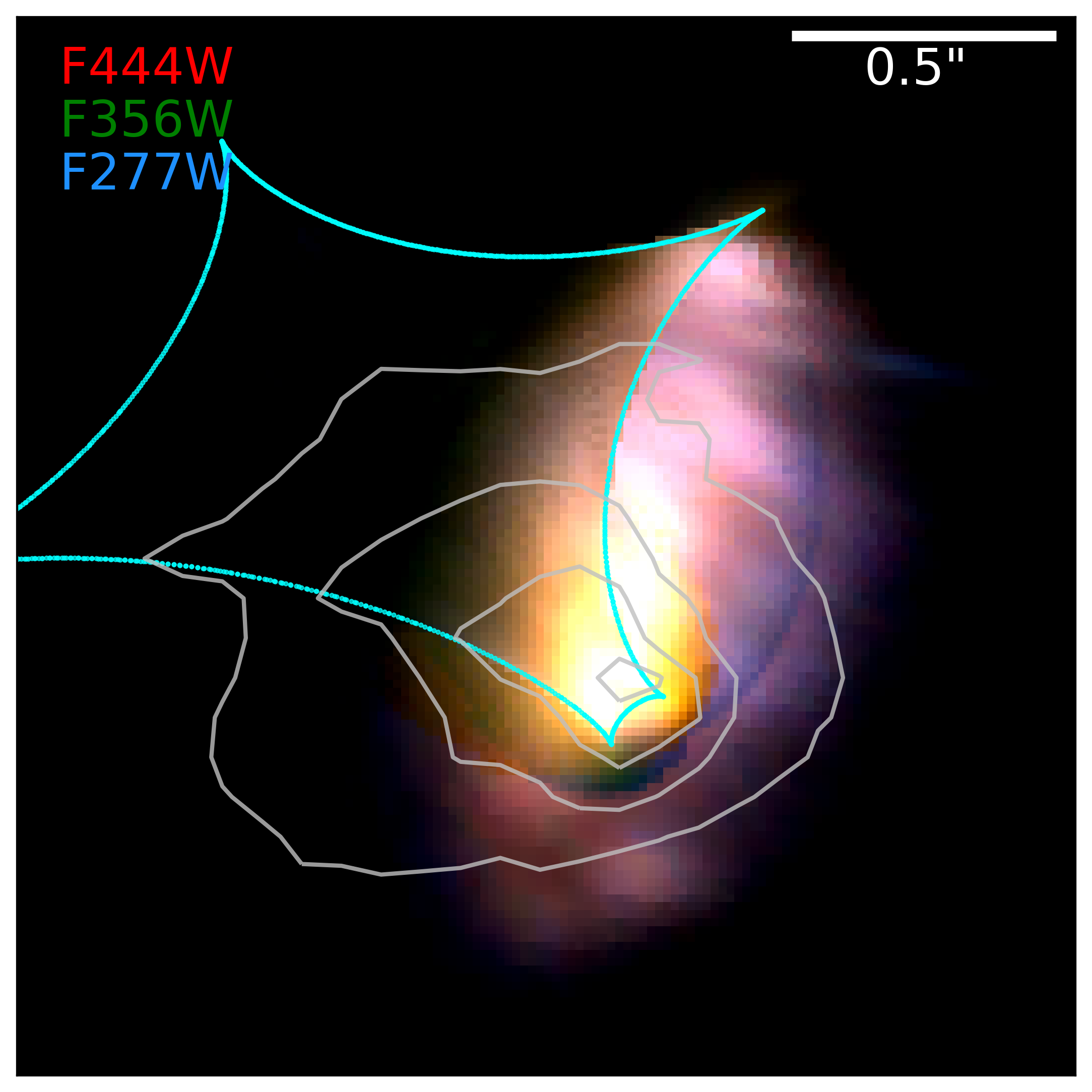}}
\caption{
    As with Fig.~\ref{fig:IP_SW_reconstruction}, but for 3 long-wavelength NIRCam filters: F444W (red), F356W (green), F277W (blue). In the observed image (left), we also show the image-plane ALMA 1 mm contours in gray. Green crosses identify the locations of apertures used for photometry, discussed in \S \ref{sec:photometry}.
    \label{fig:IP_LW_reconstruction}
}
\end{figure*}

The lensing distortion of \ElA\ is dominated primarily by a small group of cluster galaxies, and not by the cluster halo itself, as evidenced by its small Einstein radius of $\theta_{E}=1\farcs70\pm0\farcs06$. For this reason, we construct a separate lens model (in parallel to those developed recently by \citealt{Diego:2023ab} and \citealt{Frye:2023aa}) that is focused only on \ElA, so as to tailor the optimization to a small number of constraints and parameters.

Parametric gravitational lens modeling was performed with \lenstool\footnote{\url{https://projets.lam.fr/projects/lenstool/wiki}} \citep{Kneib:1993aa, Kneib:1996aa, Jullo:2007aa, Jullo:2009aa}.
\lenstool\ solves for the simple lens equation (see e.g., \citealt{Schneider:1992aa, Narayan:1996aa}), which relates the source-plane distribution of light with the observed image-plane distribution via the deflection angle, which is derived from an integral of the surface mass density in the lens plane. With a set of observed locations for a multiply-imaged family, one can constrain the foreground mass distribution (where each separate set of $n$ images provides $2n-2$ constraints). This is done completely independently of the light profile of the foreground mass, although this is used for setting priors on each parameter, and for establishing the number of foreground lenses to include (and sometimes the nature of each profile).

\lenstool\ uses a Bayesian Markov chain Monte Carlo algorithm to sample the multi-dimensional posterior distribution for the set of free parameters. The likelihood function connects the image-plane positions of multiple images used as constraints, and the resulting positions of these locations obtained by ray-tracing to the source-plane and back to the image-plane with each model iteration (as shown in Figs.~\ref{fig:IP_SW_reconstruction} and \ref{fig:IP_LW_reconstruction}). 
We use 3000 iterations to extensively sample the posterior space. The best-fit parameters are provided in Table~\ref{tab:lensmodel}, which result in an RMS deviation between observed and modeled image positions of $0.06\arcsec$. This value closely matches the FWHM of the PSF for F444W, indicating a high goodness-of-fit. While one could choose the median or mode of each parameter's posterior distribution as the solution, we opt for the highest-likelihood (minimum $\chi^2$) set of parameters, after verifying that they are consistent with median and mode values within uncertainties.

The results of our lens models are depicted in Figs.~\ref{fig:IP_SW_reconstruction} and \ref{fig:IP_LW_reconstruction}, where the panels show RGB images of the observed light distribution (after subtracting foreground light with \GALFIT, as described in Section \ref{sec:GALFIT}) for the short-wavelength and long-wavelength filters separately. The middle panels show a model-reconstructed image-plane, created by ray-tracing the original observed image into the source plane, before then ray-tracing this result back to the image plane (and smoothing by a kernel equal to half the PSF FWHM to reduce artifacts). As lensing is not a one-to-one transformation\textemdash e.g., ray-tracing one multiply-imaged feature to the source-plane and back to the image-plane should recover all members of the image family\textemdash this translation to the source-plane and back is likely to introduce unwanted features in the case of poor goodness-of-fit. 
In the right panels, we show the source-plane reconstruction itself.
In Fig.~\ref{fig:IP_ALMA_reconstruction}, we show the same results for the ALMA 870 $\mu$m and 1 mm continuum images.

\startlongtable
\begin{deluxetable}{c|cc}
\tablecaption{Locations of secure multiple image family members that were used in the lens model, which are also labeled in Figs.~\ref{fig:IP_SW_reconstruction} and \ref{fig:model_constraints}. 
\label{tab:lens_constraints}}
\tablehead{
\colhead{Image ID} & \colhead{RA} & \colhead{Dec}\\
\colhead{} & \colhead{[h:m:s]} & \colhead{[d:m:s]} 
}
\startdata
1a &  1:02:49.188   &   -49:15:08.77	\\
1b &  1:02:49.298  &  -49:15:04.58		\\
1c &  1:02:49.431  &  -49:15:06.12	\\
1d &  1:02:49.332  &  -49:15:04.92 	\\
1e & 1:02:49.320  &  -49:15:05.49 	\\
\hline
2a &  1:02:49.122   &   -49:15:06.04	\\
2b &  1:02:49.130  &  -49:15:05.50		\\
\hline
3a &  1:02:49.411   &   -49:15:05.80	\\
3b &  1:02:49.375  &  -49:15:05.45		\\
\enddata
\end{deluxetable}

\subsubsection{Lens model parameters} 
\label{sec:lens_params}

Here, we have used a common approach where each foreground elliptical is parameterized as a singular isothermal ellipsoid (SIE) or spheroid (SIS), which is efficient for describing galaxy- or group-scale lensing (e.g., \citealt{Kormann:1994aa, Treu:2004aa}).
We also include an external shear term to account for any weak lensing effect from the \ElG\ cluster. However, recent work by \citet{Etherington:2023aa} suggests that the vast majority of models with external shear are not actually affected by weak lensing, and this added factor only compensates for insufficient model complexity with the chosen parameterization (albeit, this result is primarily based on isolated field galaxies). 
We note that the best-fit orientation of the external shear term is PA$\approx$99$^\circ$ (counterclockwise from east), which is not closely aligned with the major or minor axis of the most massive foreground galaxy (PA$\approx$27$^\circ$), so we preserve this as a free parameter, with the acknowledgment that this is not necessarily representing actual external shear.

\begin{figure*}[ht!]
\includegraphics[width=\textwidth]{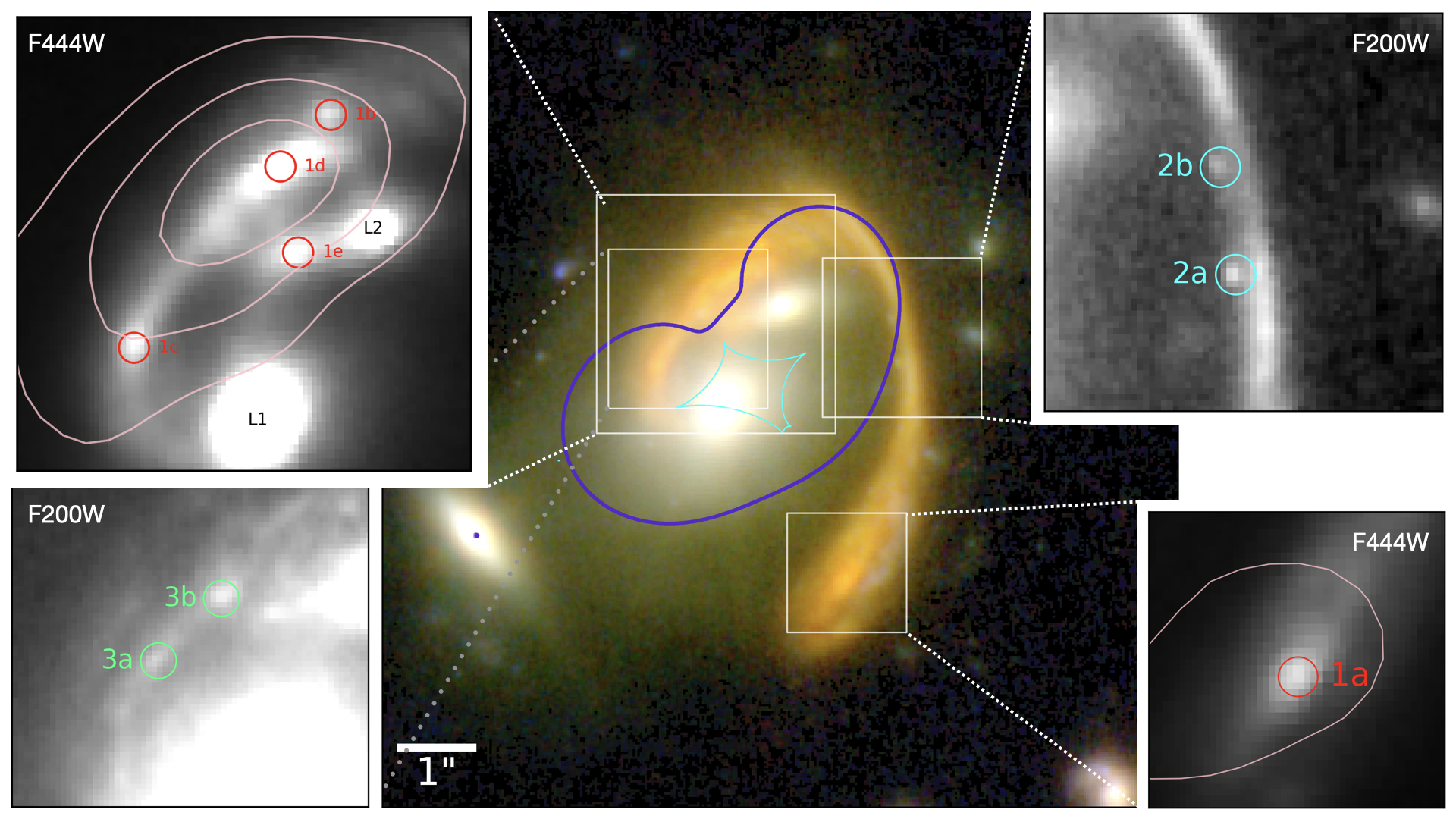}
\caption{
    \label{fig:model_constraints}
    Image families used as constraints for the lens model. The center panel shows an RGB image of all 8 NIRCam filters, with the lens model-derived caustic and critical curves shown in cyan and purple. Insets show the identified image families $1abcde$ (most prominent in F444W) and $2ab$ and $3ab$ (prominent in F200W). Pink contours in the $1abcde$ cutouts show ALMA 1mm emission at levels of $[5,10,15]\sigma$. 
}
\end{figure*}

We identify 3 image families (containing multiplicities of 2, 2, and 5 images, respectively) labeled in Fig.~\ref{fig:model_constraints}. 
With $2(n-1)$ constraints provided by each family of $n$ images, this results in 8 constraints from image set $1abcde$ and 2 constraints each from $2ab$ and $3ab$.
The locations of each image are provided in Table~\ref{tab:lens_constraints}. 
The observed dust continuum structure suggests an overall double-image morphology with a southwest and northeast component, but precise image locations are chosen carefully from NIRCam imaging. 
In particular, the structure in F444W has a prominent peak in the southwest image ($1a$), which appears as multiple peaks with a similar color in the northeast side of the arc, which we identify as multiple images of the same region. An initial, oversimplified mass model of the two primary foreground galaxies\textemdash with centroids, ellipticities, and orientations held fixed to the best-fit \GALFIT\ results (and only velocity dispersions as free parameters)\textemdash predicted a similar five-image morphology, helping to confirm our identification.

For image families $2ab$ and $3ab$, F200W in particular reveals an observable symmetry that arises from images in close proximity on opposite sides of the critical curve. The pairs of prominent clumps of similar size and luminosity are circled as $2ab$ and $3ab$ in Fig.~\ref{fig:model_constraints}. These are incorporated into the model iteratively, by first using the initial mass model to determine that these pairs of clumps could logically lie in the vicinity of the critical curve. Their eventual inclusion in the model helps to refine and further constrain the mass distribution beyond just the sightlines probed by images $1abcde$, but adding them does not significantly alter the best-fit parameters. 
It may be feasible to identify additional multiply-imaged features, but this set of secure constraints is conservative and more than sufficient given the simplicity of our lens model parameterization, in contrast with the models of the full \ElG\ cluster.
In the future, resolved spectral line observations would add further confirmation of these image identifications (which should share similar line profiles). However, the fidelity of the image-plane reconstructions in Figs.~\ref{fig:IP_SW_reconstruction} and \ref{fig:IP_LW_reconstruction} (even in regions not directly constrained as an image family, such as the highly-magnified structure between images $1a$ and $2a$) lends further credence that our image identifications are appropriate.
%
%

The two primary foreground elliptical galaxies are modeled as singular isothermal ellipsoids, as is the third foreground at greater projected distance (but with ellipticity and position angle held fixed to \GALFIT\ fitting results). The $x$ and $y$ (RA and Dec.) centers of each lens are held fixed to the median light centroid of the \JWST\ filters, as otherwise large degeneracies between the different lensing galaxies can be introduced. These galaxies are labeled in Fig.~\ref{fig:cutout}, all of which are spectroscopically-confirmed cluster members at $z=0.87$ \citep{Caminha:2022ab}. A perturbative shear factor is added to account for any missing complexity from our model, as we don't account for the halo potential of the \ElG\ cluster. In total, this combines to 9 free parameters (Table~\ref{tab:lensmodel}), leaving 3 degrees of freedom in the model. While additional free parameters could be introduced, they are not overwhelmingly physically-motivated, and so we opt for the simplest parameterization feasible.

\begin{figure*}[ht!]
\plotone{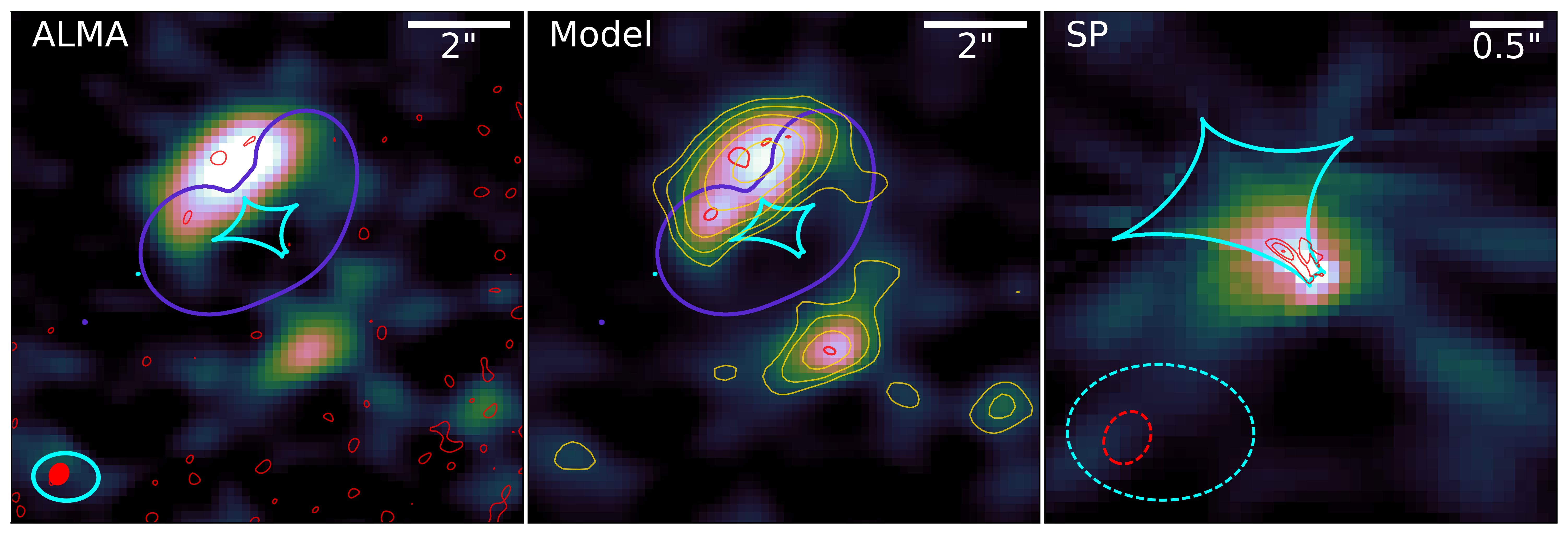}
\caption{
    As with Fig.~\ref{fig:IP_SW_reconstruction}, but for the ALMA 1mm continuum image (with the lower signal-to-noise 870 $\mu$m image shown as red contours). A different field-of-view is chosen to show more of the noise variations. The synthesized beam is shown in cyan (1 mm) and filled red (870 $\mu$m) in the lower-left corner. 
    Noise in the 870 $\mu$m image increases towards the southwest, which is closer to the edge of the primary beam.
    In the middle panel, contours of the observed image are plotted over the model reconstruction for ease of comparison. Here, the red contours now show the modeled 870 $\mu$m image-plane emission. In the last panel showing the source plane (870 $\mu$m as red contours), the beams are shown as dashed ellipses. Lensing allows one to over-resolve the structure (and the actual source-plane PSF varies due to the magnification gradient).
    It is unclear if the dust continuum actually extends further into the center of caustic curve relative to the stellar continuum (cf. Figs.~\ref{fig:IP_SW_reconstruction} and \ref{fig:IP_LW_reconstruction}), which might suggest an outflow from the plane of the disk. However, given the much larger PSF for the ALMA 1 mm image relative to \JWST\ ($\theta$ $\sim$1.1$\arcsec$ vs. $\theta$$\sim$0.03-0.15$\arcsec$), it's not possible to answer this at present.
    \label{fig:IP_ALMA_reconstruction}
}
\end{figure*}

The more disk-like lensing galaxy a little further to the southeast, L3, is somewhat poorly constrained by our model, and the best-fit velocity dispersion is an un-physical value of $\sigma\approx 27$ km s$^{-1}$. The uncertainty on this value is large and highly asymmetric, but we still include it in our model with the intention of examining more closely the amplitude of the external shear, which we find to be $\gamma = 0.36 \pm 0.08$. This value is large, but not entirely unexpected for a complex mass distribution like a group of galaxies embedded within a much larger, extremely massive cluster. Moreover, the unique morphology of the lensed background galaxy suggests a more complicated set of lensing caustics than are likely to be provided by two foreground ellipticals.

To test this further, we re-run the optimization and hold the shear amplitude fixed at $\gamma=0.2$, to be approximately consistent with the contribution from the cluster potential as estimated by models from \citet{Diego:2023ab} and \citet{Frye:2023aa}. This results in a consistent set of parameters\textemdash i.e., within 1$\sigma$ from the highest-likelihood values\textemdash except for the ellipticity of L1 and the velocity dispersion of L3. These shift to $e_1 \approx 0.32$ and $\sigma_{3} \approx 169$ km s$^{-1}$, but both of these are well within the respective 2$\sigma$ bounds from the best-fit solution, and are actually within $1\sigma$ of the median. 
This offers further evidence that the amplitude of external shear included in the best-fit model may be driven by insufficient flexibility in the parameterization of the foreground galaxies.

On the other hand, the orientation of the shear component is consistent with what would be expected for that contributed by the cluster. The center of mass adopted by \citet{Diego:2023ab} between the two peaks of the bimodal mass distribution is oriented at PA=$27\degrees$ (CCW from east) from \ElA, while the nearest, northern mass component of the cluster is centered on a luminous galaxy that is oriented at PA=$6\degrees$. This central galaxy also coincides with the surface mass density peak of the northwest component from the independent lens model of \citet{Frye:2023aa}. 
External shear at PA=$99 \pm 6\degrees$ is thus very close to perpendicular to the direction of the cluster's mass center, and consistent within uncertainty with the direction of the northern component's mass center. With this context, it seems likely that the external shear we find may be due in part to weak lensing from the cluster, while the large amplitude may be explained by some level of additional shear that is required to best describe the mass distribution of the galaxy group lensing the background DSFG.

\subsubsection{\GALFIT\ foreground subtraction} 
\label{sec:GALFIT}
Before reconstructing the observed data in the source-plane, we subtract a S{\'e}rsic model of the foreground light of the massive lensing ellipticals using \GALFIT\ \citep{Peng:2002aa}. We note that a detailed assessment of the light profiles is not directly related to the aims of this paper, and the central regions of each foreground lens (where there might be large residuals from \GALFIT\ subtraction) are masked out before ray-tracing to the source plane, 
and thus don't significantly impact our analysis.

Two S{\'e}rsic profiles were used to model the light from the two foreground elliptical cluster galaxies (L1 and L2) closest to the light from \ElA\ (see Fig.~\ref{fig:model_constraints}) The position, flux, effective radius, S{\'e}rsic index, axis ratio, and position angle were all kept as free parameters. Each filter (for both \JWST\ and \HST) was optimized independently of each other, but checked for consistency. For the F150W and F200W filters, it was necessary for the centroids of each S{\'e}rsic profile to be held fixed to match the average centroid of the six other \JWST\ filters in order for the models to converge. The weight map was constructed by \GALFIT, applied to a $5\arcsec \times 7\arcsec$ cutout as shown in Figs.~\ref{fig:IP_SW_reconstruction} and \ref{fig:IP_LW_reconstruction}.

There is non-negligible residual at the far eastern end of \ElA\ (the ``barb" of the fishhook) in some of the shorter-wavelength filters (visible predominantly in the left panel of Fig.~\ref{fig:IP_SW_reconstruction}). 
This is likely due to significant blending of emission from the background galaxy with the core of the foreground elliptical. In the middle panel of Fig.~\ref{fig:IP_SW_reconstruction}, we create an image-plane reconstruction by masking out this portion of the arc and ray-tracing to the source-plane and back to the image-plane. As a result, the lens model predicts that there is significant highly-magnified structure from a small portion of the background galaxy in this region, which means that simply adding a second S{\'e}rsic component to the foreground model is unlikely to improve the residual meaningfully. For our analysis in this work, we are able to mask out this region and exclude it from our measurements, as the southwestern image ($1a$) that we focus on contains the entire background object and is significantly less affected by foreground contamination. However, future work investigating this highly-magnified structure would warrant a more careful removal of the foreground light to reveal the lensed structure.

\subsubsection{Magnification} 
\label{sec:mag_by_filter}

The total magnification at each optical/near-IR wavelength is determined by computing the flux-weighted average of the magnification map. To measure uncertainties, we randomly draw 100 different models from each parameter's full posterior distribution from MCMC optimization. For each iteration, we reconstruct the source-plane structure, and a map of source-plane magnification at the same pixel scale. Since no assumption is made about the distribution of each parameter's posterior distribution, the resultant distribution of magnifications should accurately capture random uncertainties (although there may still be systematic uncertainties). In most cases, the magnifications are normally-distributed, so we cite $1\sigma$ errors on the median magnification in Table~\ref{tab:mags}. 

For the 1mm ALMA images, we take a slightly different approach, as the flux distribution between pixels is highly correlated within the beam size, which complicates a calculation of the flux-weighted pixel-based average of the magnification map. Instead, we estimate magnification to be $\mu_{1{\rm mm}} = 9.5 \pm 0.9$ by using the ratio of image-plane to source-plane areas above an equivalent noise threshold. For the 870 $\mu$m image, on the other hand, the DSFG is only marginally detected \citep{Cheng:2023aa}, so we estimate the magnification by using the range of values in a small aperture centered on the brightest pixel (see Fig.~\ref{fig:IP_ALMA_reconstruction}), resulting in $\mu_{870\mu{\rm m}} \approx 6 - 9$.

\subsection{Source-plane size and light profile} 
\label{sec:SP_size}

To reconstruct the source-plane light distribution, we ray-trace all lensed images (e.g., $1abcde$) to the source-plane simultaneously.
As the PSF varies over the extent of the source due to lensing distortion, the choice of a single kernel for \GALFIT\ is nuanced. Since the PSF is broadest and varies the least for image $1a$, it is the best choice for describing the galaxy's overall spatial extent. For our purpose, we place a simulated PSF from \textsc{WebbPSF} v.1.1.0 \citep{Perrin:2012aa, Perrin:2014aa} at 10 random locations within a $2\arcsec \times 2\arcsec$ box centered at the peak of the F444W emission in the counterimage ($1a$), then reconstruct in the source-plane with our lensing model. This creates a set of representative PSFs in the vicinity of the region of greatest interest. Next, we convolve the model-reconstructed source-plane structure with this source-plane kernel in order to mitigate small-scale substructure within the galaxy (primarily arising from the higher-magnification images, $1b - 1d$) and to ensure a more uniform spatial resolution. We use a single-component S{\'e}rsic profile and a sky background component in our models for each filter. We note that the masking of foreground objects when reconstructing in the source plane leads to only a small area of blank sky being included. As the S{\'e}rsic indices we derive are generally very small, this sky component likely has little effect on the estimation of the effective radius. This is more of a concern for objects with large best-fit $n$ indices, which have significant signal out to large radii.

Of primary interest are the effective radius along the semi-major axis $R_e$ (encapsulating half the total flux) and the S{\'e}rsic index (describing the radial concentration of light). The results for each filter are provided in Table \ref{tab:mags}. We find minimal scatter in the axis ratio $b/a = 0.49 \pm 0.05$, which corresponds to an estimated inclination\footnote{We use an intrinsic axial ratio of $q=0.2$ (e.g., \citealt{Unterborn:2008aa}), and estimate inclination from $\cos^2{i} = [(b/a)^2 - q^2]/(1-q^2)$.} of $i = 63\degrees \pm 7\degrees$. The uncertainties reported from \GALFIT\ $\chi^2$ optimization are likely severe underestimates of the actual, empirical dispersion in $R_e$ and $n$. For this reason, we instead adopt the relative uncertainties calculated for the lensing magnification in each filter (ranging from $\Delta \mu / \mu \sim 10 - 30\%$; \S \ref{sec:mag_by_filter}), which are likely to dominate the uncertainty budget\footnote{Lensing magnifies each dimension by approximately $\mu^{0.5}$, so the relative uncertainty in size from lensing would actually propagate to be half the relative uncertainty in magnification. However, given the substantial deviations from a smooth light profile, we maintain the full $10 - 30\%$ fractional uncertainties to be conservative.}. Importantly, the scatter in measured properties from the set of representative source-plane PSFs is also smaller than this adopted uncertainty.

In some cases, a single-component S{\'e}rsic profile does not adequately capture the various spatial frequencies of the galaxy's light. In these scenarios, a second component is added (and often held fixed at the center of the galaxy's nucleus when necessary), as in the case of a bulge $+$ disk model (e.g. \citealt{Gao:2017aa}). 
Our criterion for requiring this extra component is a $>5\sigma$ peak in the residual near the galaxy's center.
An equivalent S{\'e}rsic index and effective radius in these cases are determined by first performing optimization with two components plus a sky background, followed by a second iteration reducing to a single S{\'e}rsic component but holding the sky parameters fixed to the best-fit results. These effective values are shown separately in Table~\ref{tab:mags}. In all cases, the effective radius remains constant, but there is some very slight modification in S{\'e}rsic indices to smaller values (but well within the uncertainties of the original, one-component values).

We calculate the effective radius of the ALMA 1 mm continuum image by reconstructing in the source plane for a number of model iterations (as described in \S \ref{sec:mag_by_filter} for estimating magnification. We use the \textsc{casa} (v6.2.1) \textsc{imfit} algorithm, finding $R_e = 0.53 \pm 0.03\arcsec$. This size is substantially smaller than the image-plane beam size $1.28\arcsec \times 0.93\arcsec$, due to the lensing magnification. A more careful estimation of the dust continuum size may be warranted in the future with higher-resolution submillimeter data (i.e. with unconvolved, forward modeling of the $uv$-plane visibilities). However, at present, the source-plane effective radius is smaller than the beam to a degree that such an analysis is not likely to yield much improved results.

\begin{deluxetable*}{c|cccccc}
\tablecaption{Total magnification, spatial extent, and photometry for each filter. \label{tab:mags}}
\tablehead{
\colhead{Filter} & \colhead{$\lambda_{\rm eff, rest}$} & \colhead{Magnification $\mu$} & \colhead{$R_e$} & \colhead{$R_e$} & \colhead{S{\'e}rsic $n$} & \colhead{$m_{\rm AB} - 2.5\log{\mu_{1a}}$} \\
\colhead{} & \colhead{[nm]} & \colhead{} & \colhead{[\arcsec]} & \colhead{[kpc]} & \colhead{} & \colhead{[mag]}
}
\startdata
HST/F606W & 180    &   2.3		$\pm$ 	0.3		 & 0.24 $\pm$ 0.03 & 1.9 $\pm$ 0.2 & 0.23 $\pm$ 0.03 & 24.49 $\pm$ 0.04 	\\
HST/F625W & 192    &   2.5		$\pm$ 	0.4		 & 0.24 $\pm$ 0.04 & 2.0 $\pm$ 0.3 & 0.23 $\pm$ 0.04 & \textemdash	\\
HST/F775W & 234    &   2.6		$\pm$ 	0.4		 & 0.30 $\pm$ 0.05 & 2.5 $\pm$ 0.4 & 0.31 $\pm$ 0.05 & \textemdash	\\
HST/F814W\tablenotemark{a} & 244    &   3.1		$\pm$ 	0.9		 & \textemdash & \textemdash & \textemdash & 23.87 $\pm$ 0.04 	\\
HST/F850LP& 274    &   2.7		$\pm$ 	0.3		 & 0.35 $\pm$ 0.04  & 2.9 $\pm$ 0.3 & 0.23 $\pm$ 0.03 & \textemdash	\\
JWST/F090W& 274    &   3.8		$\pm$ 	0.8		 & 0.29 $\pm$ 0.06  & 2.4 $\pm$	0.5 & 0.30 $\pm$ 0.06 &	23.66 $\pm$ 0.01\\
HST/F105W & 321    &   8.4		$\pm$ 	1.5		 & \textemdash & \textemdash & \textemdash & \textemdash	\\
JWST/F115W& 351    &   6.9		$\pm$ 	1.6		 & 0.37 $\pm$ 0.09 & 3.1 $\pm$ 0.7 & 0.16 $\pm$ 0.04 & 22.85 $\pm$ 0.01 	\\
HST/F125W & 379    &   5.8		$\pm$ 	1.5		 & \textemdash  & \textemdash & \textemdash & \textemdash  \\
HST/F140W & 423    &   7.8		$\pm$ 	1.5		 & \textemdash  & \textemdash & \textemdash & \textemdash	\\
JWST/F150W& 456    &   7.6		$\pm$ 	1.7		 & 0.38 $\pm$ 0.08 & 3.1 $\pm$ 0.7 & 0.23 $\pm$ 0.05 & 21.886 $\pm$ 0.005	\\
HST/F160W & 467    &   8.4		$\pm$ 	2.3		 & \textemdash & \textemdash & \textemdash & \textemdash	\\
JWST/F200W& 604    &   7.8		$\pm$ 	2.3		 & 0.41 $\pm$ 0.12 & 3.4 $\pm$ 1.0 & 0.25 $\pm$ 0.08 & 21.227 $\pm$ 0.004	\\
JWST/F277W& 839    &   8.3		$\pm$ 	2.0		 & 0.41 $\pm$ 0.10 & 3.4 $\pm$ 0.8 & 0.22 $\pm$ 0.06 & 20.807 $\pm$ 0.004	\\
JWST/F356W& 1084   &   10.9	$\pm$ 	2.5		 & 0.40 $\pm$ 0.09 & 3.3 $\pm$ 0.7  & 0.42 $\pm$ 0.10 & 20.373 $\pm$ 0.004 	\\
\multicolumn{1}{r|}{({\it 2-comp.}\tablenotemark{$\dagger$})} &\textemdash &\textemdash & (0.40 $\pm$ 0.09) & (3.3 $\pm$ 0.7) & (0.42 $\pm$ 0.10) & \textemdash \\
JWST/F410M& 1240   &   8.3		$\pm$ 	2.0		 & 0.39 $\pm$ 0.09 & 3.2 $\pm$ 0.8 & 0.32 $\pm$ 0.08 & 20.202 $\pm$ 0.004 	\\
\multicolumn{1}{r|}{({\it 2-comp.}\tablenotemark{$\dagger$})} &\textemdash &\textemdash & (0.39 $\pm$ 0.09) & (3.2 $\pm$ 0.8) & (0.32 $\pm$ 0.08) & \textemdash \\
JWST/F444W& 1349   &   9.8		$\pm$ 	2.2		 & 0.39 $\pm$ 0.09  & 3.2 $\pm$ 0.7  & 0.45 $\pm$ 0.11 & 20.146 $\pm$ 0.004	\\
\multicolumn{1}{r|}{({\it 2-comp.}\tablenotemark{$\dagger$})} &\textemdash &\textemdash & (0.39 $\pm$ 0.09) & (3.2 $\pm$ 0.7) & (0.44 $\pm$ 0.10) & \textemdash \\
\hline
& $\lambda_{\rm eff, rest}$ & \textemdash &\textemdash & \textemdash\textemdash & \textemdash & Flux $\mu S_\nu$\\
& [$\mu$m] &  &  & &  & [mJy]\\
\hline
ALMA/870$\mu$m\tablenotemark{b} & 264   &   $\approx 6 - 9$	 & \textemdash  & \textemdash  & \textemdash & 1.1 $\pm$ 0.7	\\
ALMA/1100$\mu$m& 334   &   9.5		$\pm$ 	0.9	 & 0.53 $\pm$ 0.03  & 4.4 $\pm$ 0.2  & 0.5 (fixed\tablenotemark{c}) & 3.8 $\pm$ 0.2
\enddata
\tablenotetext{a}{For the F814W filter, a detector gap obscures the primary lensing galaxy in the image. While the lensed structure is largely still visible, we exclude this from our source size analysis, as a comparable initial \GALFIT\ subtraction of the foreground is not possible, and the F775W/F850LP filters can effectively be used as proxy.}
\tablenotetext{b}{The signal-to-noise ratio of the Briggs-weighted (\textsc{robust}$=2$, close to natural weighting) 870 $\mu$m image is low, so we do not attempt to derive an effective radius or S{\'e}rsic index.}
\tablenotetext{c}{The ALMA data is fit with a 2-dimensional Gaussian, which is nearly equivalent to a $n=0.5$ S{\'e}rsic profile.}
\tablenotetext{\dagger}{For the longest-wavelength filters (F356W, F410M, F444W), an additional S{\'e}rsic component is added to account for large residuals in the galaxy center. As discussed in \S \ref{sec:SP_size}, we also calculate effective values for the combined components. In all cases, the effective radius remains the same, but there is some slight change in $n$.}
\tablecomments{
Reported image-plane photometry values are only for image $1a$, and are not corrected for magnification $\mu_{1a}$, the values for which are reported in Table~\ref{tab:mags_1a}. For this image, differential magnification is minimized and approximately consistent from rest-frame far-UV to near-IR. We also tabulate the effective (or pivot) wavelength of each filter converted to the rest-frame at $z=2.291$.
For the WFC3/IR filters of \HST, we do not measure effective radii or S{\'e}rsic index, as superior imaging is available from \JWST\ at comparable wavelengths.}
\end{deluxetable*}

\subsection{Photometry of \JWST, \HST, and ALMA data} 
\label{sec:photometry}

In this study, we require photometry of the rest-frame near-UV, optical, and near-IR regimes in order to confirm the redshift of \ElA, and to measure some initial basic properties.
Image-plane photometry for \ElA\ was performed for all \JWST/NIRCam filters, in addition to some of the bluer \HST\ filters, F606W and F814W, which provide constraints on the rest-frame UV not possible with \JWST. We refer the reader to \citet{Frye:2023aa} for more details on the careful photometric measurements for the entire \ElG\ field, which we make use of here. 
Briefly, \textsc{WebbPSF} is used to generate PSF models for the short- and long-wavelength filters, and modeled ACS filter PSFs are used for the \HST\ images. All images are registered astrometrically onto the Gaia Data Release 3 reference frame \citep{Gaia-Collaboration:2016aa, Gaia-Collaboration:2021aa}. The PSF-convolved matched-aperture photometry (see \citealt{Pascale:2022aa}) was collected using {\sc SExtractor} \citep{Bertin:1996aa}, with F200W serving as the detection band. Automatic aperture fluxes from {\sc SExtractor} are measured using Kron-like elliptical apertures \citep{Kron:1980aa}.
All measurements are PSF-matched to the reference F200W filter: for longer-wavelength filters with broader PSFs, the F200W filter is convolved to each respective PSF to determine how much flux would be lost within the aperture\footnote{See \url{https://www.stsci.edu/~dcoe/ColorPro/color} for details.}. This flux correction factor is then applied to each filter.
Magnitudes are reported in Table~\ref{tab:mags}, without correction for lensing magnification, for apertures located at 
$(\alpha, \delta) =$ (01$^h$02$^m$49 \fs 169, $-49^d$15$^m$07 \fs 99)\footnote{For comparison, we also include photometry from \citet{Frye:2023aa} at an independent aperture located at (01$^h$02$^m$49 \fs 202, $-49^h$15$^m$04\fs 32), part of the highly-distorted northern portion of the arc, as shown in the right panel of Fig.~\ref{fig:EAZY}. This is labeled Aperture 2 in Fig.~\ref{fig:IP_LW_reconstruction}.}, shown as Aperture 1 in Fig.~\ref{fig:IP_LW_reconstruction}. We also measure the flux of the 870 $\mu$m and 1 mm continuum with {\sc blobcat} \citep{Hales:2012aa}, a flood-filling detection algorithm designed for radio images.

For this initial study, we opt to use only image-plane photometry for fitting the SED, as source-plane photometry is plagued by a number of issues. For example, the magnification gradient of \ElA\ is rather steep in the region of the background DSFG, which means that the source-plane PSF varies significantly over the extent of the source (although this primarily impacts the highest-magnification multiple images towards the northeast). This is because the regular, consistent PSF for the observed data becomes distorted and magnified inconsistently in the source plane. This makes it very difficult (or impossible) to perform PSF-convolved aperture photometry directly on the demagnified source-plane emission. Moreover, it is much more difficult to incorporate any local sky subtraction in the source plane, as foreground objects must be masked before ray-tracing with the lens model (and doing this for too large an area becomes computationally prohibitive).

\subsection{Inner vs. outer disk decomposition}
\label{sec:SED_demag}

To analyze the SED of {\it El Anzuelo} for the inner bulge (or central kiloparsec) relative to the outer disk, we simultaneously de-magnify the image-plane photometry and estimate the bulge-to-total fraction of light ($B/T$) in the source-plane using {\sc photutils} \citep{Bradley:2022aa}. 
Before performing this analysis, we first convolve all filters to match the broadest PSF of the F444W filter, to ensure that our spatially-resolved color analysis is not influenced by the PSF.
By correcting the image-plane photometry rather than performing the measurements directly on the source-plane, we can avoid the effects of a varying PSF and poor sky subtraction.

To determine a sensible threshold for the inner bulge versus outer disk, we follow the method employed by \citet{Cutler:2023aa} to determine appropriate sizes for a fixed aperture and annulus.
\citeauthor{Cutler:2023aa} used $z_{\rm F850LP}$ and $H_{\rm F160W}$ in order to span the 4000\AA-break at $z\sim2.3$ as an age indicator.
Here we instead use F277W and F444W (spanning the wavelength at which point the light profile begins to steepen for \ElA, which is suggestive of a bulge/spheroidal component being assembled). 
The [2.8 - 4.4] color is measured in apertures of varying radius from 0.03$\arcsec$ to 0.6$\arcsec$, in over-sampled increments of 4 milliarcseconds. 
The smallest aperture that encloses the reddest emission (i.e. maximizes [2.8 - 4.4]) is chosen, under the condition that the next local minimum at a larger radius is less than 1\% of the local maximum's color (to avoid secondary peaks). 
In our case, there is only one local maximum, which is at 0.13$\arcsec$, or approximately 1.1 kpc.
While this is comparable to the scale of the NIRCam PSF FWHM in F444W \citep{Rigby:2023ab}, lensing magnification makes it so that this inner region is safely resolved in all filters.
The fraction of light inside this aperture to the total flux of the disk (or bulge-to-total fraction, $B/T$) is given for each filter in Table~\ref{tab:mags_1a}.
These fractions are then applied to the de-magnified photometry, and \EAZY\ is run for the inner, outer, and total disk.
Coincidentally, this inner vs. outer threshold of $\sim$1 kpc is consistent with that used often to measure the central surface mass density, as an alternative to a S{\'e}rsic index (e.g., \citealt{Cheung:2012aa, Ji:2022aa}).

The available photometric coverage helps to break the degeneracy between reddening from stellar population age vs. from dust attenuation, in particular through the well-studied $UVJ$ diagram (e.g., \citealt{Labbe:2005aa, Wuyts:2007aa, Patel:2012aa}). The F115W, F150W, and F200W filters together offer strong constraints on the rest-frame $U$ and $V$ fluxes, while F356W now offers comparable spatial-resolution coverage of rest-frame $J$ flux for the first time \citep{Miller:2022ab}. As discussed by \citet{Leja:2019aa}, however, SED-fitting is often more precise and accurate than color-color diagrams in identifying properties. This may be especially true for spatially-resolved $UVJ$ diagrams, which have only recently become feasible at $z\sim2$ and have not yet been fully calibrated.

Regarding the demagnification itself, we carry this out in two different ways. For the first approach, we correct image-plane fluxes by the magnification that we derive on a filter-by-filter basis, using the values given in Table~\ref{tab:mags_1a}. However, given the expected correlation in magnifications for adjacent filters, this may not be the most physically-motivated approach. Instead, it may make more sense to adopt a regime-specific magnification (i.e. separate values for UV, for optical, and for near-IR). Yet, for image $1a$, we observe very little variation in magnification for all filters, so we instead adopt a median magnification that we apply to all filters for our second approach. Moreover, we consider that the uncertainties for each filter might be overly-conservative estimates, as the empirical scatter in magnification from far-UV to near-IR is notably smaller than any of the quoted $1\sigma$ uncertainties. 

For this approach, we adopt the dispersion in $\mu_{1a}$ as the $1\sigma$ uncertainty in magnification.
While this may be an under-estimate of any systematic uncertainty, we consider it likely that these would essentially affect all filters equally, which would modify only the amplitude of the SED and not its shape. Lastly, we note that the derived properties in Table \ref{tab:properties} are consistent between the two methods within uncertainties, so the difference has minimal impact on our interpretations.

\begin{figure}[ht!]
\includegraphics[width=0.47\textwidth]{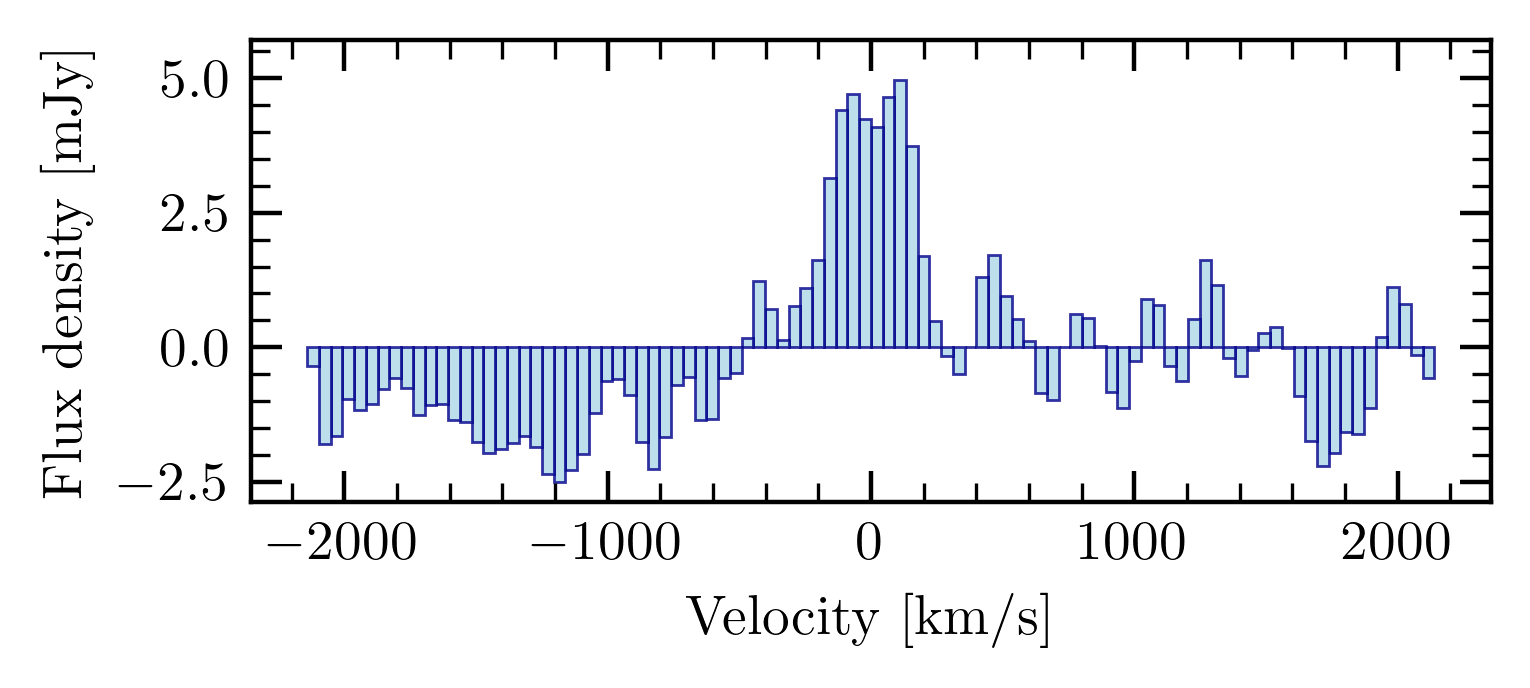}
\includegraphics[width=0.47\textwidth]{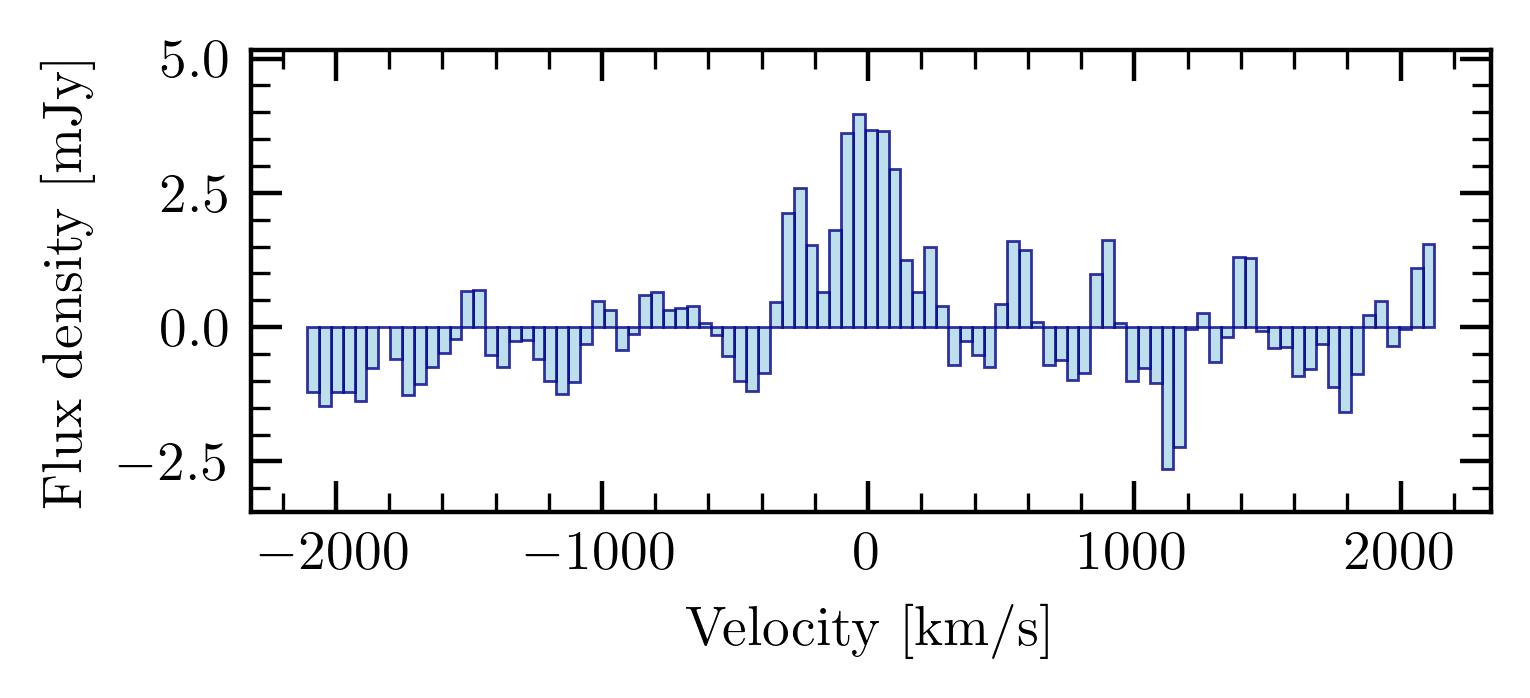}
\caption{
    The assumed CO(3-2) line detection for \ElA\ ({\it top}) from ALMA program 2015.1.01187.S \citep{Basu:2016aa} at observed frequency 105.07 GHz ($z=2.291$). While the target is near the edge of the primary beam of this pointing, an independent detection of the same line from a separate pointing (program 2017.1.01621.S, {\it bottom}) confirms that it is very likely not a spurious artifact (although this pointing is even further from the primary beam center).
    \label{fig:specz}
}
\end{figure}

\begin{figure*}[ht!]
\includegraphics[width=0.5\textwidth]{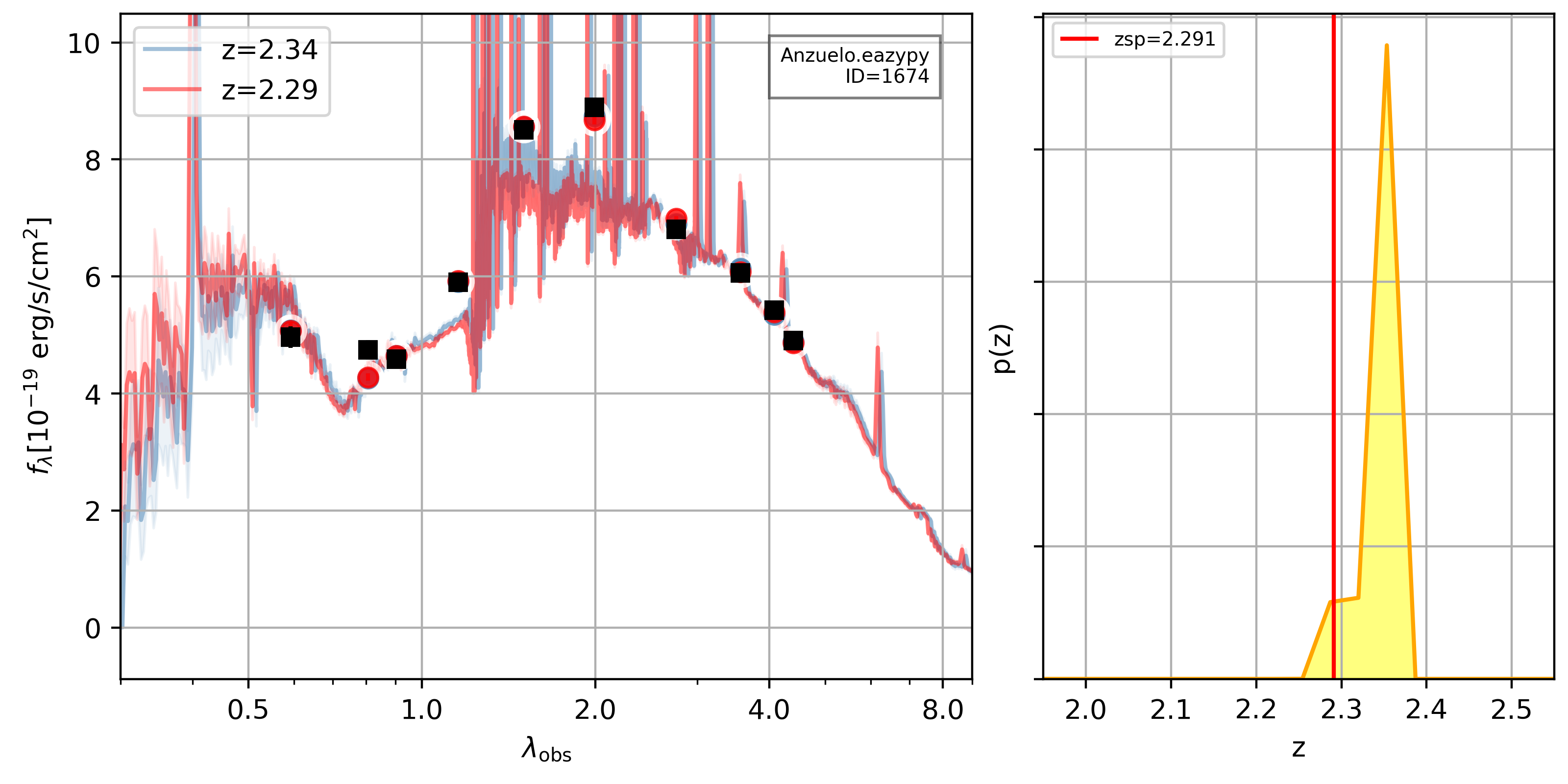}\includegraphics[width=0.5\textwidth]{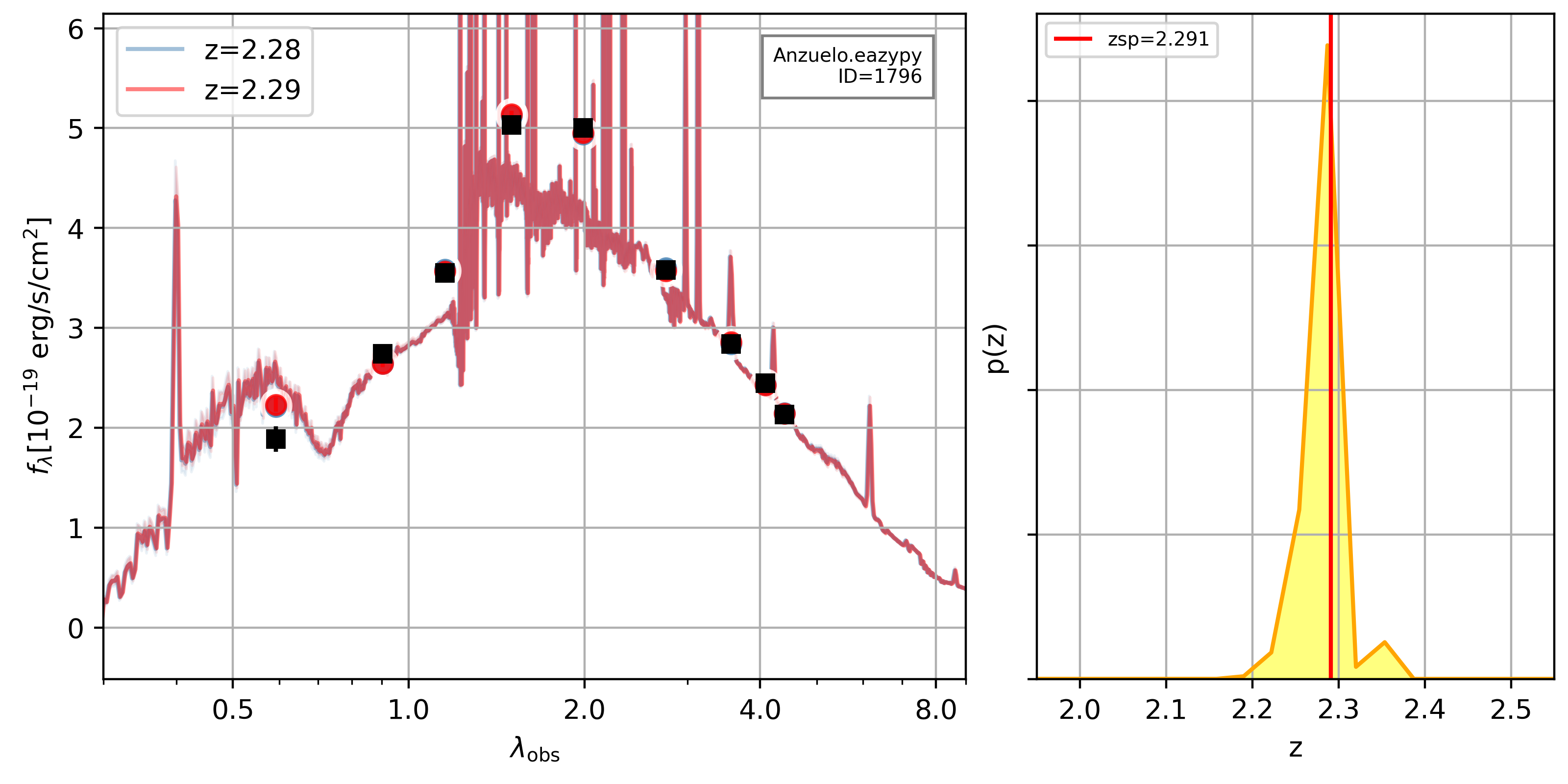}
\caption{
    SED-fitting results with \EAZY\ for image $1a$ ({\it left}) and the northernmost part of the arc where there is significant distortion ({\it right}), which we use as independent verification of the photometric redshift estimate.
    In the left part of each panel, the SED templates at the best-fit and spectroscopically-confirmed redshifts (in blue and red, respectively) are shown: $z\approx 2.34$ (left) and  $z\approx 2.28$ (right) vs. spectroscopic $z=2.291$. The measured photometry points are shown as black squares, while the red circles show the expected fluxes (including continuum and line emission) given the filter throughput. In the right part of each panel, the probability distribution function of redshift given the SED fitting is shown, with the spectroscopic redshift shown as a red vertical line. In both cases, the redshift solution is extremely narrow (with uncertainties more typical of spectroscopic redshifts), thanks to the careful sampling of the rest-frame UV through near-IR spectrum. In particular, the 4000\AA\ break and/or Balmer break (at $\sim$3600\AA) is well-constrained with the F115W and F150W filters, as the break falls right in the small gap between these wideband filters.
    \label{fig:EAZY}
}
\end{figure*}

\subsection{Source redshift confirmation and measurement of basic properties with \EAZY} 
\label{sec:EAZY}

\citet{Cheng:2023aa} included \ElA\ in their preliminary analysis of objects with joint ALMA/\JWST\ detections. 
Their preliminary photometry with only \JWST\ NIRCam filters resulted in photometric redshift solutions of $z=3.54 \pm 0.15$ and $z=3.03 \pm 0.06$ for the primary southwest and northeast lensed components. 
However, this photometry did not specifically account for the contribution of 
light from the foreground elliptical galaxies. Instead, \citeauthor{Cheng:2023aa} employed small 0.6\arcsec\ apertures centered on the peak pixel in F444W in order to minimize the contamination. As there is a spatial offset in the continuum structure between the short- and long-wavelength filters of NIRCam, this had the effect of underestimating the flux below 2 $\mu$m, thereby biasing to higher-redshift solutions. 

For the SED-fitting and photometric-redshift estimation in this work, we use \EAZY\ \citep{Brammer:2008aa, Brammer:2021ab} to optimize the rest-frame UV through near-IR SED ($0.18 - 1.35 \mu$m; observed-frame $0.59 - 4.44 \mu$m). We note, however, the need for future work using software with a complex treatment of star-formation histories, such as {\sc prospector} \citep{Johnson:2021aa}.
We used the 
{\tt tweak\_fsps\_QSF\_12\_v3.param} \EAZY\ templates, with a uniform prior on the redshift from $z=0.01-6$, which encompass a range of realistic star formation histories and levels of dust attenuation (up to $A_V\sim 3$). We test several other template sets, including the {\tt eazy\_v1.x.param} files (which include a dusty galaxy template), but find minimal observable differences in the results.

The photometric redshift for \ElA\ is remarkably precise ($z=2.34\pm0.01$), although the redshift uncertainty is very likely underestimated, since it disagrees with the spectroscopic-redshift, $z=2.291$, as we discuss in the next paragraph. We interpret the reason for this precision to be the fortuitous alignment of the Balmer break at $\lambda = 1.2 \mu$m and the 4000\AA\ break at $\lambda = 1.3 \mu$m, such that F115W exclusively samples continuum short of the 4000\AA\ break, and F150W samples only continuum longward of the break.

With this precise prediction for the redshift, we re-examined ALMA archival data that might provide coverage of bright lines at this redshift. The 3 mm continuum data (program 2015.1.01187.S, PI: K. Basu) has low spectral resolution and coarse angular resolution; moreover, the location of \ElA\ falls near the edge of the primary beam, where the sensitivity is degraded. However, we are still able to detect what we infer to be the CO(3-2) transition at 105.07 GHz with high significance at the center of the bandpass (see Fig.~\ref{fig:specz}). Additionally, a separate pointing (also at 3 mm; program 2017.1.01621.S, PI: K. Basu) has a different phase center, but the line is again detected (despite falling once again near the very edge of the usable primary beam response). We consider this to be a secure 
detection from these 3 independent sets of observations, but we will pursue follow-up observations in the future to offer further confirmation.
It may also be possible that the detected line is CO(4-3), which would indicate a redshift solution of $z=3.388$, although this is in worse agreement with the photometric evidence. For this work, we adopt $z=2.291$ as the most likely solution, and simply note that the angular-to-physical size ratio changes only modestly from 8.21 kpc arcsec$^{-1}$ at $z=$2.291 to 7.41 kpc arcsec$^{-1}$ at $z$=3.388. Moreover, most results discussed in this work pertain to relative sizes as a function of wavelength, so this has little impact.

Unfortunately, it is not possible to derive geometric redshifts based on the lens model (see, e.g., \S 4.3 of \citealt{Diego:2023ab}). This is because the lens model is only constrained by \ElA\ itself, and the background redshift is more or less degenerate with the mass of the foreground lens. One could hypothetically use Fundamental Plane relations (e.g. \citealt{Dressler:1987aa}) and the spectroscopic redshift of the cluster members to estimate the mass enclosed within the Einstein radius of the lens (thereby allowing for a determination of the background redshift, as this radius depends only on the foreground mass and the redshift geometry of lens and source planes). However, given that the lens consists of a group of galaxies and not a single massive elliptical, the dark matter distribution is more uncertain, and this method may be too tenuous.

We adopt henceforth a redshift of $z$=2.291, and
re-run \EAZY\ with the redshift held fixed, in order to estimate more accurate uncertainties on the derived properties, which we discuss in \S \ref{sec:discussion} and present in Table \ref{tab:properties}. In short, the star-formation rate (SFR) and stellar mass ($M_\star$) are inferred using only the far-UV-to-near-IR photometry, as the angular resolution of the far-IR observations is much coarser, and thus insufficient for a spatially-resolved SED decomposition. 
In the future, we intend to incorporate far-IR information at a more comparable angular resolution with a more flexible approach to SED modeling. For this present analysis, properties derived from the SED are used primarily to support inferences from the surface brightness distributions.

\section{Discussion} 
\label{sec:discussion}

Recently-collected \JWST/NIRCam imaging reveals \ElA\ to be an excellent candidate for follow-up studies, especially with targeted, higher-resolution interferometric radio/millimeter imaging and spectroscopy. This will allow for a more complete analysis of the interstellar medium driving the active star formation. For this initial case study, we focus primarily on the distribution of light in the rest-frame UV, optical, near-infrared, and far-infrared, in order to understand the distributions of new star formation and existing stellar mass from older populations. As part of this, we must take into account the fact that the effect of gravitational lensing is not nearly spatially uniform across \ElA.

\subsection{Large differential magnification between rest-frame UV, near-IR, and far-IR}
\label{sec:diff_mag}

\begin{figure*}[ht!]
\includegraphics[width=\textwidth]{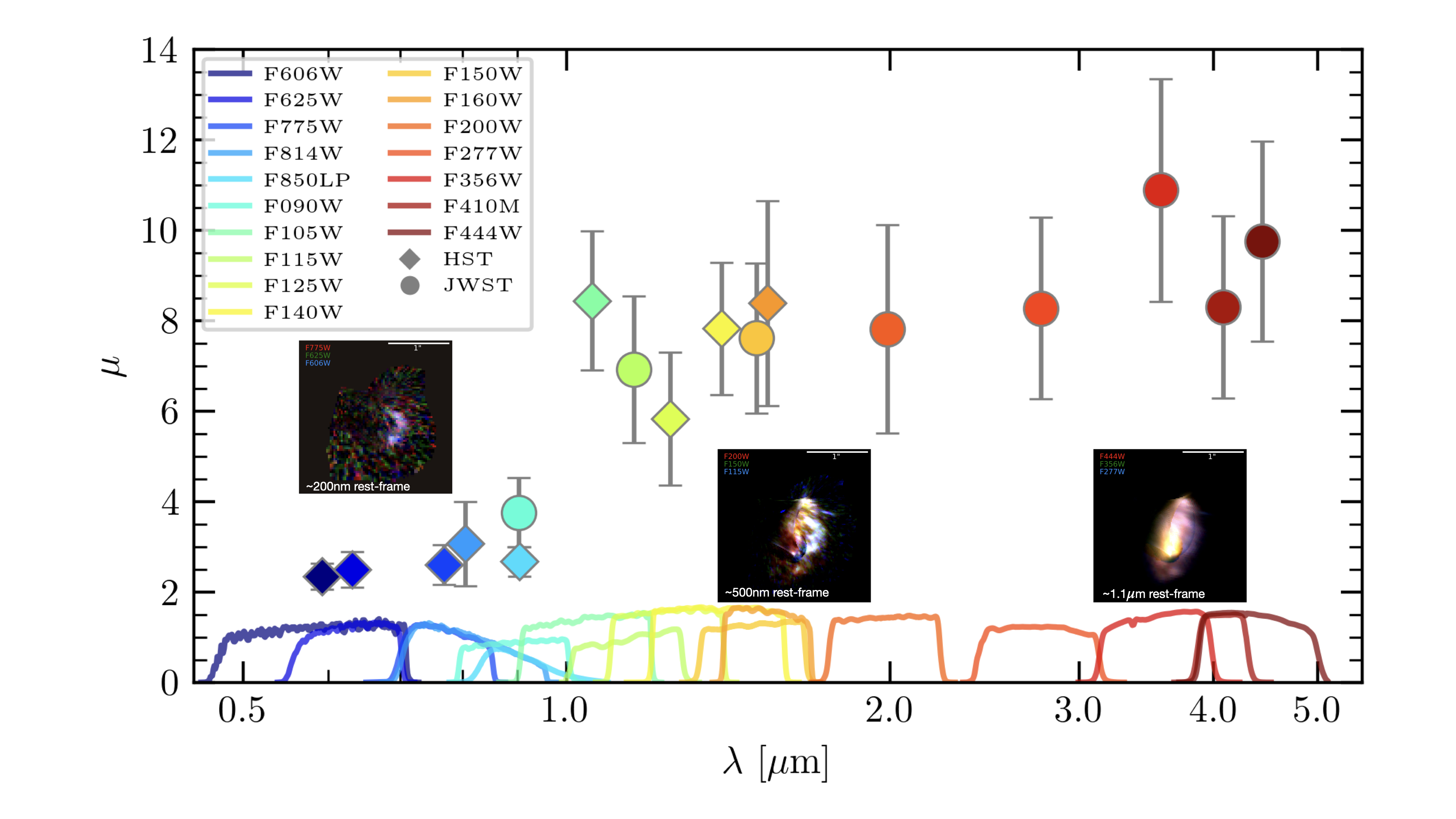}
\caption{
    Derived magnification $\mu_{\rm total}$ is shown for each \HST\ and \JWST\ filter available for \ElA. Corresponding filter throughput curves are shown at the bottom. A dramatic rise in magnification with wavelength is due to a steepening in the light profile, as shown also with the RGB inset images showing rest-frame UV, optical, and near-IR images of the source-plane reconstruction. There is a notable discontinuity in magnification around $1 \mu$m, which is due primarily to the lack of signal in the central core, where magnification is greatest. This is likely due to some combination of a dust attenuation and an age gradient, especially given the location of the 4000\AA\ break, redshifted to $1.3\mu$m for $z=2.291$. 
    \label{fig:mag_by_filter}
}
\end{figure*}

\begin{figure}[ht!]
\includegraphics[width=0.935\columnwidth]{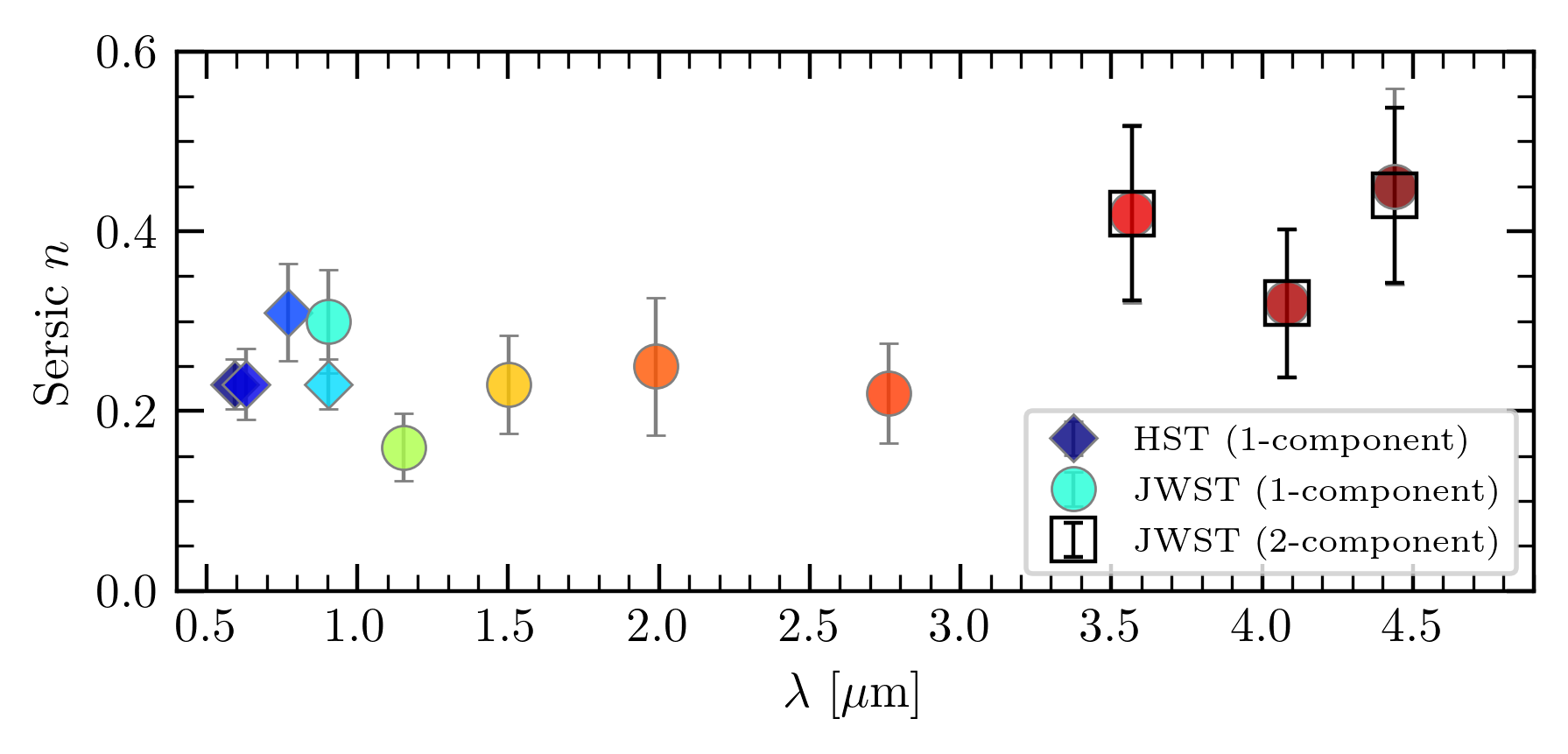}
\includegraphics[width=\columnwidth]{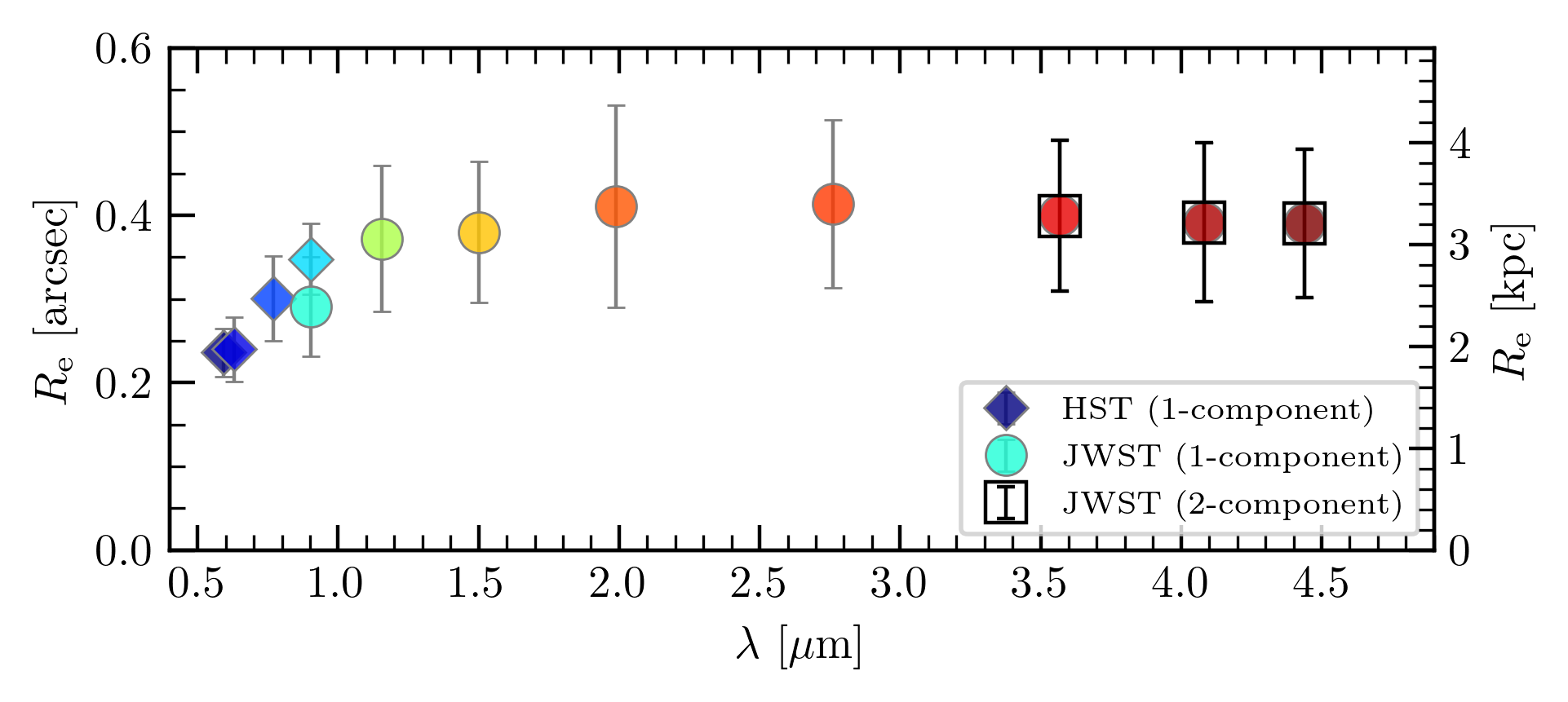}
\caption{
    \label{fig:R_e_by_filter}
    Variation in the reconstructed source-plane light profile for the \HST/ACS and \JWST/NIRCam filters.
    {\it Top:} Best-fit S{\'e}rsic index $n$ determined with \GALFIT\ for each filter. For the three longest-wavelength filters, excess model residuals near the galaxy nucleus suggests the need for a 2-component (bulge and disk) model. Here, an effective S{\'e}rsic index is shown, as discussed in \S \ref{sec:SP_size}.
    Filters are denoted by color
    as in Fig.~\ref{fig:mag_by_filter}.
    As the uncertainties reported by \GALFIT\ are likely underestimations, we scale them up to match the fractional uncertainties in magnification for each filter (Table~\ref{tab:mags}).
    {\it Bottom:} As above, but showing effective radii for the S{\'e}rsic profiles for each filter, in angular and physical units (at the redshift of the source, $z=2.291$). 
}
\end{figure}

\begin{figure*}[ht!]
\includegraphics[width=0.34\textwidth]{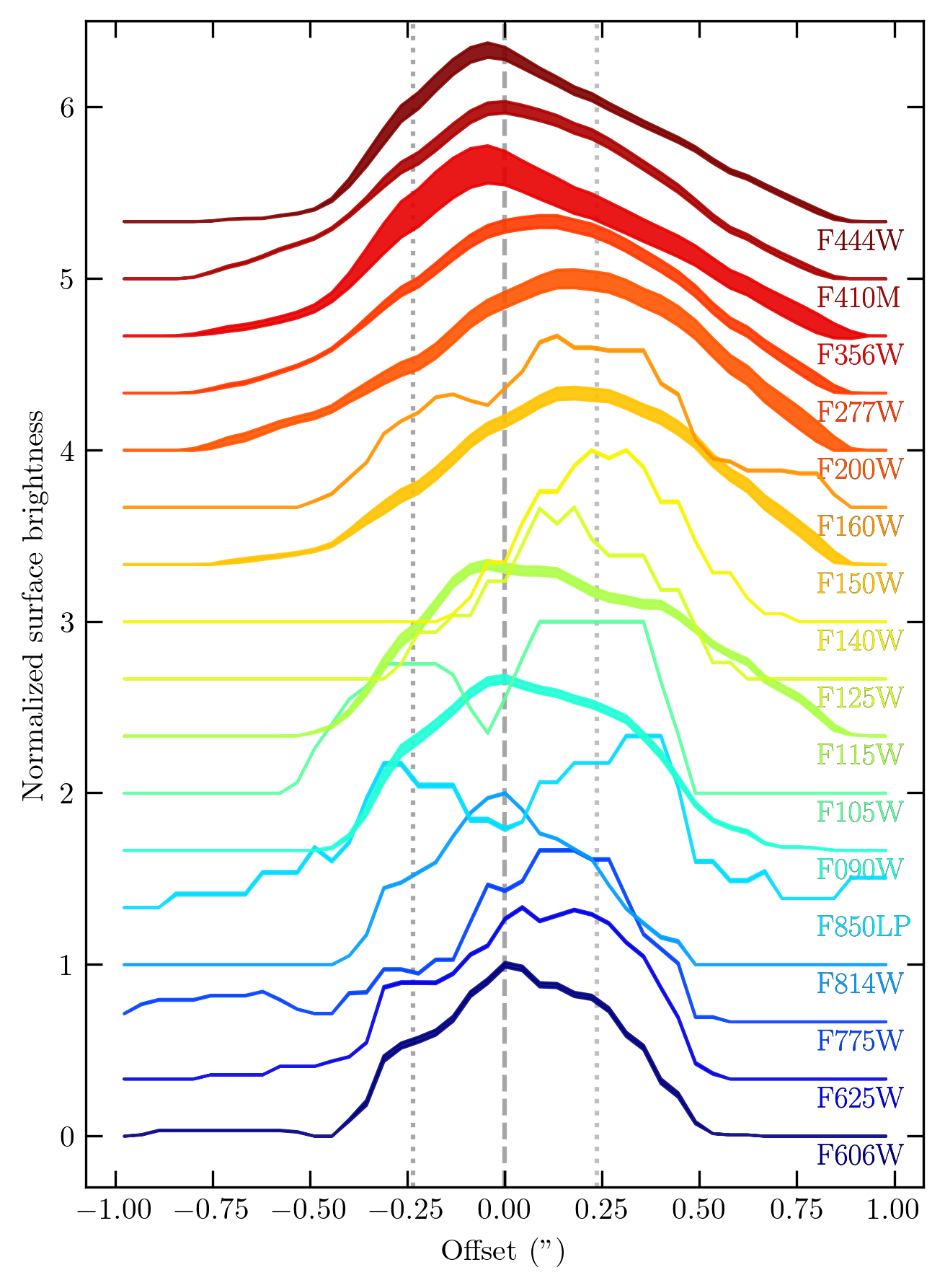}
\raisebox{0.32\height}{\includegraphics[width=0.3\textwidth]{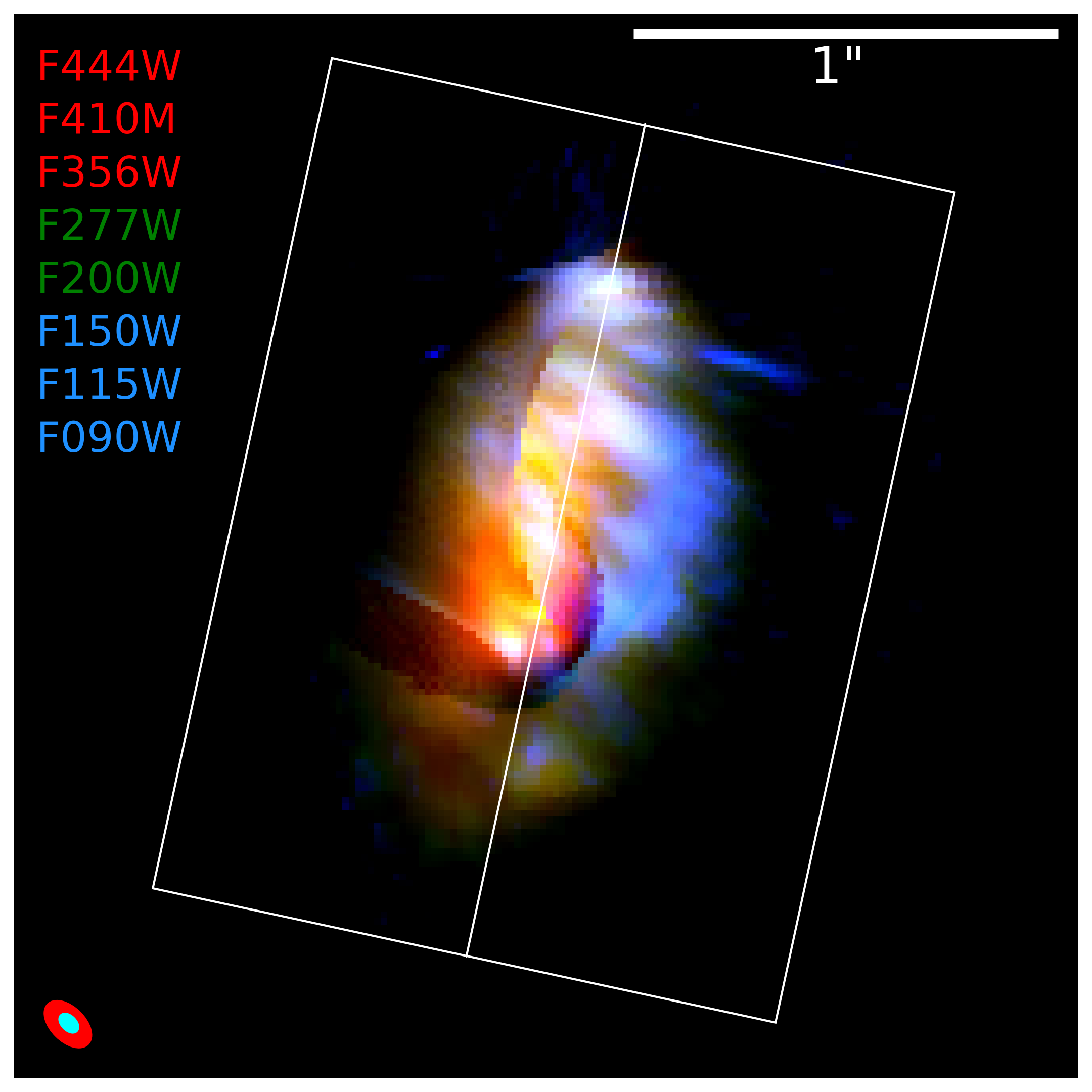}}
\includegraphics[width=0.34\textwidth]{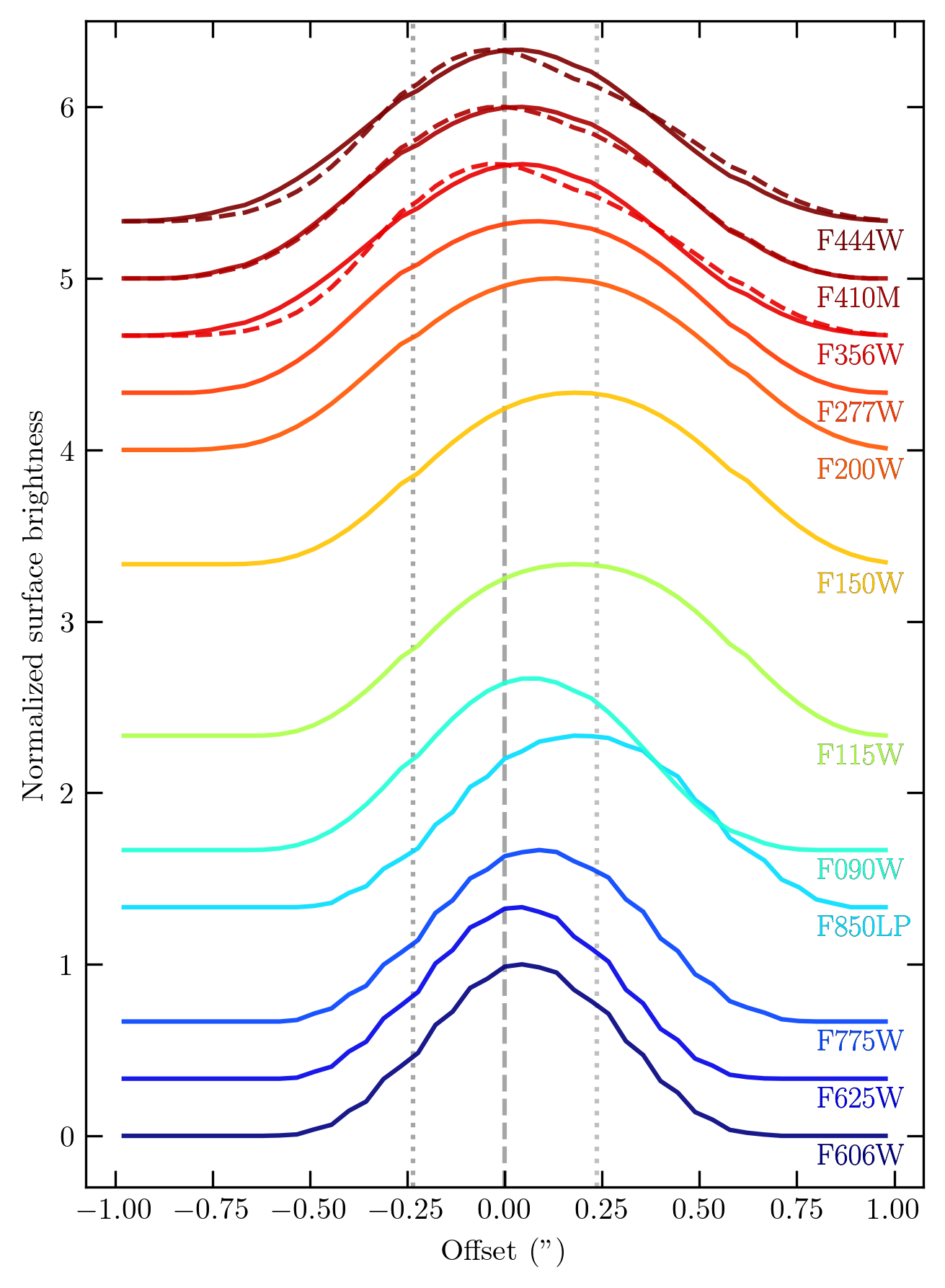}
\caption{
    \label{fig:SB_by_filter}
    Major axis 1-dimensional surface brightness profiles of the source-plane reconstruction for each filter.
    The large inferred inclination, clear asymmetry, 
    and 
    lack of an obvious center for all filters
    make this a more robust approach than a radial profile, as tested by \citet{Devour:2017aa, Devour:2019aa}.
    {\it Left:}
    Filter-by-filter surface brightness profiles along the major axis (as determined by the median position angle from \GALFIT\ models for all filters). Each bin along the profile, which shows the sum of pixels above a $2\sigma$-clipped threshold, is extracted from a 1.5$\arcsec$-wide box perpendicular to the major axis. The profile for each filter (each convolved to match the F444W PSF) is normalized between 0 and 1 and has arbitrary flux units, and they are shifted vertically for the purpose of legibility.
    Bins are spaced at intervals of $\approx 0.045\arcsec$ along the $2\arcsec$-long slit, comparable to the pixel size.
    The thickness of the curve indicates signal-to-noise, which is determined from the signal-to-noise ratio of the 95th percentile value of pixels in each bin. 
    Differing pixel scales and depths likely play a role in discrepancies between adjacent filters.
    Vertical dashed lines show the smallest effective radius (for F606W).
    {\it Center:} An RGB image showing all 8 NIRCam filters (red: F356W + F410M + F444W, green: F200W + F277W, blue: F090W + F115W + F150W). The white box shows the slit used to compute the surface brightness profiles, with the middle line showing the major axis orientation.
    The beams in the lower left show the FWHM and orientation of representative source-plane PSFs for F115W (cyan) and F444W (red). These are created by ray-tracing a model PSF placed at the location of the galaxy centroid in the less-magnified counter-image ($1a$) into the source-plane, which has the least spatial variation in PSF. While the more magnified images can offer better resolution on their own, this source-plane reconstruction combines all images together, effectively smearing the structure to the resolution of the counter-image.
    {\it Right:} Inferred surface brightness profiles in the same manner as the left panel, but applied to our \GALFIT-modeled light profiles. 
    Dashed curves for the three reddest NIRCAM filters show the 2-component models, which shift the profiles slightly.
}
\end{figure*}

While gravitational lensing is nominally achromatic (i.e., all wavelengths originating from the same location of the source-plane are deflected in the same manner), differential magnification can arise due to different source-plane distributions of light at different wavelengths. This effect is easily observable for \ElA, as the redder, central core of the galaxy (which is significantly more prominent beyond 2 $\mu$m) is precisely aligned with some of the highest-magnification regions of the source-plane near the caustic curves (as we discuss in more detail in \S \ref{sec:5th_image}). As the total magnification for each filter is a flux-weighted average of the magnification map, the long-wavelength NIRCam filters with this prominent core are amplified much more strongly.

In Fig.~\ref{fig:mag_by_filter}, we illustrate this trend towards larger magnifications at longer wavelengths between 0.5 and 5 $\mu$m. Shorter than 1 $\mu$m (observed-frame), the emission is confined to an outer region of the disk, with a substantially different morphology from that of the long-wavelength bands of NIRCam. Given this, it is not surprising that the magnification for rest-frame UV differs from that for rest-frame near-IR. The magnitude of this difference is notable: the near-IR stellar continuum is amplified by a magnification $\mu \sim 8 - 12$, 
while the rest-frame UV is amplified by only 
$\mu \sim 2 - 3$. This factor of $\sim 4$ difference in magnification appears to be driven primarily by the alignment of the 
red galaxy core 
with two caustic cusps in very close proximity (see Fig.~\ref{fig:IP_LW_reconstruction}). As total magnification is weighted by flux, the lack of signal in this region for shorter-wavelength filters dilutes their magnification and weights more highly the lower magnification ($\mu\sim2$) from the outer regions of the galaxy.
One can readily observe this in Figs.~\ref{fig:IP_SW_reconstruction} and \ref{fig:IP_LW_reconstruction}, where the northeast image is dominated by emission at wavelengths longer than $\sim$1.5$\mu$m.
This is effectively a manifestation of the well-known lensing size bias, in which more compact (or more highly concentrated) objects are often magnified more strongly (e.g., \citealt{Hezaveh:2012aa, Robertson:2020ac}).  

At present, there are relatively few robust measurements of differential magnification to which we can compare.
In simulations of lensed DSFG samples, \citet{Serjeant:2012aa} predicted differential magnifications between older to younger stellar populations of log[$\mu_{\star}$/$\mu_{\rm SFR}$] = $0 \pm 0.26$ in the case of $\mu > 10$, or log[$\mu_{\star}$/$\mu_{\rm SFR}$] = $0 \pm 0.14$ for modest magnifications $2 < \mu < 5$. 
For observed lensed DSFGs, \citet{Calanog:2014aa} found a typical differential magnification between near-IR and far-IR (880 $\mu$m) of 
$\mu_{\rm FIR}$/$\mu_{\rm NIR} \sim 1.5$ (i.e. +0.18 dex).
Thus, the $\sim$0.6 dex difference in our case clearly represents an extreme case of differential magnification.
By assuming a median magnification of $\mu\sim 7$ for all filters, for example, this would have the effect of underestimating the UV flux by a factor of $\sim 2-3$, and overestimating the IR flux by a factor of $\sim 1.5$, the combined effect of which could severely impact the inferred galaxy properties.
However, this is only the case for the full galaxy-integrated photometry. For the aperture we use at the location of image $1a$, as we discuss in \S \ref{sec:SED_demag}, the median magnification for all \HST/\JWST\ filters is $\mu_{1a,{\rm med}} = 2.7 \pm 0.2$. This smaller uncertainty reflects the empirical scatter as a function of wavelength, but may underestimate systematic uncertainties that affect all filters.

\subsection{A dust-obscured nucleus or inside-out growth\textemdash or both?} 
\label{sec:dusty_nucleus}

As shown in Figs.~\ref{fig:R_e_by_filter} and \ref{fig:SB_by_filter}, there is actually not a high degree of variation in the light profile from $\sim 1.5 - 4.4 \mu$m, all of which is longward of the 4000\AA-break (observed-frame $1.3~\mu$m). 
At first, this consistency in light profile from rest-frame near-UV to near-IR might appear to rule out large gradients in dust attenuation, which would push the half-light radii of shorter wavelengths to larger values\footnote{This is because the expected central peak in surface brightness becomes increasingly suppressed for bluer filters when there is a radially-decreasing profile of dust attenuation, such that an increasingly large aperture is required to contain half of the galaxy's observed flux.} \citep{Nelson:2016aa, Bondi:2018aa, Tacchella:2018aa, Suess:2019aa, Popping:2022ac}. This runs somewhat contrary
to the expectation that IR-luminous galaxies would have high concentrations of dust (and high IR luminosity surface densities) within the central kiloparsec (as for ultra-luminous infrared galaxies, or LIRGs and ULIRGs, e.g., \citealt{Sanders:1996aa, Lonsdale:2006aa, Armus:2009aa, Rujopakarn:2011aa}). 
As we point out later in 
this section, this could also be the result of inclination effects.
Recently, \citet{Miller:2022ab} found that 70\% of star-forming galaxies at $z\sim2$ demonstrated $U-V$/$V-J$ color gradients caused primarily by gradients in dust attenuation (also \citealt{Liu:2017aa, Wang:2017ab}). For the same sample, 23\% of the objects showed color gradients that were in line with gradients in stellar population properties (such as specific star formation rate).

At wavelengths shorter than the Balmer and 4000$\AA$ breaks ($<1.2\mu$m observed-frame), we find that the effective radius decreases gradually to $0.24\arcsec$, about 40\% smaller than the sizes in the rest-frame near-IR.
In an analysis of the resolved dust continuum of star-forming galaxies simulated with FIRE (Feedback in Realistic Environments; \citealt{Hopkins:2014aa,Hopkins:2018aa}), \citet{Cochrane:2019aa} showed that rest-frame UV was significantly correlated spatially with both cool, dense gas (i.e. regions of active star formation) and holes in the dust distribution. In both cases, this leads to a very clumpy spatial distribution, often offset from the near-IR structure. Similarly, \citet{Ma:2018ab} found that these UV-bright clumps (which can dominate the emission) are likely poor, biased tracers of stellar mass. 
On the other hand, while only for $z>7$ galaxies, \citet{Yang:2022ac} found only a modest decrease in size from rest-frame optical to UV. 
\citet{Shibuya:2015aa} likewise found close agreement in median UV and optical sizes for $z\sim1-2$ and stellar mass $\log(M_\star/M_\odot) = 9 - 11$, although with large scatter.
However, \citet{Ma:2018ab} also point out that parametric size measurements for very clumpy, irregular galaxies will break down and will primarily capture only the separations between individual bright clumps.
Given the patchy, nearly non-contiguous structure of \ElA\ in the UV, we consider this to be a likely explanation for the suppressed UV sizes.

In Table~\ref{tab:properties}, we report basic (magnification-corrected) properties derived from fitting the SED of \ElA\ as a whole, while also repeating this for the inner and outer disk regions separately. 
In correcting the photometry for lensing magnification, we compare two options: $a)$ applying filter-by-filter magnification corrections, versus $b)$ applying a median magnification factor,
for which uncertainty is given by the empirical scatter for the set of filters,
as discussed in \S \ref{sec:SED_demag}. 
For our purposes, we primarily use the 
median magnification, which is more physically motivated in that magnification is unlikely to change significantly between adjacent filters for image $1a$ (given the extensive photometric sampling available, and given the minimal magnification gradient in this region of the image plane).

A similar analysis of an un-lensed galaxy would likely be significantly impeded by the PSF, but the magnification of \ElA\ means that this central region is resolved by NIRCam (such that the observed color gradient is not driven only by variation in PSF with wavelength). The characterization of PSFs for NIRCam is still ongoing; so far it has been found that the PSFs can vary spatially across the field-of-view (e.g. \citealt{Zhuang:2023aa}) and temporally over timescales of a month (e.g. \citealt{Nardiello:2022aa}), both of order $\sim5\%$. 
For this reason, it is desirable to examine and calibrate color gradients in objects with angular sizes much larger than the instrumental resolution, such as lensed galaxies. While gravitational lens modeling introduces its own systematic uncertainties, some quantities (such as color variations) can actually be interpreted in the image plane, requiring no assumptions about the intervening mass.

When we perform SED fitting on the inner vs. outer disk in the prescribed manner,
we find that the implied attenuation is larger in the galaxy center than in the outskirts ($A_V = 2.0$ vs. $A_V = 1.4$). We also observe that the half-light radius for the dust continuum is $R_e \approx 0.5\arcsec$, which is reasonably consistent with the UV/optical sizes. This may be due in part to the coarser resolution of the 1 mm ALMA continuum imaging relative to \JWST, but there is not an obvious factor of $2-4\times$ difference in effective radius for the UV vs. far-IR, which has been observed in many other studies of DSFGs (including \citealt{Barro:2013aa, Barro:2016aa, Calanog:2014aa, Chen:2015aa, Ikarashi:2015aa, Simpson:2015ab, Tadaki:2015aa, Hodge:2016aa, Lutz:2016aa,  Massardi:2018aa, Talia:2018aa, Gullberg:2019aa, Tadaki:2020aa, Pantoni:2021ab}, and references therein, for a sampling).
Moreover, \citet{Lang:2019aa} found that the rest-frame far-IR continuum emission was not just more compact than the rest-frame optical light, but also more compact than the stellar mass distribution itself (described by the half-mass radius), indicating a higher central concentration of star formation relative to existing stellar mass.

The central light profile for $\lambda = 3 - 5 ~\mu$m (S{\'e}rsic $n \sim 0.4$) is also steeper than shorter wavelengths (Fig.~\ref{fig:R_e_by_filter}), 
implying a greater concentration in stellar mass than in the distribution of more recent star formation (probed by UV continuum).
We caution still that these values might be subject to systematic effects arising from applying \GALFIT\ to the source plane.
Given the highly asymmetric nature of the disk, with the red central bulge appearing offset to one end, we also examine the surface brightness profiles in a non-parametric fashion in Fig.~\ref{fig:SB_by_filter}. 
This approach is similar to the inclination-independent linear surface brightness profile developed by \citep{Devour:2017aa, Devour:2019aa}.
We extract the 1-dimensional surface brightness profiles along a slit $2\arcsec$ in length for each filter (oriented at the median position angle from \GALFIT\ and centered at the F444W peak, with a width of $1.5\arcsec$ perpendicular to the major axis).
Each image was first convolved in the image-plane with a kernel to match the PSF of the F444W filter, before reconstructing in the source plane.
Pixels in each bin are sigma-clipped at a conservative $2\sigma$-level and then summed. The curve thickness denotes signal-to-noise, determined by taking the 95th-percentile value of the distribution of S/N pixel values in each bin (which indirectly incorporates the number of pixels above the noise threshold). In the right panel of Fig.~\ref{fig:SB_by_filter}, we repeat this process exactly, but for the \GALFIT\ modeling results (including the 2-component models for F356W, F410M, and F444W) to show the inferred, modeled profiles.
We can visualize the increase in galaxy size from $\sim 0.6 - 1.2~\mu$m (observed), but relative consistency at longer wavelengths, although the peaks are not co-spatial. 

As discussed previously, a number of issues plague the measurement of photometry in the source plane. These include the need for a proper accounting of the sky background (which becomes distorted and limited in scope after ray-tracing the lensed structure to the source plane). In capturing one-dimensional surface brightness profiles along a slit in Fig.~\ref{fig:SB_by_filter}, we can examine consistency between filters while re-normalizing the profiles so that the sky background does not need to be removed. Moreover, while the source-plane PSF may vary across the object, this effect may be mitigated by averaging within each bin along the length of the slit. 
Variation in the PSF from filter to filter can be observed in the smoothness of each profile, but the structure is resolved for all filters and so the overall profile shape is not hindered by the PSF.

\begin{figure}[ht!]
\includegraphics[width=0.43\textwidth]{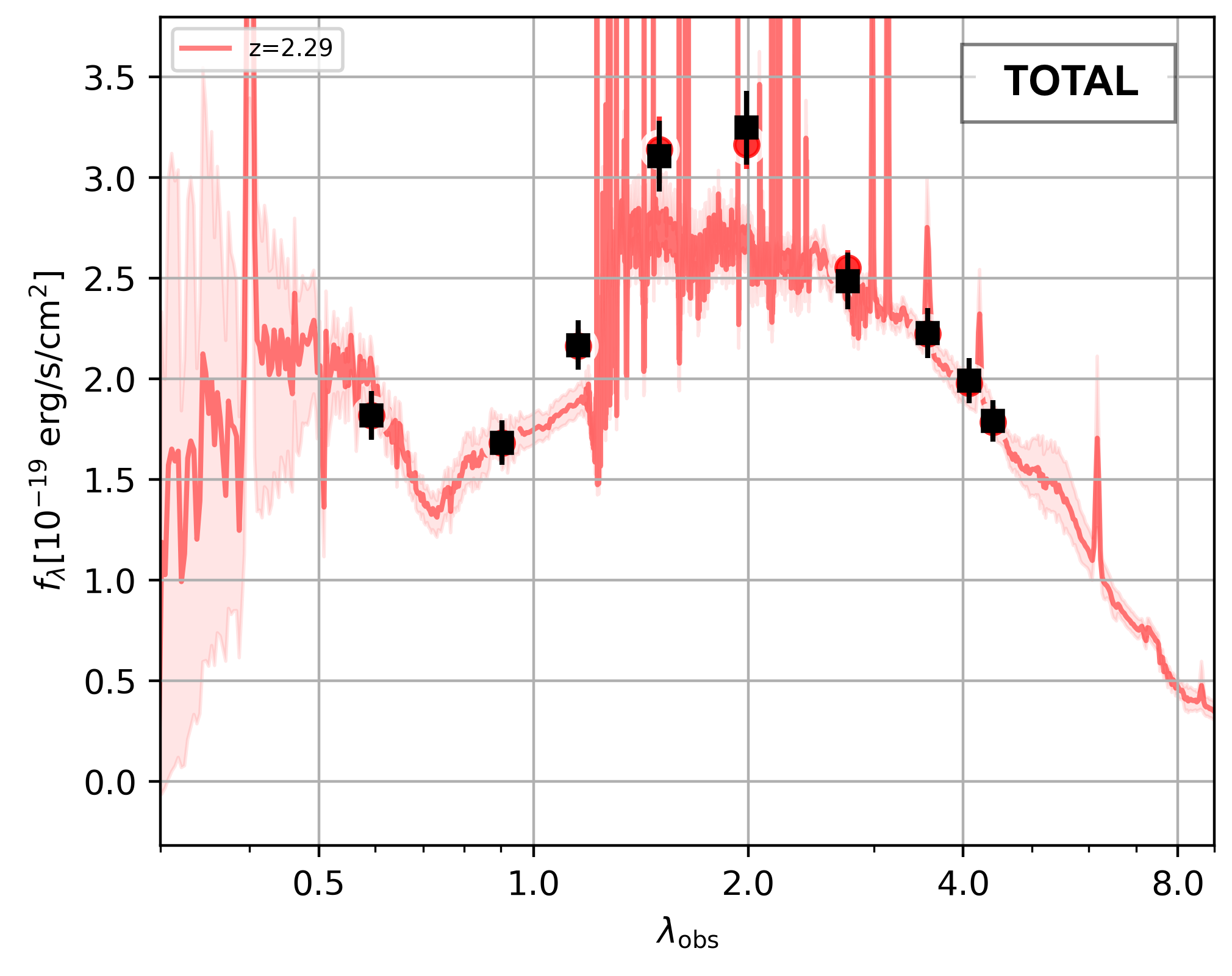}
\includegraphics[width=0.43\textwidth]{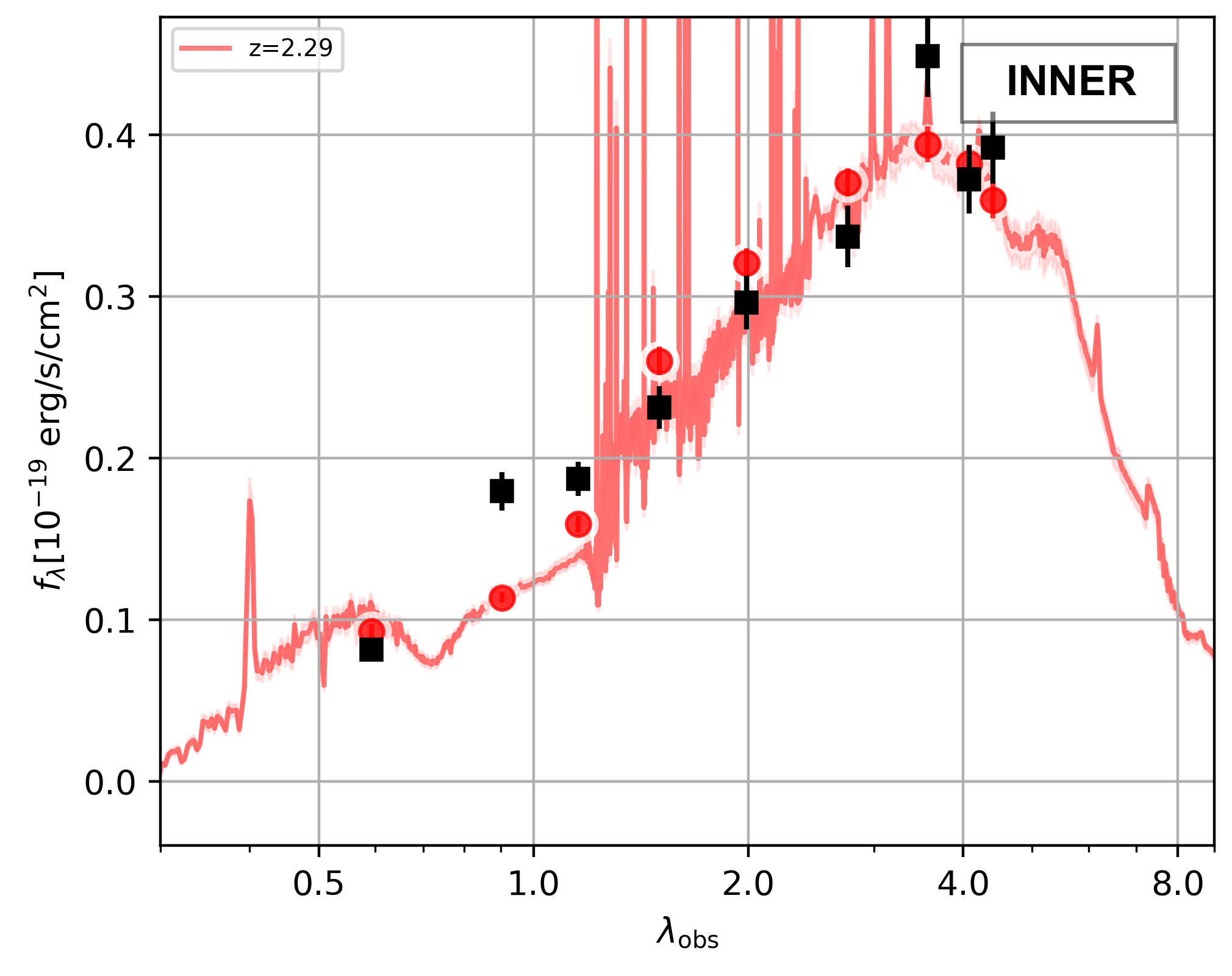}
\includegraphics[width=0.43\textwidth]{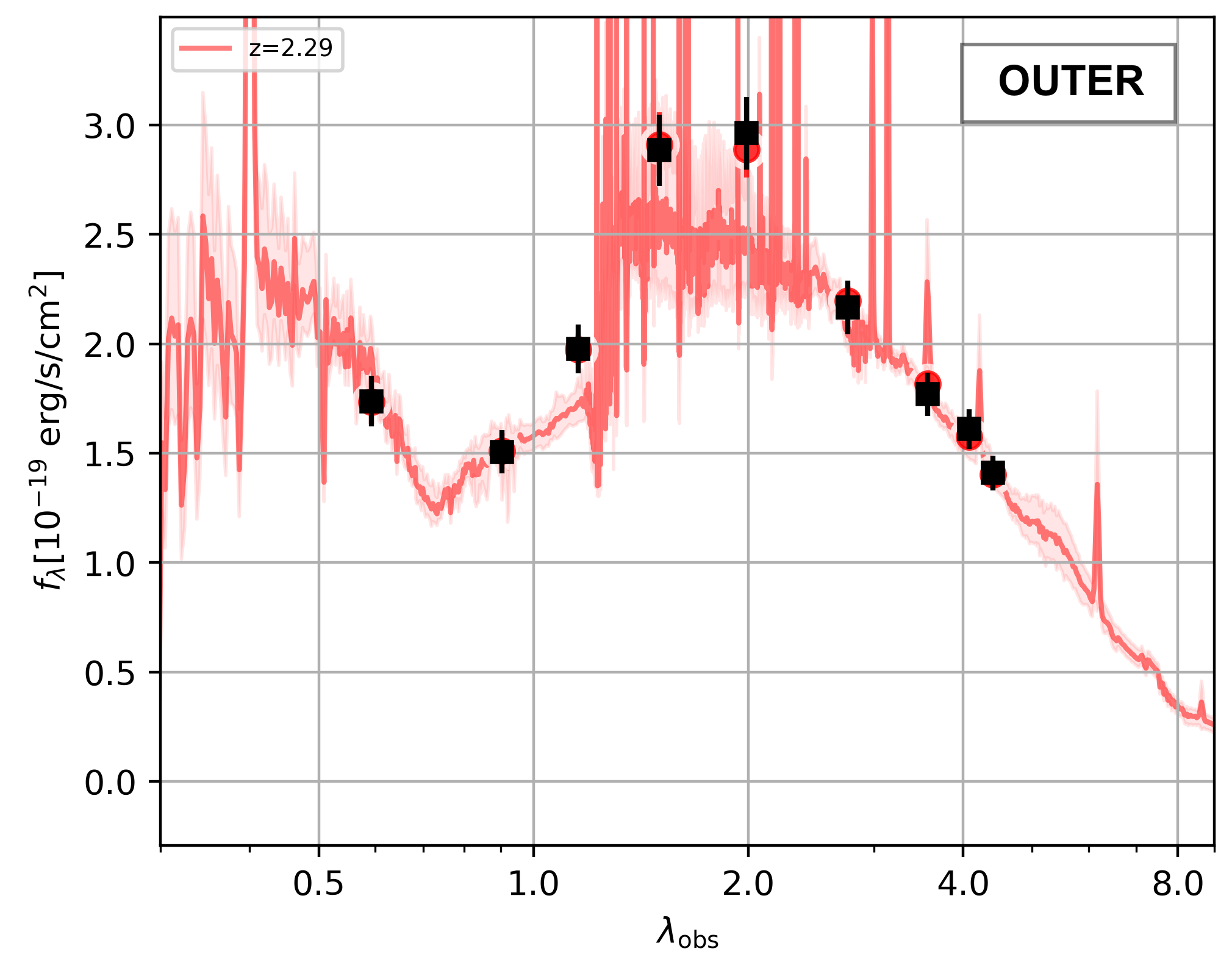}
\caption{
    SED-fitting results for image $1a$, as with Fig.~\ref{fig:EAZY}, but fluxes from Table~\ref{tab:properties} are de-magnified using the median magnification $\mu_{\rm 1a} = 2.7 \pm 0.2$ (see Table~\ref{tab:mags_1a}). 
    The SED is spatially decomposed into the total disk ({\it top}), inner $\sim$1 kpc ({\it middle}), and outer disk ({\it bottom}) according to the method described in \S \ref{sec:SED_demag}.
    \label{fig:EAZY_demag}
}
\end{figure}

The spatial offsets we observe might be due to a patchy and uneven dust distribution (e.g., \citealt{Liu:2013aa, Casey:2014ab, Popping:2017ab, Narayanan:2018ab, Ma:2019ad}). Likewise, the western portion of the source appearing bluer than the eastern side (see e.g., middle panel of Fig.~\ref{fig:SB_by_filter}) could be the result of a nearly edge-on inclination ($i = 63\degrees \pm 7 \degrees$; \S \ref{sec:SP_size}) coupled with a dust lane. 
This interpretation would indicate that the western portion is the far side of the disk, for which less stellar light is attenuated (e.g., \citealt{Hubble:1943aa, de-Vaucouleurs:1958aa}).
In this scenario, it is possible that the size measured in the bluer filters is dominated by the light profile of the far side of the disk.
However, given that the inclination-independent 1-dimensional surface brightness profiles in Fig.~\ref{fig:SB_by_filter} are consistent in width beyond $\sim$1.2 $\mu$m (rest-frame $\sim$350 $\mu$m), this explanation appears to be unlikely.

Another interpretation might be that the spatial offset is the result of
dust driven off the plane of the disk by being entrained within outflows and strong stellar winds, as seen locally in the starburst galaxy M82 (\citealt{Sanders:1971aa, Alton:1999aa, Roussel:2010aa, Hutton:2014aa}).
Local LIRGs and ULIRGs also contain a significant fraction of active galactic nuclei (AGN), of order 20\% \citep{Goulding:2010aa,Petric:2011aa} and 40-60\% \citep{Armus:2007aa, Desai:2007aa, Veilleux:2009aa}, respectively.
Similarly, this could also be reminiscent of star formation triggered by jet-driven shocks compressing the interstellar medium, as in the case of Centaurus A \citep{Crockett:2012aa}.
Additionally, such color asymmetries and spatial offsets can be seen in the local Universe in (ultra-)luminous infrared galaxies (LIRGs and ULIRGs).
One particularly illustrative example is the mid-stage merger, VV 114 (e.g., \citealt{Knop:1994aa, Yun:1994aa, Iono:2004aa, Saito:2015aa, Evans:2022ac}), for which the gas kinematics are dominated by the remnants of orbiting disks. In this case, a dust lane from one of the disk remnants reddens the other component, which may be quite similar to what we are seeing with \ElA.

\subsubsection{SED-derived SFR and $M_\star$} 
\label{sec:SED_SFR}

In order to have an effective radius consistent with that of near-IR, but with a shallower core from dust attenuation, the rest-frame UV light would need to truncate more steeply at large radii.
This itself would suggest an age gradient with massive, more short-lived and more UV-emitting stellar populations in the center than in the outskirts. 
A radially-increasing age gradient would also manifest as a  truncation in the extent of UV emission, from the lack of hot, young stars in the galaxy outskirts and from metal absorption in the cooler stars that are present. 

Yet,
this presumed age gradient is not reflected in the SED-derived specific star formation rate (sSFR $\equiv {\rm SFR}/M_\star$), which is greater (i.e. closer to a starburst) for the outer disk relative to the inner disk
by a factor of 3 ($1.5 \pm 0.9~{\rm Gyr}^{-1}$ 
versus $0.51 \pm 0.04~{\rm Gyr}^{-1}$). The respective SED fits are shown in Fig.~\ref{fig:EAZY_demag}; see also \citet{Tacchella:2018aa}, for example. 
At the redshift of \ElA, the specific SFR at the main sequence is 
${\rm sSFR}_{\rm MS} \sim 2~{\rm Gyr}^{-1}$ \citep{Speagle:2014aa}.
We find the total sSFR for the galaxy to be consistent with the main sequence at $1.7 \pm 0.4~{\rm Gyr}^{-1}$, but we caution that SFR derived from the UV-to-NIR SED can be underestimated without incorporating the far-IR that probes the dust SED (e.g., \citealt{Elbaz:2018aa}).
Likewise, stellar masses from the UV-to-NIR SED alone can also be underestimated by up to $\sim$0.3 dex for massive DSFGs ($M_\star \gtrsim 10^{10}~M_\odot$; \citealt{Battisti:2019aa}).
Unfortunately, the expected differential magnification and the low angular resolution of existing millimeter imaging (or low sensitivity, for 870 $\mu$m) makes a more robust measurement unfeasible at present.
Still, it is clear that this object does not match the hyper-luminous regime ($L_{IR} > 10^{13}~L_\odot$) seen for the most extreme DSFGs and the AGN-driven, hot Dust-Obscured Galaxies (DOGs; \citealt{Dey:2008aa, Pope:2008aa, Eisenhardt:2012aa, Wu:2012aa, Wu:2014aa, Tsai:2015aa, Penney:2020aa}) identified with the {\it Wide-field IR Survey Explorer} (\WISE; \citealt{Wright:2010aa}) or the {\it Spitzer Space Telescope} \citep{Werner:2004aa}.

To test the robustness of the rising specific star-formation rate that we find in the source plane, especially given the influence of the lensing-distorted PSF, we repeat this analysis in the image plane. We use a fixed elliptical aperture ($0.3\arcsec \times 0.2\arcsec$, PA$=50\degrees$) centered on the F444W peak for the inner component, and an elliptical annulus (outer extent $1.5\arcsec \times 1.0\arcsec$, PA$=50\degrees$) for the outer component, and perform PSF-matched photometry on images convolved to the F444W PSF. These elliptical apertures roughly account for the distortion by lensing of the circular apertures used in the source plane when mapped to the image plane.
Since specific star-formation rate is the ratio of SFR to $M_\star$, both of which are approximately equally scaled by lensing magnification $\mu$, one would expect the sSFR to remain consistent in the image plane. With this approach, we find an inner sSFR $= 0.44\pm0.03$ Gyr$^{-1}$ and outer sSFR $= 1.3\pm0.2$ Gyr$^{-1}$,
both slightly lower than the values measured in the source-plane, but in agreement within the $1\sigma$ uncertainties. The lower sSFR estimates could also simply be the result of the elliptical image-plane apertures not matching the exact source-plane regions covered by the circular apertures. Thus, since the rise in sSFR towards outer radii that we observe in the source plane is also recovered through PSF-matched image-plane photometry, we have reason to believe this trend is real.

Similarly, we find relative consistency in the surface density of star formation\footnote{$\Sigma_{\rm SFR} \equiv {\rm SFR} / A_d$, where $A_d$ is the disk surface area, $2 \pi (R_{\rm out}^2 - R_{\rm in}^2)$ for the inner and outer radial bounds of each region, $R_{\rm in}$ and $R_{\rm out}$.} ($\Sigma_{\rm SFR}$) 
between the inner 1 kpc and outer disk, at
$\Sigma_{\rm SFR} = 0.88 \pm 0.03~M_\odot~{\rm yr}^{-1}~{\rm kpc}^{-2}$, versus $0.7 \pm 0.3~M_\odot~{\rm yr}^{-1}~{\rm kpc}^{-2}$, respectively. 
These are also in line with dust-corrected SFR surface densities found by \citet{Tacchella:2018aa} for $z$$\sim$2 galaxies ($10 < \log[M_\star/M_\odot] < 11$) on the star-forming main sequence 
For context, \citet{Heckman:2001ab} suggest that $0.1~M_\odot~{\rm yr}^{-1}~{\rm kpc}^{-2}$ is the threshold beyond which starburst-driven galactic winds become ubiquitous (see also \citealt{Sharma:2016aa}).

In the context of the inside-out growth framework of stellar mass assembly (an extensive set of literature, including \citealt{Carrasco:2010aa, van-Dokkum:2010aa, van-Dokkum:2014aa, Oser:2010aa, La-Barbera:2012aa, Nelson:2012aa, Nelson:2016ab, van-der-Wel:2014ab, Wuyts:2013aa, Dimauro:2022aa}),
these findings support a scenario where \ElA\ has begun quenching its star formation in an inner bulge. With a mass $M_\star = [1.3 \pm 0.1] \times 10^{10}~M_\odot$, this is roughly a quarter of the total stellar mass, $M_\star = [4.9 \pm 1.4] \times 10^{10}~M_\odot$.
This is %
in line with
the observation that galaxies above $M_\star = 10^{10}~M_\odot$ 
seem to have an increasing percentage of active star-forming regions with increasing radius (e.g., \citealt{Rowlands:2018aa}), interpreted as a signpost of inside-out quenching.
It may also be the case that feedback from a (likely buried) AGN is contributing to quenching in the galaxy center, but this is challenging to confirm directly (e.g., review by \citealt{Fabian:2012aa}; also \citealt{Smethurst:2016aa,Smethurst:2017aa, Luo:2022ab}).
Recent findings have also shown that
the most strongly star-forming systems at high-$z$ can host active star formation widely distributed on kiloparsec scales throughout the disk (e.g., \citealt{Bussmann:2012aa, Bussmann:2013aa,Hodge:2019aa, Kamieneski:2023aa}).
These often-clumpy sites of star formation may be driven by gravitational instabilities in the outer gas disk, and may eventually migrate radially inwards and add to the growth of a central bulge (e.g., \citealt{Ceverino:2010aa, Ceverino:2012aa, Wuyts:2012aa}). This results in morphological quenching \citep{Martig:2009aa}, by which the bulge stabilizes gas in the galaxy from gravitational collapse.

Still, central starbursts are not uncommon (e.g., \citealt{Ellison:2018aa,Ellison:2020aa, Wang:2019aa}), and 
galaxies above the main sequence are more likely to have amplified star formation rate surface densities in their central region (out to $\sim$0.5 $R_e$). 
Likewise, the ratio of inner-to-outer sSFR is expected to have a positive correlation with distance above the main sequence \citep{Tacchella:2016aa, Tacchella:2016ab, Tacchella:2018aa, Zolotov:2015aa}.
In a decomposition of the bulges vs. disks of main sequence galaxies (also at $z$$\sim$2.3), \citet{Cutler:2023aa} found that central regions had star formation histories (SFH) dominated by recent ($<100$ Myr) bursts, whereas outer portions of the disk showed much steadier SFHs. They suggested that this could be driven by dissipative gas compaction events
driven by disk instabilities 
(e.g., \citealt{Dekel:2014aa})
or accretion of star-forming clumps \citep{Dekel:2009ab, van-Dokkum:2013aa}.
In the case of \ElA, major or minor mergers might have been responsible for tidal disruption of gas towards the center \citep{Bournaud:2011aa}, as is the case with ULIRGs in the local Universe.
However, unlike local ULIRGs, strongly star-forming systems at Cosmic Noon are a much more heterogeneous sample beyond only major mergers (e.g., \citealt{Hayward:2011aa}).
We presume that \ElA\ is not yet
a dynamically-settled disk, although follow-up kinematic observations would provide helpful context as to its level of rotational support.

We now revisit the observation that rest-frame near-UV sizes are similar to those at near-IR, which seems to stand in contrast to the simple picture of widely-distributed star formation and a central concentration of greater dust attenuation (which would actually broaden the UV sizes relative to longer wavelengths). 
Perhaps the simplest observation comes from the shifting peak along the major axis from far-UV to near-IR, seen in Fig.~\ref{fig:SB_by_filter}, with offsets of order $\sim$0.3$\arcsec$, or $\sim$2.5 kpc in physical units. This separation is substantial, as it is comparable to (or even exceeds) the effective radius in the far-UV (Table~\ref{tab:mags}).
This suggests that the far-UV and near-IR are primarily tracing spatially-distinct components\textemdash similar to the common offsets in UV versus far-IR\textemdash as might be the case for the late stages of a galaxy merger like VV 114. 
We conclude that it's likely that the complex distribution of dust and ongoing star formation in \ElA\ is not easily characterized by a 1-dimensional surface brightness profile on its own, and a full spatially-resolved mapping of dust, gas, and star formation (e.g., through integral field spectroscopy) is a better-suited observational approach.

\begin{deluxetable*}{ccccccc}
\tablecaption{Basic galaxy properties derived with \EAZY\ SED-fitting, including an inner vs. outer decomposition.
\label{tab:properties}}
\tablehead{
\colhead{Component} & \colhead{SFR} & \colhead{$M_\star$} & \colhead{sSFR $\equiv$ SFR / $M_\star$} & \colhead{$\Sigma_{\rm SFR}$} & \colhead{$\Sigma_{M_\star}$} & \colhead{$A_V$} \\
\colhead{} & \colhead{[$M_\odot~{\rm yr}^{-1}$]} &  \colhead{[$10^{10}~M_\odot$]} &  \colhead{[Gyr$^{-1}$]} & \colhead{[$M_\odot~{\rm yr}^{-1}~{\rm kpc}^{-2}$]} & \colhead{[$10^8~M_\odot~{\rm kpc}^{-2}$]} & \colhead{}
}
\startdata
{\it Median}\tablenotemark{$a$} & & & & &  & \\
Total &  $81_{-2}^{+7}$    &   4.9 $\pm$ 1.4 & 1.7 $\pm$ 0.4 & 1.3	$\pm$ 0.1 & 7.6 $\pm$	2.2 & $1.6 \pm0.3$	\\
Inner disk &  $6.7_{-0.2}^{+0.2}$   &   1.3 $\pm$ 0.1 & 0.51 $\pm$ 0.04 & 0.88	$\pm$ 0.03 & 17 $\pm$ 1 & $2.04 \pm 0.03$\\
Outer disk &  $71_{-33}^{+16}$   &   4.9 $\pm$ 2.1 & 1.5 $\pm$ 0.9 & 0.7 $\pm$ 0.3 & 5.0 $\pm$ 2.1 & $1.4 \pm0.3$	\\
\hline
{\it Filter-by-filter}\tablenotemark{$b$}  & &  & & & & \\
Total &  $80_{-4}^{+9}$    &   5.2 $\pm$ 2.0 & 1.6 $\pm$ 0.5 & 1.2 $\pm$	0.1 & 8.1 $\pm$ 3.1 & $1.8 \pm 0.3$	\\
Inner disk &  $6.6_{-0.5}^{+0.5}$   &   1.3 $\pm$ 0.2 & 0.5 $\pm$ 0.1 & 0.87	$\pm$ 0.07 & 17 $\pm$ 3 & $2.0 \pm 0.1$	\\
Outer disk &  $73_{-3}^{+10}$   &   4.7 $\pm$ 1.6 & 1.7 $\pm$ 0.5  & 0.74	$\pm$ 0.07 & 5 $\pm$ 2 & $1.6 \pm 0.4$	\\
\enddata
\tablenotetext{a}{Median magnification 
for Aperture 1 (see Fig.~\ref{fig:IP_LW_reconstruction}): 
$\mu_{\rm med} = 2.7 \pm 0.2$.
}
\tablenotetext{b}{See magnifications in Table~\ref{tab:mags_1a}.}
\tablecomments{
Inner vs. outer decomposition is performed at projected radius $0.13\arcsec \approx 1.1$ kpc. The outer bound for the outer disk is taken to be at 4.1 kpc, consistent with the effective radius of the dust continuum. As the SED fit for the inner vs. outer disk is completely independent, summed values may not agree with total values.
All values are corrected for lensing magnification, but using different methods\textemdash i.e. corrected with a single median value, corrected by regime (UV, optical, or near-IR), or corrected filter-by-filter. We don't use the filter-based method for the decomposition due to the large uncertainties. 
}
\end{deluxetable*}

\begin{deluxetable}{c|cc}
\tablecaption{Magnification and bulge-to-total fraction ($B/T$) for image $1a$ only, used to correct the photometry in Table \ref{tab:mags}. \label{tab:mags_1a}}
\tablehead{
\colhead{Filter} &  \colhead{Magnification $\mu_{1a}$} & \colhead{$B/T$} 
}
\startdata
HST/F606W  &    2.7	$\pm$	0.4	& 0.04		\\
HST/F625W  &    2.7	$\pm$	0.4	& 0.10		\\
HST/F775W  &    2.7	$\pm$	0.4	& 0.10		\\
HST/F814W  &    3.0	$\pm$	0.6	& \textemdash\tablenotemark{$a$}		\\
HST/F850LP &    2.9	$\pm$	0.5	& 0.19		\\
JWST/F090W &    2.6	$\pm$	0.4	& 0.11		\\
HST/F105W  &    3.0	$\pm$	0.6	& 0.15		\\
JWST/F115W &    2.6	$\pm$	0.4	& 0.09		\\
HST/F125W  &    2.8	$\pm$	0.5	& 0.07		\\
HST/F140W  &    3.1	$\pm$	0.7	& 0.06		\\
JWST/F150W &    2.8	$\pm$	0.5	& 0.07		\\
HST/F160W  &    2.9	$\pm$	0.5	& 0.10		\\
JWST/F200W &    2.8	$\pm$	0.5	& 0.09		\\
JWST/F277W &    2.7	$\pm$	0.4	& 0.13		\\
JWST/F356W &    2.7	$\pm$	0.4	& 0.20		\\
JWST/F410M &    2.6	$\pm$	0.4	& 0.19		\\
JWST/F444W &    2.6	$\pm$	0.4	& 0.22		\\
\enddata
\tablenotetext{a}{F814W is excluded from the SED decomposition due to its coincidence with a detector gap.}
\tablecomments{
%
In using an aperture centered on this lensed counter-image, differential magnification is minimized, while still effectively capturing flux from the entire background object.}
\end{deluxetable}

\subsection{Five-image morphology from a bimodal mass distribution} 
\label{sec:5th_image}

One of the more remarkable aspects of \ElA's lensing morphology is the complex structure apparent primarily in the longest wavelength filters of NIRCam ($2.7-4.4~\mu$m), which reveal that the bright central, reddish core of the background galaxy seen in the southwest image (1a) is multiply imaged an additional 4 times towards the northeast ($1b$ through $1e$). This was identified in the F444W image, which showed 5 distinct peaks. Using predictions from our lens model, this appears to be due to a ``beak-to-beak" caustic metamorphosis, where the usual diamond-shaped caustics of each foreground mass merges together into a single contiguous caustic network. The position of \ElA's core at a short fold region between two closely-separated cusps (see Figs.~\ref{fig:IP_SW_reconstruction} and \ref{fig:IP_LW_reconstruction}) closely mirrors 
Fig. 6 of \citet{Schneider:1986aa}. 
There is a point-like source (most prominent in the F200W filter) that lies barely on the interior of the southernmost caustic (something akin to an intermediate location of what is shown in Figs. 6b and 6d of \citeauthor{Schneider:1986aa}). 
This alignment results in 5 multiple images, but it is not obvious to what degree the core image is demagnified (in fact, it is likely magnified).
This analysis was extended from the binary point masses of \citet{Schneider:1986aa}
to be applied to binary isothermal potentials by
\citet{Shin:2008aa}, who found that the image configurations could include 3, 5, or even 7 multiple images.
Here, the 5-image morphology is closest to what \citeauthor{Shin:2008aa} label as configuration 5B-1 (their Fig. 11). 

Based on our current lens model, we estimate the magnification near the core image (labeled $1e$) to be $\mu \approx$1.5\textendash 8. This is seemingly confirmed by the reasonably comparable flux observed in the demagnified image vs. the other magnified images. However, there are likely some parts of the background DSFG that are lensed to have a demagnified image, in particular near the core of the secondary lensing galaxy. Pursuing this would benefit greatly from spectroscopic confirmation with other tracers (e.g. molecular gas), or using dust continuum at high-resolution.
For now,
we consider this interpretation of a magnified 5th image to be the most likely, given the 5-image morphology that we observe and the bimodal (nearly equal) mass distribution of the foreground. 

However, there is still a chance that the core image is demagnified (although the demagnification would need to be very weak).
While central demagnified images have long been predicted \citep{Dyer:1980aa, Burke:1981aa}, they are quite rarely observed, for three primary reasons: $a)$ they are typically expected near the bright central cores of galaxies, where very high dynamic range would be required to identify the small contrast of the light of the lensed object (except at long wavelengths where foreground contamination is low); $b)$ the images are demagnified, sometimes very strongly, so that they often fall below the limiting magnitudes of observations; and $c)$ they require a central surface mass density slope in the deflector that is shallower than $\Sigma \propto r^{-1}$ \citep{Rusin:2001ab}.
Few have been discovered to date 
(e.g.,
\citealt{Colley:1996aa, Winn:2004aa, Inada:2005aa, Frye:2007aa, Sharon:2012aa, Collett:2017aa, Ostrovski:2018aa, Muller:2020aa}), but 
they are uniquely valuable tools for probing the inner density profiles of galaxies
\citep{Wallington:1993aa, Rusin:2001ab, Keeton:2003aa, Tamura:2015aa, Wong:2015aa}.
Whether the 5th central image is demagnified or not, the discovery of this exotic morphology is exciting and will provide observational tests of rare lensing geometries, which in turn help constrain the mass density profiles of the lensing (foreground) galaxies (e.g., \citealt{Bozza:2016aa, Bozza:2020aa}).

The lensing morphology also lends some clues to what we might expect for the resolved distribution of dust continuum. 
The northeast component arises from only the small inner portion of the disk interior to the lensing caustic ($\sim 0.1\arcsec$, or $\sim800$pc, in radius), yet it is actually $\sim2$ times brighter than the southwest component at 1mm (see \citealt{Cheng:2023aa}). 
This implies that a significant fraction of the total 1mm flux arises from this high-magnification region.
It is not possible to constrain this fraction at present, however, as the northeast component consists of multiple images blended together (see e.g. the 870$\mu$m structure in Fig.~\ref{fig:IP_ALMA_reconstruction}), each of varying magnifications.
Higher-resolution follow-up observations could test this and directly map the extent of dust-obscured star formation in \ElA, which will have important implications for its current evolutionary stage and relation to the main sequence.

Finally, we also remark that the best-fit caustic network is asymmetric, in that there are two cusps in close proximity near the core of \ElA, but only a single cusp on the opposite end, towards the northwest. This is possibly due to the external shear included in the model, or to the orientation of the two SIE lenses relative to each other.
In future work, it may be worthwhile to employ the latest lensing models of the entire \ElG\ cluster (e.g. \citealt{Diego:2023ab, Frye:2023aa}) to make predictions for the external shear that the cluster potential contributes at this location, and to compare with the best-fit external shear amplitude of this work ($\gamma = 0.36 \pm 0.08$), which is quite substantial. 
We will refrain from drawing definitive 
conclusions in this regard,
as recent results from \citet{Etherington:2023aa} show that this added shear might only be to compensate for insufficient model complexity.
Yet, as discussed in \S \ref{sec:lens_params}, the orientation of the best-fit shear is what would be expected for weak lensing from the \ElG\ cluster.
We reserve a more in-depth analysis of these finer details for a future work, as additional constraints will help to refine and further test our lens model.

\section{Conclusions} 
\label{sec:conclusions}

\ElA\ is a $z$=2.291 ALMA-detected dusty star-forming galaxy that is strongly lensed by the massive $z=0.87$ galaxy cluster, \ElG. New \JWST/NIRCam imaging reveals an extraordinary lensing morphology, with some portions of the background source being multiply-imaged five times. After developing a new parametric lens model specifically tailored to this object using \lenstool, we take advantage of the magnification and amplification to examine the surface brightness distribution of the rest-frame UV through near-IR ($0.18 - 1.3~\mu$m) in this pilot study. The near-IR is expected to be a better proxy for stellar mass, as it 
is more sensitive to light from redder, lower-mass stars that account for most of a galaxy's stellar mass.
This is in contrast to rest-frame UV or blue optical imaging, which is both more affected by dust obscuration and more easily biased by the light from short-lived massive (and more luminous) stars.

We find the peculiar result that the source-plane surface brightness distribution is largely consistent between 1.2 and $4.4~\mu$m (or rest-frame $0.35 - 1.3~\mu$m).We interpret the ostensible match of the near-UV light 
and the stellar distribution 
as the combined effect of a central concentration of dust and widespread recent star formation. Dust attenuation may truncate the peak of UV light that would otherwise be seen at relatively smaller radii than longer wavelengths, 
which should have the effect of broadening the UV sizes. Since this is not the case, we interpret that the UV and near-IR are tracing different components (suggested also by the $\sim$2.5 kpc offset in their peaks), perhaps from an ongoing late-stage coalescence of a galaxy merger.

In a decomposition of the SED for an inner, bulge-like component for the central kiloparsec versus the outskirts of the disk, we find elevated specific star formation rates at larger radii: $1.5 \pm 0.9~{\rm Gyr}^{-1}$ in the outer disk and $0.51 \pm 0.04~{\rm Gyr}^{-1}$ for the inner disk.
We interpret the suppressed sSFR in the galaxy center to be a possible indication of the early stages of inside-out galaxy quenching, 
possibly driven by the inward radial migration of star-forming clumps distributed throughout the disk.
The overall (magnification-corrected) intrinsic SFR is estimated to be $81^{+7}_{-2}~M_\odot~{\rm yr}^{-1}$, consistent with the star-forming main sequence at $z=2.3$ (and suggesting an inferred intrinsic IR luminosity of $L_{IR} \approx 5 \times 10^{11}~L_\odot$). This may be an underestimate, however, as it is derived only from the UV, optical, and near-IR SED. Higher-resolution far-IR measurements are needed to test if the DSFG might lie above the main sequence, as the peak of archival low-resolution imaging coincides with the region of greatest magnification, such that the intrinsic (de-magnified) far-IR SED is poorly constrained.

We note the large differential magnification indicated by our lens model, by which the near-IR ($\mu_{\rm tot}\sim$ 8\textendash12) is magnified approximately four times more strongly than the UV ($\mu_{\rm tot}\sim$ 2\textendash3). This is directly related to the surface brightness profiles we observe, as dust in the galaxy center (which is coincidentally aligned with the high-magnification region of the source plane) preferentially attenuates shorter wavelengths. Finally, we briefly discuss the implications of a 5-image lensing morphology arising from a bimodal foreground mass distribution. These include the possibility of detecting central demagnified images, which are a powerful probe of the central mass densities of lensing galaxies and groups, but which have so far proven quite elusive to discover. The serendipitous alignment of the galaxy nucleus with a part of the source-plane that is both highly-magnified and multiply-imaged 5 times makes \ElA\ 
a very attractive target
for future study (e.g., in examining possible variability in an AGN).\\ \\


We thank the anonymous referee for their suggestions to improve this manuscript.
PSK would like to thank Ian Smail, Allison Noble, and Alex Pigarelli for helpful discussion in framing this work.
This work is based on observations made with the NASA/ESA/CSA James Webb Space Telescope. The data were obtained from the Mikulski Archive for Space Telescopes at the Space Telescope Science Institute, which is operated by the Association of Universities for Research in Astronomy, Inc., under NASA contract NAS 5-03127 for JWST. These observations are associated with JWST program \# 1176 (PEARLS; PI: R. Windhorst).
RAW, SHC, and RAJ acknowledge support from NASA JWST Interdisciplinary Scientist grants NAG5-12460, NNX14AN10G and 80NSSC18K0200 from GSFC.
This research is also based on observations made with the NASA/ESA Hubble Space Telescope obtained from the Space Telescope Science Institute, which is operated by the Association of Universities for Research in Astronomy, Inc., under NASA contract NAS 5—26555. These observations are associated with programs GO-12477, GO-12755, and GO-14096.
The specific observations from JWST and HST are available at: \dataset[doi:10.17909/7cnj-5y94]{https://doi.org/10.17909/7cnj-5y94}.
This paper makes use of the following ALMA data: ADS/JAO.ALMA\# 2013.1.01051.S, 2013.1.01358.S, 2015.1.01187.S, 2017.1.01621.S, and 2018.1.00035.L.
ALMA is a partnership of ESO (representing its member states), NSF (USA) and NINS (Japan), together with NRC (Canada), MOST and ASIAA (Taiwan), and KASI (Republic of Korea), in cooperation with the Republic of Chile. The Joint ALMA Observatory is operated by ESO, AUI/NRAO and NAOJ. The National Radio Astronomy Observatory is a facility of the National Science Foundation operated under cooperative agreement by Associated Universities, Inc.
This research has made extensive use of NASA’s Astrophysics Data System.

We also acknowledge the indigenous peoples of Arizona, including the Akimel O'odham (Pima) and Pee Posh (Maricopa) Indian Communities, whose care and keeping of the land has enabled us to be at ASU's Tempe campus in the Salt River Valley, where much of our work was conducted.


%

\facilities{JWST(NIRCam), HST(ACS,WFC3), ALMA}


\software{{\sc APLpy} \citep{Robitaille:2012aa,Robitaille:2019aa}
          {\sc astropy} \citep{Astropy-Collaboration:2013aa},
          {\sc blobcat} \citep{Hales:2012aa},
          {\sc casa} \citep{McMullin:2007aa},
          {\sc eazy} \citep{Brammer:2008aa, Brammer:2021ab},
          \GALFIT \citep{Peng:2002aa},
          \lenstool\ \citep{Kneib:1993aa, Kneib:1996aa, Jullo:2007aa, Jullo:2009aa}, 
          Ned Wright's Cosmology Calculator \citep{Wright:2006aa},
          {\sc photutils} \citep{Bradley:2022aa}
          {\sc SExtractor} \citep{Bertin:1996aa}
          }

\bibliography{RESEARCH_2017}

\begin{thebibliography}{}
\expandafter\ifx\csname natexlab\endcsname\relax\def\natexlab#1{#1}\fi
\providecommand{\url}[1]{\href{#1}{#1}}
\providecommand{\dodoi}[1]{doi:~\href{http://doi.org/#1}{\nolinkurl{#1}}}
\providecommand{\doeprint}[1]{\href{http://ascl.net/#1}{\nolinkurl{http://ascl.net/#1}}}
\providecommand{\doarXiv}[1]{\href{https://arxiv.org/abs/#1}{\nolinkurl{https://arxiv.org/abs/#1}}}

\bibitem[{{Alton} {et~al.}(1999){Alton}, {Davies}, \& {Bianchi}}]{Alton:1999aa}
{Alton}, P.~B., {Davies}, J.~I., \& {Bianchi}, S. 1999, \aap, 343, 51

\bibitem[{{Armus} {et~al.}(2007){Armus}, {Charmandaris}, {Bernard-Salas},
  {Spoon}, {Marshall}, {Higdon}, {Desai}, {Teplitz}, {Hao}, {Devost}, {Brandl},
  {Wu}, {Sloan}, {Soifer}, {Houck}, \& {Herter}}]{Armus:2007aa}
{Armus}, L., {Charmandaris}, V., {Bernard-Salas}, J., {et~al.} 2007, \apj, 656,
  148, \dodoi{10.1086/510107}

\bibitem[{{Armus} {et~al.}(2009){Armus}, {Mazzarella}, {Evans}, {Surace},
  {Sanders}, {Iwasawa}, {Frayer}, {Howell}, {Chan}, {Petric}, {Vavilkin},
  {Kim}, {Haan}, {Inami}, {Murphy}, {Appleton}, {Barnes}, {Bothun}, {Bridge},
  {Charmandaris}, {Jensen}, {Kewley}, {Lord}, {Madore}, {Marshall},
  {Melbourne}, {Rich}, {Satyapal}, {Schulz}, {Spoon}, {Sturm}, {U}, {Veilleux},
  \& {Xu}}]{Armus:2009aa}
{Armus}, L., {Mazzarella}, J.~M., {Evans}, A.~S., {et~al.} 2009, \pasp, 121,
  559, \dodoi{10.1086/600092}

\bibitem[{{Astropy Collaboration} {et~al.}(2013){Astropy Collaboration},
  {Robitaille}, {Tollerud}, {Greenfield}, {Droettboom}, {Bray}, {Aldcroft},
  {Davis}, {Ginsburg}, {Price-Whelan}, {Kerzendorf}, {Conley}, {Crighton},
  {Barbary}, {Muna}, {Ferguson}, {Grollier}, {Parikh}, {Nair}, {Unther},
  {Deil}, {Woillez}, {Conseil}, {Kramer}, {Turner}, {Singer}, {Fox}, {Weaver},
  {Zabalza}, {Edwards}, {Azalee Bostroem}, {Burke}, {Casey}, {Crawford},
  {Dencheva}, {Ely}, {Jenness}, {Labrie}, {Lim}, {Pierfederici}, {Pontzen},
  {Ptak}, {Refsdal}, {Servillat}, \&
  {Streicher}}]{Astropy-Collaboration:2013aa}
{Astropy Collaboration}, {Robitaille}, T.~P., {Tollerud}, E.~J., {et~al.} 2013,
  \aap, 558, A33, \dodoi{10.1051/0004-6361/201322068}

\bibitem[{{Barro} {et~al.}(2013){Barro}, {Faber}, {P{\'e}rez-Gonz{\'a}lez},
  {Koo}, {Williams}, {Kocevski}, {Trump}, {Mozena}, {McGrath}, {van der Wel},
  {Wuyts}, {Bell}, {Croton}, {Ceverino}, {Dekel}, {Ashby}, {Cheung},
  {Ferguson}, {Fontana}, {Fang}, {Giavalisco}, {Grogin}, {Guo}, {Hathi},
  {Hopkins}, {Huang}, {Koekemoer}, {Kartaltepe}, {Lee}, {Newman}, {Porter},
  {Primack}, {Ryan}, {Rosario}, {Somerville}, {Salvato}, \&
  {Hsu}}]{Barro:2013aa}
{Barro}, G., {Faber}, S.~M., {P{\'e}rez-Gonz{\'a}lez}, P.~G., {et~al.} 2013,
  \apj, 765, 104, \dodoi{10.1088/0004-637X/765/2/104}

\bibitem[{{Barro} {et~al.}(2016){Barro}, {Kriek}, {P{\'e}rez-Gonz{\'a}lez},
  {Trump}, {Koo}, {Faber}, {Dekel}, {Primack}, {Guo}, {Kocevski},
  {Mu{\~n}oz-Mateos}, {Rujopakarn}, \& {Seth}}]{Barro:2016aa}
{Barro}, G., {Kriek}, M., {P{\'e}rez-Gonz{\'a}lez}, P.~G., {et~al.} 2016,
  \apjl, 827, L32, \dodoi{10.3847/2041-8205/827/2/L32}

\bibitem[{{Barrufet} {et~al.}(2023){Barrufet}, {Oesch}, {Weibel}, {Brammer},
  {Bezanson}, {Bouwens}, {Fudamoto}, {Gonzalez}, {Gottumukkala}, {Illingworth},
  {Heintz}, {Holden}, {Labbe}, {Magee}, {Naidu}, {Nelson}, {Stefanon}, {Smit},
  {van Dokkum}, {Weaver}, \& {Williams}}]{Barrufet:2023ab}
{Barrufet}, L., {Oesch}, P.~A., {Weibel}, A., {et~al.} 2023, \mnras, 522, 449,
  \dodoi{10.1093/mnras/stad947}

\bibitem[{{Basu} {et~al.}(2016){Basu}, {Sommer}, {Erler}, {Eckert}, {Vazza},
  {Magnelli}, {Bertoldi}, \& {Tozzi}}]{Basu:2016aa}
{Basu}, K., {Sommer}, M., {Erler}, J., {et~al.} 2016, \apjl, 829, L23,
  \dodoi{10.3847/2041-8205/829/2/L23}

\bibitem[{{Battisti} {et~al.}(2019){Battisti}, {da Cunha}, {Grasha}, {Salvato},
  {Daddi}, {Davies}, {Jin}, {Liu}, {Schinnerer}, {Vaccari}, \& {COSMOS
  Collaboration}}]{Battisti:2019aa}
{Battisti}, A.~J., {da Cunha}, E., {Grasha}, K., {et~al.} 2019, \apj, 882, 61,
  \dodoi{10.3847/1538-4357/ab345d}

\bibitem[{{Bertin} \& {Arnouts}(1996)}]{Bertin:1996aa}
{Bertin}, E., \& {Arnouts}, S. 1996, \aaps, 117, 393,
  \dodoi{10.1051/aas:1996164}

\bibitem[{{Blain} {et~al.}(2002){Blain}, {Smail}, {Ivison}, {Kneib}, \&
  {Frayer}}]{Blain:2002aa}
{Blain}, A.~W., {Smail}, I., {Ivison}, R.~J., {Kneib}, J.-P., \& {Frayer},
  D.~T. 2002, \physrep, 369, 111, \dodoi{10.1016/S0370-1573(02)00134-5}

\bibitem[{{Bondi} {et~al.}(2018){Bondi}, {Zamorani}, {Ciliegi},
  {Smol{\v{c}}i{\'c}}, {Schinnerer}, {Delvecchio}, {Jim{\'e}nez-Andrade},
  {Liu}, {Lang}, {Magnelli}, {Murphy}, \& {Vardoulaki}}]{Bondi:2018aa}
{Bondi}, M., {Zamorani}, G., {Ciliegi}, P., {et~al.} 2018, \aap, 618, L8,
  \dodoi{10.1051/0004-6361/201834243}

\bibitem[{{Bournaud} {et~al.}(2011){Bournaud}, {Chapon}, {Teyssier}, {Powell},
  {Elmegreen}, {Elmegreen}, {Duc}, {Contini}, {Epinat}, \&
  {Shapiro}}]{Bournaud:2011aa}
{Bournaud}, F., {Chapon}, D., {Teyssier}, R., {et~al.} 2011, \apj, 730, 4,
  \dodoi{10.1088/0004-637X/730/1/4}

\bibitem[{{Bozza} \& {Melchiorre}(2016)}]{Bozza:2016aa}
{Bozza}, V., \& {Melchiorre}, C. 2016, \jcap, 2016, 040,
  \dodoi{10.1088/1475-7516/2016/03/040}

\bibitem[{{Bozza} {et~al.}(2020){Bozza}, {Pietroni}, \&
  {Melchiorre}}]{Bozza:2020aa}
{Bozza}, V., {Pietroni}, S., \& {Melchiorre}, C. 2020, Universe, 6, 106,
  \dodoi{10.3390/universe6080106}

\bibitem[{{Bradley} {et~al.}(2022){Bradley}, {Sip{\H{o}}cz}, {Robitaille},
  {Tollerud}, {Vin{\'\i}cius}, {Deil}, {Barbary}, {Wilson}, {Busko}, {Donath},
  {G{\"u}nther}, {Cara}, {Lim}, {Me{\ss}linger}, {Conseil}, {Bostroem},
  {Droettboom}, {Bray}, {Andersen Bratholm}, {Barentsen}, {Craig}, {Ginsburg},
  {Rathi}, {Pascual}, {Perren}, {Georgiev}, {De Val-Borro}, {Kerzendorf},
  {Bach}, \& {Quint}}]{Bradley:2022aa}
{Bradley}, L., {Sip{\H{o}}cz}, B., {Robitaille}, T., {et~al.} 2022,
  {astropy/photutils: 1.6.0}, 1.6.0, Zenodo,  Zenodo,
  \dodoi{10.5281/zenodo.7419741}

\bibitem[{{Brammer}(2021)}]{Brammer:2021ab}
{Brammer}, G. 2021, {gbrammer/eazy-py: Tagged release 2021}, 0.5.2, Zenodo,
  Zenodo, \dodoi{10.5281/zenodo.5012705}

\bibitem[{{Brammer} {et~al.}(2008){Brammer}, {van Dokkum}, \&
  {Coppi}}]{Brammer:2008aa}
{Brammer}, G.~B., {van Dokkum}, P.~G., \& {Coppi}, P. 2008, \apj, 686, 1503,
  \dodoi{10.1086/591786}

\bibitem[{{Burke}(1981)}]{Burke:1981aa}
{Burke}, W.~L. 1981, \apjl, 244, L1, \dodoi{10.1086/183466}

\bibitem[{{Bussmann} {et~al.}(2012){Bussmann}, {Gurwell}, {Fu}, {Smith}, {Dye},
  {Auld}, {Baes}, {Baker}, {Bonfield}, {Cava}, {Clements}, {Cooray}, {Coppin},
  {Dannerbauer}, {Dariush}, {De Zotti}, {Dunne}, {Eales}, {Fritz}, {Hopwood},
  {Ibar}, {Ivison}, {Jarvis}, {Kim}, {Leeuw}, {Maddox}, {Micha{\l}owski},
  {Negrello}, {Pascale}, {Pohlen}, {Riechers}, {Rigby}, {Scott}, {Temi}, {Van
  der Werf}, {Wardlow}, {Wilner}, \& {Verma}}]{Bussmann:2012aa}
{Bussmann}, R.~S., {Gurwell}, M.~A., {Fu}, H., {et~al.} 2012, \apj, 756, 134,
  \dodoi{10.1088/0004-637X/756/2/134}

\bibitem[{{Bussmann} {et~al.}(2013){Bussmann}, {P{\'e}rez-Fournon}, {Amber},
  {Calanog}, {Gurwell}, {Dannerbauer}, {De Bernardis}, {Fu}, {Harris}, {Krips},
  {Lapi}, {Maiolino}, {Omont}, {Riechers}, {Wardlow}, {Baker}, {Birkinshaw},
  {Bock}, {Bourne}, {Clements}, {Cooray}, {De Zotti}, {Dunne}, {Dye}, {Eales},
  {Farrah}, {Gavazzi}, {Gonz{\'a}lez Nuevo}, {Hopwood}, {Ibar}, {Ivison},
  {Laporte}, {Maddox}, {Mart{\'{\i}}nez-Navajas}, {Michalowski}, {Negrello},
  {Oliver}, {Roseboom}, {Scott}, {Serjeant}, {Smith}, {Smith}, {Streblyanska},
  {Valiante}, {van der Werf}, {Verma}, {Vieira}, {Wang}, \&
  {Wilner}}]{Bussmann:2013aa}
{Bussmann}, R.~S., {P{\'e}rez-Fournon}, I., {Amber}, S., {et~al.} 2013, \apj,
  779, 25, \dodoi{10.1088/0004-637X/779/1/25}

\bibitem[{{Calanog} {et~al.}(2014){Calanog}, {Fu}, {Cooray}, {Wardlow}, {Ma},
  {Amber}, {Baker}, {Baes}, {Bock}, {Bourne}, {Bussmann}, {Casey}, {Chapman},
  {Clements}, {Conley}, {Dannerbauer}, {De Zotti}, {Dunne}, {Dye}, {Eales},
  {Farrah}, {Furlanetto}, {Harris}, {Ivison}, {Kim}, {Maddox}, {Magdis},
  {Messias}, {Micha{\l}owski}, {Negrello}, {Nightingale}, {O'Bryan}, {Oliver},
  {Riechers}, {Scott}, {Serjeant}, {Simpson}, {Smith}, {Timmons}, {Thacker},
  {Valiante}, \& {Vieira}}]{Calanog:2014aa}
{Calanog}, J.~A., {Fu}, H., {Cooray}, A., {et~al.} 2014, \apj, 797, 138,
  \dodoi{10.1088/0004-637X/797/2/138}

\bibitem[{{Calistro Rivera} {et~al.}(2018){Calistro Rivera}, {Hodge}, {Smail},
  {Swinbank}, {Weiss}, {Wardlow}, {Walter}, {Rybak}, {Chen}, {Brandt},
  {Coppin}, {da Cunha}, {Dannerbauer}, {Greve}, {Karim}, {Knudsen},
  {Schinnerer}, {Simpson}, {Venemans}, \& {van der
  Werf}}]{Calistro-Rivera:2018aa}
{Calistro Rivera}, G., {Hodge}, J.~A., {Smail}, I., {et~al.} 2018, \apj, 863,
  56, \dodoi{10.3847/1538-4357/aacffa}

\bibitem[{{Caminha} {et~al.}(2022){Caminha}, {Grillo}, {Rosati}, {Liu},
  {Acebron}, {Bergamini}, {Caputi}, {Mercurio}, {Tozzi}, {Vanzella}, {Demarco},
  {Frye}, {Rosani}, \& {Sharon}}]{Caminha:2022ab}
{Caminha}, G.~B., {Grillo}, C., {Rosati}, P., {et~al.} 2022, arXiv e-prints,
  arXiv:2209.02718.
\newblock \doarXiv{2209.02718}

\bibitem[{{Caputi} {et~al.}(2021){Caputi}, {Caminha}, {Fujimoto}, {Kohno},
  {Sun}, {Egami}, {Deshmukh}, {Tang}, {Ao}, {Bradley}, {Coe}, {Espada},
  {Grillo}, {Hatsukade}, {Knudsen}, {Lee}, {Magdis}, {Morokuma-Matsui},
  {Oesch}, {Ouchi}, {Rosati}, {Umehata}, {Valentino}, {Vanzella}, {Wang}, {Wu},
  \& {Zitrin}}]{Caputi:2021aa}
{Caputi}, K.~I., {Caminha}, G.~B., {Fujimoto}, S., {et~al.} 2021, \apj, 908,
  146, \dodoi{10.3847/1538-4357/abd4d0}

\bibitem[{{Carrasco} {et~al.}(2010){Carrasco}, {Conselice}, \&
  {Trujillo}}]{Carrasco:2010aa}
{Carrasco}, E.~R., {Conselice}, C.~J., \& {Trujillo}, I. 2010, \mnras, 405,
  2253, \dodoi{10.1111/j.1365-2966.2010.16645.x}

\bibitem[{{Casey} {et~al.}(2014{\natexlab{a}}){Casey}, {Narayanan}, \&
  {Cooray}}]{Casey:2014aa}
{Casey}, C.~M., {Narayanan}, D., \& {Cooray}, A. 2014{\natexlab{a}}, \physrep,
  541, 45, \dodoi{10.1016/j.physrep.2014.02.009}

\bibitem[{{Casey} {et~al.}(2014{\natexlab{b}}){Casey}, {Scoville}, {Sanders},
  {Lee}, {Cooray}, {Finkelstein}, {Capak}, {Conley}, {De Zotti}, {Farrah},
  {Fu}, {Le Floc'h}, {Ilbert}, {Ivison}, \& {Takeuchi}}]{Casey:2014ab}
{Casey}, C.~M., {Scoville}, N.~Z., {Sanders}, D.~B., {et~al.}
  2014{\natexlab{b}}, \apj, 796, 95, \dodoi{10.1088/0004-637X/796/2/95}

\bibitem[{{Ceverino} {et~al.}(2010){Ceverino}, {Dekel}, \&
  {Bournaud}}]{Ceverino:2010aa}
{Ceverino}, D., {Dekel}, A., \& {Bournaud}, F. 2010, \mnras, 404, 2151,
  \dodoi{10.1111/j.1365-2966.2010.16433.x}

\bibitem[{{Ceverino} {et~al.}(2012){Ceverino}, {Dekel}, {Mandelker},
  {Bournaud}, {Burkert}, {Genzel}, \& {Primack}}]{Ceverino:2012aa}
{Ceverino}, D., {Dekel}, A., {Mandelker}, N., {et~al.} 2012, \mnras, 420, 3490,
  \dodoi{10.1111/j.1365-2966.2011.20296.x}

\bibitem[{{Chapman} {et~al.}(2004){Chapman}, {Smail}, {Windhorst}, {Muxlow}, \&
  {Ivison}}]{Chapman:2004aa}
{Chapman}, S.~C., {Smail}, I., {Windhorst}, R., {Muxlow}, T., \& {Ivison},
  R.~J. 2004, \apj, 611, 732, \dodoi{10.1086/422383}

\bibitem[{{Chen} {et~al.}(2015){Chen}, {Smail}, {Swinbank}, {Simpson}, {Ma},
  {Alexander}, {Biggs}, {Brandt}, {Chapman}, {Coppin}, {Danielson},
  {Dannerbauer}, {Edge}, {Greve}, {Ivison}, {Karim}, {Menten}, {Schinnerer},
  {Walter}, {Wardlow}, {Wei{\ss}}, \& {van der Werf}}]{Chen:2015aa}
{Chen}, C.-C., {Smail}, I., {Swinbank}, A.~M., {et~al.} 2015, \apj, 799, 194,
  \dodoi{10.1088/0004-637X/799/2/194}

\bibitem[{{Chen} {et~al.}(2017){Chen}, {Hodge}, {Smail}, {Swinbank}, {Walter},
  {Simpson}, {Calistro Rivera}, {Bertoldi}, {Brandt}, {Chapman}, {da Cunha},
  {Dannerbauer}, {De Breuck}, {Harrison}, {Ivison}, {Karim}, {Knudsen},
  {Wardlow}, {Wei{\ss}}, \& {van der Werf}}]{Chen:2017aa}
{Chen}, C.-C., {Hodge}, J.~A., {Smail}, I., {et~al.} 2017, \apj, 846, 108,
  \dodoi{10.3847/1538-4357/aa863a}

\bibitem[{{Cheng} {et~al.}(2023){Cheng}, {Huang}, {Smail}, {Yan}, {Cohen},
  {Jansen}, {Windhorst}, {Ma}, {Koekemoer}, {Willmer}, {Willner}, {Diego},
  {Frye}, {Conselice}, {Ferreira}, {Petric}, {Yun}, {Gim}, {Polletta},
  {Duncan}, {Holwerda}, {R{\"o}ttgering}, {Honor}, {Hathi}, {Kamieneski},
  {Adams}, {Coe}, {Broadhurst}, {Summers}, {Tompkins}, {Driver}, {Grogin},
  {Marshall}, {Pirzkal}, {Robotham}, \& {Ryan}}]{Cheng:2023aa}
{Cheng}, C., {Huang}, J.-S., {Smail}, I., {et~al.} 2023, \apjl, 942, L19,
  \dodoi{10.3847/2041-8213/aca9d0}

\bibitem[{{Cheung} {et~al.}(2012){Cheung}, {Faber}, {Koo}, {Dutton}, {Simard},
  {McGrath}, {Huang}, {Bell}, {Dekel}, {Fang}, {Salim}, {Barro}, {Bundy},
  {Coil}, {Cooper}, {Conselice}, {Davis}, {Dom{\'\i}nguez}, {Kassin},
  {Kocevski}, {Koekemoer}, {Lin}, {Lotz}, {Newman}, {Phillips}, {Rosario},
  {Weiner}, \& {Willmer}}]{Cheung:2012aa}
{Cheung}, E., {Faber}, S.~M., {Koo}, D.~C., {et~al.} 2012, \apj, 760, 131,
  \dodoi{10.1088/0004-637X/760/2/131}

\bibitem[{{Cochrane} {et~al.}(2019){Cochrane}, {Hayward},
  {Angl{\'e}s-Alc{\'a}zar}, {Lotz}, {Parsotan}, {Ma}, {Kere{\v{s}}},
  {Feldmann}, {Faucher-Gigu{\`e}re}, \& {Hopkins}}]{Cochrane:2019aa}
{Cochrane}, R.~K., {Hayward}, C.~C., {Angl{\'e}s-Alc{\'a}zar}, D., {et~al.}
  2019, \mnras, 488, 1779, \dodoi{10.1093/mnras/stz1736}

\bibitem[{{Cochrane} {et~al.}(2021){Cochrane}, {Best}, {Smail}, {Ibar},
  {Cheng}, {Swinbank}, {Molina}, {Sobral}, \&
  {Dudzevi{\v{c}}i{\={u}}t{\.{e}}}}]{Cochrane:2021aa}
{Cochrane}, R.~K., {Best}, P.~N., {Smail}, I., {et~al.} 2021, \mnras, 503,
  2622, \dodoi{10.1093/mnras/stab467}

\bibitem[{{Coe} {et~al.}(2019){Coe}, {Salmon}, {Brada{\v{c}}}, {Bradley},
  {Sharon}, {Zitrin}, {Acebron}, {Cerny}, {Cibirka}, {Strait},
  {Paterno-Mahler}, {Mahler}, {Avila}, {Ogaz}, {Huang}, {Pelliccia}, {Stark},
  {Mainali}, {Oesch}, {Trenti}, {Carrasco}, {Dawson}, {Rodney}, {Strolger},
  {Riess}, {Jones}, {Frye}, {Czakon}, {Umetsu}, {Vulcani}, {Graur}, {Jha},
  {Graham}, {Molino}, {Nonino}, {Hjorth}, {Selsing}, {Christensen},
  {Kikuchihara}, {Ouchi}, {Oguri}, {Welch}, {Lemaux}, {Andrade-Santos}, {Hoag},
  {Johnson}, {Peterson}, {Past}, {Fox}, {Agulli}, {Livermore}, {Ryan}, {Lam},
  {Sendra-Server}, {Toft}, {Lovisari}, \& {Su}}]{Coe:2019aa}
{Coe}, D., {Salmon}, B., {Brada{\v{c}}}, M., {et~al.} 2019, \apj, 884, 85,
  \dodoi{10.3847/1538-4357/ab412b}

\bibitem[{{Collett} {et~al.}(2017){Collett}, {Buckley-Geer}, {Lin}, {Bacon},
  {Nichol}, {Nord}, {Morice-Atkinson}, {Amara}, {Birrer}, {Kuropatkin}, {More},
  {Papovich}, {Romer}, {Tessore}, {Abbott}, {Allam}, {Annis},
  {Benoit-L{\'e}vy}, {Brooks}, {Burke}, {Carrasco Kind}, {Castander},
  {D'Andrea}, {da Costa}, {Desai}, {Diehl}, {Doel}, {Eifler}, {Flaugher},
  {Frieman}, {Gerdes}, {Goldstein}, {Gruen}, {Gschwend}, {Gutierrez}, {James},
  {Kuehn}, {Kuhlmann}, {Lahav}, {Li}, {Lima}, {Maia}, {March}, {Marshall},
  {Martini}, {Melchior}, {Miquel}, {Plazas}, {Rykoff}, {Sanchez}, {Scarpine},
  {Schindler}, {Schubnell}, {Sevilla-Noarbe}, {Smith}, {Sobreira}, {Suchyta},
  {Swanson}, {Tarle}, {Tucker}, \& {Walker}}]{Collett:2017aa}
{Collett}, T.~E., {Buckley-Geer}, E., {Lin}, H., {et~al.} 2017, \apj, 843, 148,
  \dodoi{10.3847/1538-4357/aa76e6}

\bibitem[{{Colley} {et~al.}(1996){Colley}, {Tyson}, \&
  {Turner}}]{Colley:1996aa}
{Colley}, W.~N., {Tyson}, J.~A., \& {Turner}, E.~L. 1996, \apjl, 461, L83,
  \dodoi{10.1086/310015}

\bibitem[{{Cresci} {et~al.}(2010){Cresci}, {Mannucci}, {Maiolino}, {Marconi},
  {Gnerucci}, \& {Magrini}}]{Cresci:2010aa}
{Cresci}, G., {Mannucci}, F., {Maiolino}, R., {et~al.} 2010, \nat, 467, 811,
  \dodoi{10.1038/nature09451}

\bibitem[{{Crockett} {et~al.}(2012){Crockett}, {Shabala}, {Kaviraj},
  {Antonuccio-Delogu}, {Silk}, {Mutchler}, {O'Connell}, {Rejkuba}, {Whitmore},
  \& {Windhorst}}]{Crockett:2012aa}
{Crockett}, R.~M., {Shabala}, S.~S., {Kaviraj}, S., {et~al.} 2012, \mnras, 421,
  1603, \dodoi{10.1111/j.1365-2966.2012.20418.x}

\bibitem[{{Cunow}(2001)}]{Cunow:2001aa}
{Cunow}, B. 2001, \mnras, 323, 130, \dodoi{10.1046/j.1365-8711.2001.04140.x}

\bibitem[{{Cutler} {et~al.}(2023){Cutler}, {Giavalisco}, {Ji}, \&
  {Cheng}}]{Cutler:2023aa}
{Cutler}, S.~E., {Giavalisco}, M., {Ji}, Z., \& {Cheng}, Y. 2023, \apj, 945,
  97, \dodoi{10.3847/1538-4357/acb5e9}

\bibitem[{{de Jong}(1996)}]{de-Jong:1996aa}
{de Jong}, R.~S. 1996, \aap, 313, 377, \dodoi{10.48550/arXiv.astro-ph/9604010}

\bibitem[{{de Vaucouleurs}(1958)}]{de-Vaucouleurs:1958aa}
{de Vaucouleurs}, G. 1958, \apj, 127, 487, \dodoi{10.1086/146476}

\bibitem[{{Dekel} \& {Burkert}(2014)}]{Dekel:2014aa}
{Dekel}, A., \& {Burkert}, A. 2014, \mnras, 438, 1870,
  \dodoi{10.1093/mnras/stt2331}

\bibitem[{{Dekel} {et~al.}(2009){Dekel}, {Birnboim}, {Engel}, {Freundlich},
  {Goerdt}, {Mumcuoglu}, {Neistein}, {Pichon}, {Teyssier}, \&
  {Zinger}}]{Dekel:2009ab}
{Dekel}, A., {Birnboim}, Y., {Engel}, G., {et~al.} 2009, \nat, 457, 451,
  \dodoi{10.1038/nature07648}

\bibitem[{{Desai} {et~al.}(2007){Desai}, {Armus}, {Spoon}, {Charmandaris},
  {Bernard-Salas}, {Brandl}, {Farrah}, {Soifer}, {Teplitz}, {Ogle}, {Devost},
  {Higdon}, {Marshall}, \& {Houck}}]{Desai:2007aa}
{Desai}, V., {Armus}, L., {Spoon}, H.~W.~W., {et~al.} 2007, \apj, 669, 810,
  \dodoi{10.1086/522104}

\bibitem[{{Devour} \& {Bell}(2017)}]{Devour:2017aa}
{Devour}, B.~M., \& {Bell}, E.~F. 2017, \mnras, 468, L31,
  \dodoi{10.1093/mnrasl/slx021}

\bibitem[{{Devour} \& {Bell}(2019)}]{Devour:2019aa}
---. 2019, \apjs, 244, 3, \dodoi{10.3847/1538-4365/ab339c}

\bibitem[{{Dey} {et~al.}(2008){Dey}, {Soifer}, {Desai}, {Brand}, {Le Floc'h},
  {Brown}, {Jannuzi}, {Armus}, {Bussmann}, {Brodwin}, {Bian}, {Eisenhardt},
  {Higdon}, {Weedman}, \& {Willner}}]{Dey:2008aa}
{Dey}, A., {Soifer}, B.~T., {Desai}, V., {et~al.} 2008, \apj, 677, 943,
  \dodoi{10.1086/529516}

\bibitem[{{Diego} {et~al.}(2023){Diego}, {Meena}, {Adams}, {Broadhurst}, {Dai},
  {Coe}, {Frye}, {Kelly}, {Koekemoer}, {Pascale}, {Willner}, {Zackrisson},
  {Zitrin}, {Windhorst}, {Cohen}, {Jansen}, {Summers}, {Tompkins}, {Conselice},
  {Driver}, {Yan}, {Grogin}, {Marshall}, {Pirzkal}, {Robotham}, {Ryan},
  {Willmer}, {Bradley}, {Caminha}, {Caputi}, {Carleton}, \&
  {Kamieneski}}]{Diego:2023ab}
{Diego}, J.~M., {Meena}, A.~K., {Adams}, N.~J., {et~al.} 2023, \aap, 672, A3,
  \dodoi{10.1051/0004-6361/202245238}

\bibitem[{{Dimauro} {et~al.}(2022){Dimauro}, {Daddi}, {Shankar}, {Cattaneo},
  {Huertas-Company}, {Bernardi}, {Caro}, {Dupke}, {H{\"a}u{\ss}ler},
  {Johnston}, {Cortesi}, {Mei}, \& {Peletier}}]{Dimauro:2022aa}
{Dimauro}, P., {Daddi}, E., {Shankar}, F., {et~al.} 2022, \mnras, 513, 256,
  \dodoi{10.1093/mnras/stac884}

\bibitem[{{Dressler} {et~al.}(1987){Dressler}, {Lynden-Bell}, {Burstein},
  {Davies}, {Faber}, {Terlevich}, \& {Wegner}}]{Dressler:1987aa}
{Dressler}, A., {Lynden-Bell}, D., {Burstein}, D., {et~al.} 1987, \apj, 313,
  42, \dodoi{10.1086/164947}

\bibitem[{{Driver} {et~al.}(2007{\natexlab{a}}){Driver}, {Allen}, {Liske}, \&
  {Graham}}]{Driver:2007aa}
{Driver}, S.~P., {Allen}, P.~D., {Liske}, J., \& {Graham}, A.~W.
  2007{\natexlab{a}}, \apjl, 657, L85, \dodoi{10.1086/513106}

\bibitem[{{Driver} {et~al.}(2007{\natexlab{b}}){Driver}, {Popescu}, {Tuffs},
  {Liske}, {Graham}, {Allen}, \& {de Propris}}]{Driver:2007ab}
{Driver}, S.~P., {Popescu}, C.~C., {Tuffs}, R.~J., {et~al.} 2007{\natexlab{b}},
  \mnras, 379, 1022, \dodoi{10.1111/j.1365-2966.2007.11862.x}

\bibitem[{{Dyer} \& {Roeder}(1980)}]{Dyer:1980aa}
{Dyer}, C.~C., \& {Roeder}, R.~C. 1980, \apjl, 238, L67, \dodoi{10.1086/183260}

\bibitem[{{Eisenhardt} {et~al.}(2012){Eisenhardt}, {Wu}, {Tsai}, {Assef},
  {Benford}, {Blain}, {Bridge}, {Condon}, {Cushing}, {Cutri}, {Evans},
  {Gelino}, {Griffith}, {Grillmair}, {Jarrett}, {Lonsdale}, {Masci}, {Mason},
  {Petty}, {Sayers}, {Stanford}, {Stern}, {Wright}, \&
  {Yan}}]{Eisenhardt:2012aa}
{Eisenhardt}, P. R.~M., {Wu}, J., {Tsai}, C.-W., {et~al.} 2012, \apj, 755, 173,
  \dodoi{10.1088/0004-637X/755/2/173}

\bibitem[{{Elbaz} {et~al.}(2018){Elbaz}, {Leiton}, {Nagar}, {Okumura},
  {Franco}, {Schreiber}, {Pannella}, {Wang}, {Dickinson}, {D{\'\i}az-Santos},
  {Ciesla}, {Daddi}, {Bournaud}, {Magdis}, {Zhou}, \&
  {Rujopakarn}}]{Elbaz:2018aa}
{Elbaz}, D., {Leiton}, R., {Nagar}, N., {et~al.} 2018, \aap, 616, A110,
  \dodoi{10.1051/0004-6361/201732370}

\bibitem[{{Ellison} {et~al.}(2018){Ellison}, {S{\'a}nchez}, {Ibarra-Medel},
  {Antonio}, {Mendel}, \& {Barrera-Ballesteros}}]{Ellison:2018aa}
{Ellison}, S.~L., {S{\'a}nchez}, S.~F., {Ibarra-Medel}, H., {et~al.} 2018,
  \mnras, 474, 2039, \dodoi{10.1093/mnras/stx2882}

\bibitem[{{Ellison} {et~al.}(2020){Ellison}, {Thorp}, {Pan}, {Lin}, {Scudder},
  {Bluck}, {S{\'a}nchez}, \& {Sargent}}]{Ellison:2020aa}
{Ellison}, S.~L., {Thorp}, M.~D., {Pan}, H.-A., {et~al.} 2020, \mnras, 492,
  6027, \dodoi{10.1093/mnras/staa001}

\bibitem[{{Etherington} {et~al.}(2023){Etherington}, {Nightingale}, {Massey},
  {Tam}, {Cao}, {Niemiec}, {He}, {Robertson}, {Li}, {Amvrosiadis}, {Cole},
  {Diego}, {Frenk}, {Frye}, {Harvey}, {Jauzac}, {Koekemoer}, {Lagattuta},
  {Limousin}, {Mahler}, {Sirks}, \& {Steinhardt}}]{Etherington:2023aa}
{Etherington}, A., {Nightingale}, J.~W., {Massey}, R., {et~al.} 2023, arXiv
  e-prints, arXiv:2301.05244, \dodoi{10.48550/arXiv.2301.05244}

\bibitem[{{Evans} {et~al.}(2022){Evans}, {Frayer}, {Charmandaris}, {Armus},
  {Inami}, {Surace}, {Linden}, {Soifer}, {Diaz-Santos}, {Larson}, {Rich},
  {Song}, {Barcos-Munoz}, {Mazzarella}, {Privon}, {U}, {Medling}, {B{\"o}ker},
  {Aalto}, {Iwasawa}, {Howell}, {van der Werf}, {Appleton}, {Bohn}, {Brown},
  {Hayward}, {Hoshioka}, {Kemper}, {Lai}, {Law}, {Malkan}, {Marshall},
  {Murphy}, {Sanders}, \& {Stierwalt}}]{Evans:2022ac}
{Evans}, A.~S., {Frayer}, D.~T., {Charmandaris}, V., {et~al.} 2022, \apjl, 940,
  L8, \dodoi{10.3847/2041-8213/ac9971}

\bibitem[{{Evans}(1994)}]{Evans:1994aa}
{Evans}, R. 1994, \mnras, 266, 511, \dodoi{10.1093/mnras/266.2.511}

\bibitem[{{Fabian}(2012)}]{Fabian:2012aa}
{Fabian}, A.~C. 2012, \araa, 50, 455,
  \dodoi{10.1146/annurev-astro-081811-125521}

\bibitem[{{Ferreira} {et~al.}(2022){Ferreira}, {Adams}, {Conselice},
  {Sazonova}, {Austin}, {Caruana}, {Ferrari}, {Verma}, {Trussler},
  {Broadhurst}, {Diego}, {Frye}, {Pascale}, {Wilkins}, {Windhorst}, \&
  {Zitrin}}]{Ferreira:2022aa}
{Ferreira}, L., {Adams}, N., {Conselice}, C.~J., {et~al.} 2022, \apjl, 938, L2,
  \dodoi{10.3847/2041-8213/ac947c}

\bibitem[{{Finkelstein} {et~al.}(2023){Finkelstein}, {Bagley}, {Ferguson},
  {Wilkins}, {Kartaltepe}, {Papovich}, {Yung}, {Haro}, {Behroozi}, {Dickinson},
  {Kocevski}, {Koekemoer}, {Larson}, {Le Bail}, {Morales},
  {P{\'e}rez-Gonz{\'a}lez}, {Burgarella}, {Dav{\'e}}, {Hirschmann},
  {Somerville}, {Wuyts}, {Bromm}, {Casey}, {Fontana}, {Fujimoto}, {Gardner},
  {Giavalisco}, {Grazian}, {Grogin}, {Hathi}, {Hutchison}, {Jha}, {Jogee},
  {Kewley}, {Kirkpatrick}, {Long}, {Lotz}, {Pentericci}, {Pierel}, {Pirzkal},
  {Ravindranath}, {Ryan}, {Trump}, {Yang}, {Bhatawdekar}, {Bisigello}, {Buat},
  {Calabr{\`o}}, {Castellano}, {Cleri}, {Cooper}, {Croton}, {Daddi}, {Dekel},
  {Elbaz}, {Franco}, {Gawiser}, {Holwerda}, {Huertas-Company}, {Jaskot},
  {Leung}, {Lucas}, {Mobasher}, {Pandya}, {Tacchella}, {Weiner}, \&
  {Zavala}}]{Finkelstein:2023aa}
{Finkelstein}, S.~L., {Bagley}, M.~B., {Ferguson}, H.~C., {et~al.} 2023, \apjl,
  946, L13, \dodoi{10.3847/2041-8213/acade4}

\bibitem[{{Franco} {et~al.}(2018){Franco}, {Elbaz}, {B{\'e}thermin},
  {Magnelli}, {Schreiber}, {Ciesla}, {Dickinson}, {Nagar}, {Silverman},
  {Daddi}, {Alexander}, {Wang}, {Pannella}, {Le Floc'h}, {Pope}, {Giavalisco},
  {Maury}, {Bournaud}, {Chary}, {Demarco}, {Ferguson}, {Finkelstein}, {Inami},
  {Iono}, {Juneau}, {Lagache}, {Leiton}, {Lin}, {Magdis}, {Messias},
  {Motohara}, {Mullaney}, {Okumura}, {Papovich}, {Pforr}, {Rujopakarn},
  {Sargent}, {Shu}, \& {Zhou}}]{Franco:2018aa}
{Franco}, M., {Elbaz}, D., {B{\'e}thermin}, M., {et~al.} 2018, \aap, 620, A152,
  \dodoi{10.1051/0004-6361/201832928}

\bibitem[{{Frye} {et~al.}(2007){Frye}, {Coe}, {Bowen}, {Ben{\'\i}tez},
  {Broadhurst}, {Guhathakurta}, {Illingworth}, {Menanteau}, {Sharon}, {Lupton},
  {Meylan}, {Zekser}, {Meurer}, \& {Hurley}}]{Frye:2007aa}
{Frye}, B.~L., {Coe}, D., {Bowen}, D.~V., {et~al.} 2007, \apj, 665, 921,
  \dodoi{10.1086/519244}

\bibitem[{{Frye} {et~al.}(2023){Frye}, {Pascale}, {Foo}, {Leimbach}, {Garuda},
  {Robles}, {Summers}, {Diaz}, {Kamieneski}, {Furtak}, {Cohen}, {Diego},
  {Beauchesne}, {Windhorst}, {Willner}, {Koekemoer}, {Zitrin}, {Caminha},
  {Caputi}, {Coe}, {Conselice}, {Dai}, {Dole}, {Driver}, {Grogin},
  {Harrington}, {Jansen}, {Kneib}, {Lehnert}, {Lowenthal}, {Marshall},
  {Menanteau}, {Pampliega}, {Pirzkal}, {Polletta}, {Richard}, {Robotham},
  {Ryan}, {Rutkowski}, {Sif{\'o}n}, {Tompkins}, {Wang}, {Yan}, \&
  {Yun}}]{Frye:2023aa}
{Frye}, B.~L., {Pascale}, M., {Foo}, N., {et~al.} 2023, \apj, 952, 81,
  \dodoi{10.3847/1538-4357/acd929}

\bibitem[{{Gadotti} \& {dos Anjos}(2001)}]{Gadotti:2001aa}
{Gadotti}, D.~A., \& {dos Anjos}, S. 2001, \aj, 122, 1298,
  \dodoi{10.1086/322126}

\bibitem[{{Gaia Collaboration} {et~al.}(2016){Gaia Collaboration}, {Brown},
  {Vallenari}, {Prusti}, {de Bruijne}, {Mignard}, {Drimmel}, {Babusiaux},
  {Bailer-Jones}, {Bastian}, \& et~al.}]{Gaia-Collaboration:2016aa}
{Gaia Collaboration}, {Brown}, A.~G.~A., {Vallenari}, A., {et~al.} 2016, \aap,
  595, A2, \dodoi{10.1051/0004-6361/201629512}

\bibitem[{{Gaia Collaboration} {et~al.}(2021){Gaia Collaboration}, {Brown},
  {Vallenari}, {Prusti}, {de Bruijne}, {Babusiaux}, {Biermann}, {Creevey},
  {Evans}, {Eyer}, \& et~al.}]{Gaia-Collaboration:2021aa}
---. 2021, \aap, 649, A1, \dodoi{10.1051/0004-6361/202039657}

\bibitem[{{Gao} \& {Ho}(2017)}]{Gao:2017aa}
{Gao}, H., \& {Ho}, L.~C. 2017, \apj, 845, 114,
  \dodoi{10.3847/1538-4357/aa7da4}

\bibitem[{{Goldader} {et~al.}(2002){Goldader}, {Meurer}, {Heckman}, {Seibert},
  {Sanders}, {Calzetti}, \& {Steidel}}]{Goldader:2002aa}
{Goldader}, J.~D., {Meurer}, G., {Heckman}, T.~M., {et~al.} 2002, \apj, 568,
  651, \dodoi{10.1086/339165}

\bibitem[{{Goulding} {et~al.}(2010){Goulding}, {Alexander}, {Lehmer}, \&
  {Mullaney}}]{Goulding:2010aa}
{Goulding}, A.~D., {Alexander}, D.~M., {Lehmer}, B.~D., \& {Mullaney}, J.~R.
  2010, \mnras, 406, 597, \dodoi{10.1111/j.1365-2966.2010.16700.x}

\bibitem[{{Graham} \& {Worley}(2008)}]{Graham:2008aa}
{Graham}, A.~W., \& {Worley}, C.~C. 2008, \mnras, 388, 1708,
  \dodoi{10.1111/j.1365-2966.2008.13506.x}

\bibitem[{{Gullberg} {et~al.}(2019){Gullberg}, {Smail}, {Swinbank},
  {Dudzevi{\v{c}}i{\={u}}t{\.{e}}}, {Stach}, {Thomson}, {Almaini}, {Chen},
  {Conselice}, {Cooke}, {Farrah}, {Ivison}, {Maltby}, {Micha{\l}owski},
  {Simpson}, {Scott}, {Wardlow}, \& {Weiss}}]{Gullberg:2019aa}
{Gullberg}, B., {Smail}, I., {Swinbank}, A.~M., {et~al.} 2019, \mnras, 490,
  4956, \dodoi{10.1093/mnras/stz2835}

\bibitem[{{Hales} {et~al.}(2012){Hales}, {Murphy}, {Curran}, {Middelberg},
  {Gaensler}, \& {Norris}}]{Hales:2012aa}
{Hales}, C.~A., {Murphy}, T., {Curran}, J.~R., {et~al.} 2012, \mnras, 425, 979,
  \dodoi{10.1111/j.1365-2966.2012.21373.x}

\bibitem[{{H{\"a}u{\ss}ler} {et~al.}(2013){H{\"a}u{\ss}ler}, {Bamford}, {Vika},
  {Rojas}, {Barden}, {Kelvin}, {Alpaslan}, {Robotham}, {Driver}, {Baldry},
  {Brough}, {Hopkins}, {Liske}, {Nichol}, {Popescu}, \&
  {Tuffs}}]{Haussler:2013aa}
{H{\"a}u{\ss}ler}, B., {Bamford}, S.~P., {Vika}, M., {et~al.} 2013, \mnras,
  430, 330, \dodoi{10.1093/mnras/sts633}

\bibitem[{{Hayward} {et~al.}(2011){Hayward}, {Kere{\v s}}, {Jonsson},
  {Narayanan}, {Cox}, \& {Hernquist}}]{Hayward:2011aa}
{Hayward}, C.~C., {Kere{\v s}}, D., {Jonsson}, P., {et~al.} 2011, \apj, 743,
  159, \dodoi{10.1088/0004-637X/743/2/159}

\bibitem[{{Heckman}(2001)}]{Heckman:2001ab}
{Heckman}, T.~M. 2001, in Astronomical Society of the Pacific Conference
  Series, Vol. 240, Gas and Galaxy Evolution, ed. J.~E. {Hibbard}, M.~{Rupen},
  \& J.~H. {van Gorkom}, 345, \dodoi{10.48550/arXiv.astro-ph/0009075}

\bibitem[{{Hezaveh} {et~al.}(2012){Hezaveh}, {Marrone}, \&
  {Holder}}]{Hezaveh:2012aa}
{Hezaveh}, Y.~D., {Marrone}, D.~P., \& {Holder}, G.~P. 2012, \apj, 761, 20,
  \dodoi{10.1088/0004-637X/761/1/20}

\bibitem[{{Hodge} {et~al.}(2015){Hodge}, {Riechers}, {Decarli}, {Walter},
  {Carilli}, {Daddi}, \& {Dannerbauer}}]{Hodge:2015aa}
{Hodge}, J.~A., {Riechers}, D., {Decarli}, R., {et~al.} 2015, \apjl, 798, L18,
  \dodoi{10.1088/2041-8205/798/1/L18}

\bibitem[{{Hodge} {et~al.}(2016){Hodge}, {Swinbank}, {Simpson}, {Smail},
  {Walter}, {Alexander}, {Bertoldi}, {Biggs}, {Brandt}, {Chapman}, {Chen},
  {Coppin}, {Cox}, {Dannerbauer}, {Edge}, {Greve}, {Ivison}, {Karim},
  {Knudsen}, {Menten}, {Rix}, {Schinnerer}, {Wardlow}, {Weiss}, \& {van der
  Werf}}]{Hodge:2016aa}
{Hodge}, J.~A., {Swinbank}, A.~M., {Simpson}, J.~M., {et~al.} 2016, \apj, 833,
  103, \dodoi{10.3847/1538-4357/833/1/103}

\bibitem[{{Hodge} {et~al.}(2019){Hodge}, {Smail}, {Walter}, {da Cunha},
  {Swinbank}, {Rybak}, {Venemans}, {Brandt}, {Calistro Rivera}, {Chapman},
  {Chen}, {Cox}, {Dannerbauer}, {Decarli}, {Greve}, {Knudsen}, {Menten},
  {Schinnerer}, {Simpson}, {van der Werf}, {Wardlow}, \&
  {Weiss}}]{Hodge:2019aa}
{Hodge}, J.~A., {Smail}, I., {Walter}, F., {et~al.} 2019, \apj, 876, 130,
  \dodoi{10.3847/1538-4357/ab1846}

\bibitem[{{Hopkins} {et~al.}(2010){Hopkins}, {Bundy}, {Hernquist}, {Wuyts}, \&
  {Cox}}]{Hopkins:2010ab}
{Hopkins}, P.~F., {Bundy}, K., {Hernquist}, L., {Wuyts}, S., \& {Cox}, T.~J.
  2010, \mnras, 401, 1099, \dodoi{10.1111/j.1365-2966.2009.15699.x}

\bibitem[{{Hopkins} {et~al.}(2014){Hopkins}, {Kere{\v s}}, {O{\~n}orbe},
  {Faucher-Gigu{\`e}re}, {Quataert}, {Murray}, \& {Bullock}}]{Hopkins:2014aa}
{Hopkins}, P.~F., {Kere{\v s}}, D., {O{\~n}orbe}, J., {et~al.} 2014, \mnras,
  445, 581, \dodoi{10.1093/mnras/stu1738}

\bibitem[{{Hopkins} {et~al.}(2018){Hopkins}, {Wetzel}, {Kere{\v{s}}},
  {Faucher-Gigu{\`e}re}, {Quataert}, {Boylan-Kolchin}, {Murray}, {Hayward},
  {Garrison-Kimmel}, {Hummels}, {Feldmann}, {Torrey}, {Ma},
  {Angl{\'e}s-Alc{\'a}zar}, {Su}, {Orr}, {Schmitz}, {Escala}, {Sanderson},
  {Grudi{\'c}}, {Hafen}, {Kim}, {Fitts}, {Bullock}, {Wheeler}, {Chan},
  {Elbert}, \& {Narayanan}}]{Hopkins:2018aa}
{Hopkins}, P.~F., {Wetzel}, A., {Kere{\v{s}}}, D., {et~al.} 2018, \mnras, 480,
  800, \dodoi{10.1093/mnras/sty1690}

\bibitem[{{Huang} {et~al.}(2011){Huang}, {Zheng}, {Rigopoulou}, {Magdis},
  {Fazio}, \& {Wang}}]{Huang:2011aa}
{Huang}, J.~S., {Zheng}, X.~Z., {Rigopoulou}, D., {et~al.} 2011, \apjl, 742,
  L13, \dodoi{10.1088/2041-8205/742/1/L13}

\bibitem[{{Hubble}(1943)}]{Hubble:1943aa}
{Hubble}, E. 1943, \apj, 97, 112, \dodoi{10.1086/144504}

\bibitem[{{Hutton} {et~al.}(2014){Hutton}, {Ferreras}, {Wu}, {Kuin},
  {Breeveld}, {Yershov}, {Cropper}, \& {Page}}]{Hutton:2014aa}
{Hutton}, S., {Ferreras}, I., {Wu}, K., {et~al.} 2014, \mnras, 440, 150,
  \dodoi{10.1093/mnras/stu185}

\bibitem[{{Ikarashi} {et~al.}(2015){Ikarashi}, {Ivison}, {Caputi}, {Aretxaga},
  {Dunlop}, {Hatsukade}, {Hughes}, {Iono}, {Izumi}, {Kawabe}, {Kohno}, {Lagos},
  {Motohara}, {Nakanishi}, {Ohta}, {Tamura}, {Umehata}, {Wilson}, {Yabe}, \&
  {Yun}}]{Ikarashi:2015aa}
{Ikarashi}, S., {Ivison}, R.~J., {Caputi}, K.~I., {et~al.} 2015, \apj, 810,
  133, \dodoi{10.1088/0004-637X/810/2/133}

\bibitem[{{Inada} {et~al.}(2005){Inada}, {Oguri}, {Keeton}, {Eisenstein},
  {Castander}, {Chiu}, {Hall}, {Hennawi}, {Johnston}, {Pindor}, {Richards},
  {Rix}, {Schneider}, \& {Zheng}}]{Inada:2005aa}
{Inada}, N., {Oguri}, M., {Keeton}, C.~R., {et~al.} 2005, \pasj, 57, L7,
  \dodoi{10.1093/pasj/57.3.L7}

\bibitem[{{Iono} {et~al.}(2004){Iono}, {Ho}, {Yun}, {Matsushita}, {Peck}, \&
  {Sakamoto}}]{Iono:2004aa}
{Iono}, D., {Ho}, P. T.~P., {Yun}, M.~S., {et~al.} 2004, \apjl, 616, L63,
  \dodoi{10.1086/420784}

\bibitem[{{Jansen} {et~al.}(2000){Jansen}, {Franx}, {Fabricant}, \&
  {Caldwell}}]{Jansen:2000aa}
{Jansen}, R.~A., {Franx}, M., {Fabricant}, D., \& {Caldwell}, N. 2000, \apjs,
  126, 271, \dodoi{10.1086/313303}

\bibitem[{{Jansen} {et~al.}(1994){Jansen}, {Knapen}, {Beckman}, {Peletier}, \&
  {Hes}}]{Jansen:1994aa}
{Jansen}, R.~A., {Knapen}, J.~H., {Beckman}, J.~E., {Peletier}, R.~F., \&
  {Hes}, R. 1994, \mnras, 270, 373, \dodoi{10.1093/mnras/270.2.373}

\bibitem[{{Ji} \& {Giavalisco}(2022)}]{Ji:2022aa}
{Ji}, Z., \& {Giavalisco}, M. 2022, \apj, 935, 120,
  \dodoi{10.3847/1538-4357/ac7f43}

\bibitem[{{Johnson} {et~al.}(2021){Johnson}, {Leja}, {Conroy}, \&
  {Speagle}}]{Johnson:2021aa}
{Johnson}, B.~D., {Leja}, J., {Conroy}, C., \& {Speagle}, J.~S. 2021, \apjs,
  254, 22, \dodoi{10.3847/1538-4365/abef67}

\bibitem[{{Jones} {et~al.}(2010){Jones}, {Ellis}, {Jullo}, \&
  {Richard}}]{Jones:2010ab}
{Jones}, T., {Ellis}, R., {Jullo}, E., \& {Richard}, J. 2010, \apjl, 725, L176,
  \dodoi{10.1088/2041-8205/725/2/L176}

\bibitem[{{Jullo} \& {Kneib}(2009)}]{Jullo:2009aa}
{Jullo}, E., \& {Kneib}, J.-P. 2009, \mnras, 395, 1319,
  \dodoi{10.1111/j.1365-2966.2009.14654.x}

\bibitem[{{Jullo} {et~al.}(2007){Jullo}, {Kneib}, {Limousin},
  {El{\'{\i}}asd{\'o}ttir}, {Marshall}, \& {Verdugo}}]{Jullo:2007aa}
{Jullo}, E., {Kneib}, J.-P., {Limousin}, M., {et~al.} 2007, New Journal of
  Physics, 9, 447, \dodoi{10.1088/1367-2630/9/12/447}

\bibitem[{{Kamieneski} {et~al.}(2023){Kamieneski}, {Yun}, {Harrington},
  {Lowenthal}, {Wang}, {Frye}, {Jimenez-Andrade}, {Vishwas}, {Cooper},
  {Pascale}, {Foo}, {Berman}, {Englert}, \& {Garcia Diaz}}]{Kamieneski:2023aa}
{Kamieneski}, P.~S., {Yun}, M.~S., {Harrington}, K.~C., {et~al.} 2023, arXiv
  e-prints, arXiv:2301.09746.
\newblock \doarXiv{2301.09746}

\bibitem[{{Keeton}(2003)}]{Keeton:2003aa}
{Keeton}, C.~R. 2003, \apj, 582, 17, \dodoi{10.1086/344539}

\bibitem[{{Kelvin} {et~al.}(2012){Kelvin}, {Driver}, {Robotham}, {Hill},
  {Alpaslan}, {Baldry}, {Bamford}, {Bland-Hawthorn}, {Brough}, {Graham},
  {H{\"a}ussler}, {Hopkins}, {Liske}, {Loveday}, {Norberg}, {Phillipps},
  {Popescu}, {Prescott}, {Taylor}, \& {Tuffs}}]{Kelvin:2012aa}
{Kelvin}, L.~S., {Driver}, S.~P., {Robotham}, A. S.~G., {et~al.} 2012, \mnras,
  421, 1007, \dodoi{10.1111/j.1365-2966.2012.20355.x}

\bibitem[{{Kennedy} {et~al.}(2015){Kennedy}, {Bamford}, {Baldry},
  {H{\"a}u{\ss}ler}, {Holwerda}, {Hopkins}, {Kelvin}, {Lange}, {Moffett},
  {Popescu}, {Taylor}, {Tuffs}, {Vika}, \& {Vulcani}}]{Kennedy:2015aa}
{Kennedy}, R., {Bamford}, S.~P., {Baldry}, I., {et~al.} 2015, \mnras, 454, 806,
  \dodoi{10.1093/mnras/stv2032}

\bibitem[{{Kennedy} {et~al.}(2016){Kennedy}, {Bamford}, {H{\"a}u{\ss}ler},
  {Baldry}, {Bremer}, {Brough}, {Brown}, {Driver}, {Duncan}, {Graham},
  {Holwerda}, {Hopkins}, {Kelvin}, {Lange}, {Phillipps}, {Vika}, \&
  {Vulcani}}]{Kennedy:2016aa}
{Kennedy}, R., {Bamford}, S.~P., {H{\"a}u{\ss}ler}, B., {et~al.} 2016, \mnras,
  460, 3458, \dodoi{10.1093/mnras/stw1176}

\bibitem[{{Kneib} {et~al.}(1996){Kneib}, {Ellis}, {Smail}, {Couch}, \&
  {Sharples}}]{Kneib:1996aa}
{Kneib}, J.-P., {Ellis}, R.~S., {Smail}, I., {Couch}, W.~J., \& {Sharples},
  R.~M. 1996, \apj, 471, 643, \dodoi{10.1086/177995}

\bibitem[{{Kneib} {et~al.}(1993){Kneib}, {Mellier}, {Fort}, \&
  {Mathez}}]{Kneib:1993aa}
{Kneib}, J.~P., {Mellier}, Y., {Fort}, B., \& {Mathez}, G. 1993, \aap, 273, 367

\bibitem[{{Knop} {et~al.}(1994){Knop}, {Soifer}, {Graham}, {Matthews},
  {Sanders}, \& {Scoville}}]{Knop:1994aa}
{Knop}, R.~A., {Soifer}, B.~T., {Graham}, J.~R., {et~al.} 1994, \aj, 107, 920,
  \dodoi{10.1086/116906}

\bibitem[{{Kormann} {et~al.}(1994){Kormann}, {Schneider}, \&
  {Bartelmann}}]{Kormann:1994aa}
{Kormann}, R., {Schneider}, P., \& {Bartelmann}, M. 1994, \aap, 284, 285

\bibitem[{{Kron}(1980)}]{Kron:1980aa}
{Kron}, R.~G. 1980, \apjs, 43, 305, \dodoi{10.1086/190669}

\bibitem[{{La Barbera} {et~al.}(2002){La Barbera}, {Busarello}, {Merluzzi},
  {Massarotti}, \& {Capaccioli}}]{La-Barbera:2002aa}
{La Barbera}, F., {Busarello}, G., {Merluzzi}, P., {Massarotti}, M., \&
  {Capaccioli}, M. 2002, \apj, 571, 790, \dodoi{10.1086/340021}

\bibitem[{{La Barbera} {et~al.}(2010){La Barbera}, {de Carvalho}, {de La Rosa},
  {Lopes}, {Kohl-Moreira}, \& {Capelato}}]{La-Barbera:2010aa}
{La Barbera}, F., {de Carvalho}, R.~R., {de La Rosa}, I.~G., {et~al.} 2010,
  \mnras, 408, 1313, \dodoi{10.1111/j.1365-2966.2010.16850.x}

\bibitem[{{La Barbera} {et~al.}(2012){La Barbera}, {Ferreras}, {de Carvalho},
  {Bruzual}, {Charlot}, {Pasquali}, \& {Merlin}}]{La-Barbera:2012aa}
{La Barbera}, F., {Ferreras}, I., {de Carvalho}, R.~R., {et~al.} 2012, \mnras,
  426, 2300, \dodoi{10.1111/j.1365-2966.2012.21848.x}

\bibitem[{{Labb{\'e}} {et~al.}(2005){Labb{\'e}}, {Huang}, {Franx}, {Rudnick},
  {Barmby}, {Daddi}, {van Dokkum}, {Fazio}, {F{\"o}rster Schreiber},
  {Moorwood}, {Rix}, {R{\"o}ttgering}, {Trujillo}, \& {van der
  Werf}}]{Labbe:2005aa}
{Labb{\'e}}, I., {Huang}, J., {Franx}, M., {et~al.} 2005, \apjl, 624, L81,
  \dodoi{10.1086/430700}

\bibitem[{{Lang} {et~al.}(2019){Lang}, {Schinnerer}, {Smail},
  {Dudzevi{\v{c}}i{\={u}}t{\.{e}}}, {Swinbank}, {Liu}, {Leslie}, {Almaini},
  {An}, {Bertoldi}, {Blain}, {Chapman}, {Chen}, {Conselice}, {Cooke}, {Coppin},
  {Dunlop}, {Farrah}, {Fudamoto}, {Geach}, {Gullberg}, {Harrington}, {Hodge},
  {Ivison}, {Jim{\'e}nez-Andrade}, {Magnelli}, {Micha{\l}owski}, {Oesch},
  {Scott}, {Simpson}, {Smol{\v{c}}i{\'c}}, {Stach}, {Thomson}, {Toft},
  {Vardoulaki}, {Wardlow}, {Weiss}, \& {van der Werf}}]{Lang:2019aa}
{Lang}, P., {Schinnerer}, E., {Smail}, I., {et~al.} 2019, \apj, 879, 54,
  \dodoi{10.3847/1538-4357/ab1f77}

\bibitem[{{Leja} {et~al.}(2019){Leja}, {Tacchella}, \& {Conroy}}]{Leja:2019aa}
{Leja}, J., {Tacchella}, S., \& {Conroy}, C. 2019, \apjl, 880, L9,
  \dodoi{10.3847/2041-8213/ab2f8c}

\bibitem[{{Li} \& {Draine}(2001)}]{Li:2001aa}
{Li}, A., \& {Draine}, B.~T. 2001, \apj, 554, 778, \dodoi{10.1086/323147}

\bibitem[{{Liu} {et~al.}(2017){Liu}, {Jiang}, {Faber}, {Koo}, {Yesuf},
  {Tacchella}, {Mao}, {Wang}, {Guo}, {Fang}, {Barro}, {Zheng}, {Jia}, {Tong},
  {Liu}, \& {Meng}}]{Liu:2017aa}
{Liu}, F.~S., {Jiang}, D., {Faber}, S.~M., {et~al.} 2017, \apjl, 844, L2,
  \dodoi{10.3847/2041-8213/aa7cf5}

\bibitem[{{Liu} {et~al.}(2013){Liu}, {Calzetti}, {Hong}, {Whitmore}, {Chandar},
  {O'Connell}, {Blair}, {Cohen}, {Frogel}, \& {Kim}}]{Liu:2013aa}
{Liu}, G., {Calzetti}, D., {Hong}, S., {et~al.} 2013, \apjl, 778, L41,
  \dodoi{10.1088/2041-8205/778/2/L41}

\bibitem[{{Lonsdale} {et~al.}(2006){Lonsdale}, {Farrah}, \&
  {Smith}}]{Lonsdale:2006aa}
{Lonsdale}, C.~J., {Farrah}, D., \& {Smith}, H.~E. 2006, {Ultraluminous
  Infrared Galaxies} (Springer-Verlag), 285, \dodoi{10.1007/3-540-30313-8_9}

\bibitem[{{Luo} {et~al.}(2022){Luo}, {Rowlands}, {Alatalo}, {Sazonova},
  {Abdurro'uf}, {Heckman}, {Medling}, {Deustua}, {Nyland}, {Lanz}, {Petric},
  {Otter}, {Aalto}, {Dimassimo}, {French}, {Gallagher}, {Roediger}, \&
  {Stepanoff}}]{Luo:2022ab}
{Luo}, Y., {Rowlands}, K., {Alatalo}, K., {et~al.} 2022, \apj, 938, 63,
  \dodoi{10.3847/1538-4357/ac8b7d}

\bibitem[{{Lutz} {et~al.}(2016){Lutz}, {Berta}, {Contursi}, {F{\"o}rster
  Schreiber}, {Genzel}, {Graci{\'a}-Carpio}, {Herrera-Camus}, {Netzer},
  {Sturm}, {Tacconi}, {Tadaki}, \& {Veilleux}}]{Lutz:2016aa}
{Lutz}, D., {Berta}, S., {Contursi}, A., {et~al.} 2016, \aap, 591, A136,
  \dodoi{10.1051/0004-6361/201527706}

\bibitem[{{Ma} {et~al.}(2018){Ma}, {Hopkins}, {Boylan-Kolchin},
  {Faucher-Gigu{\`e}re}, {Quataert}, {Feldmann}, {Garrison-Kimmel}, {Hayward},
  {Kere{\v{s}}}, \& {Wetzel}}]{Ma:2018ab}
{Ma}, X., {Hopkins}, P.~F., {Boylan-Kolchin}, M., {et~al.} 2018, \mnras, 477,
  219, \dodoi{10.1093/mnras/sty684}

\bibitem[{{Ma} {et~al.}(2019){Ma}, {Hayward}, {Casey}, {Hopkins}, {Quataert},
  {Liang}, {Faucher-Gigu{\`e}re}, {Feldmann}, \& {Kere{\v{s}}}}]{Ma:2019ad}
{Ma}, X., {Hayward}, C.~C., {Casey}, C.~M., {et~al.} 2019, \mnras, 487, 1844,
  \dodoi{10.1093/mnras/stz1324}

\bibitem[{{Madau} \& {Dickinson}(2014)}]{Madau:2014aa}
{Madau}, P., \& {Dickinson}, M. 2014, \araa, 52, 415,
  \dodoi{10.1146/annurev-astro-081811-125615}

\bibitem[{{Marian} {et~al.}(2018){Marian}, {Ziegler}, {Kuchner}, \&
  {Verdugo}}]{Marian:2018aa}
{Marian}, V., {Ziegler}, B., {Kuchner}, U., \& {Verdugo}, M. 2018, \aap, 617,
  A34, \dodoi{10.1051/0004-6361/201832750}

\bibitem[{{Martig} {et~al.}(2009){Martig}, {Bournaud}, {Teyssier}, \&
  {Dekel}}]{Martig:2009aa}
{Martig}, M., {Bournaud}, F., {Teyssier}, R., \& {Dekel}, A. 2009, \apj, 707,
  250, \dodoi{10.1088/0004-637X/707/1/250}

\bibitem[{{Massardi} {et~al.}(2018){Massardi}, {Enia}, {Negrello}, {Mancuso},
  {Lapi}, {Vignali}, {Gilli}, {Burkutean}, {Danese}, \&
  {Zotti}}]{Massardi:2018aa}
{Massardi}, M., {Enia}, A.~F.~M., {Negrello}, M., {et~al.} 2018, \aap, 610,
  A53, \dodoi{10.1051/0004-6361/201731751}

\bibitem[{{McMullin} {et~al.}(2007){McMullin}, {Waters}, {Schiebel}, {Young},
  \& {Golap}}]{McMullin:2007aa}
{McMullin}, J.~P., {Waters}, B., {Schiebel}, D., {Young}, W., \& {Golap}, K.
  2007, in Astronomical Society of the Pacific Conference Series, Vol. 376,
  Astronomical Data Analysis Software and Systems XVI, ed. R.~A. {Shaw},
  F.~{Hill}, \& D.~J. {Bell}, 127

\bibitem[{{Menanteau} {et~al.}(2012){Menanteau}, {Hughes}, {Sif{\'o}n},
  {Hilton}, {Gonz{\'a}lez}, {Infante}, {Barrientos}, {Baker}, {Bond}, {Das},
  {Devlin}, {Dunkley}, {Hajian}, {Hincks}, {Kosowsky}, {Marsden}, {Marriage},
  {Moodley}, {Niemack}, {Nolta}, {Page}, {Reese}, {Sehgal}, {Sievers},
  {Spergel}, {Staggs}, \& {Wollack}}]{Menanteau:2012aa}
{Menanteau}, F., {Hughes}, J.~P., {Sif{\'o}n}, C., {et~al.} 2012, \apj, 748, 7,
  \dodoi{10.1088/0004-637X/748/1/7}

\bibitem[{{Miller} {et~al.}(2022){Miller}, {Whitaker}, {Nelson}, {van Dokkum},
  {Bezanson}, {Brammer}, {Heintz}, {Leja}, {Suess}, \&
  {Weaver}}]{Miller:2022ab}
{Miller}, T.~B., {Whitaker}, K.~E., {Nelson}, E.~J., {et~al.} 2022, \apjl, 941,
  L37, \dodoi{10.3847/2041-8213/aca675}

\bibitem[{{Mo} {et~al.}(1998){Mo}, {Mao}, \& {White}}]{Mo:1998aa}
{Mo}, H.~J., {Mao}, S., \& {White}, S. D.~M. 1998, \mnras, 295, 319,
  \dodoi{10.1046/j.1365-8711.1998.01227.x}

\bibitem[{{M{\"o}llenhoff} {et~al.}(2006){M{\"o}llenhoff}, {Popescu}, \&
  {Tuffs}}]{Mollenhoff:2006aa}
{M{\"o}llenhoff}, C., {Popescu}, C.~C., \& {Tuffs}, R.~J. 2006, \aap, 456, 941,
  \dodoi{10.1051/0004-6361:20054727}

\bibitem[{{Mowla} {et~al.}(2019){Mowla}, {van Dokkum}, {Brammer}, {Momcheva},
  {van der Wel}, {Whitaker}, {Nelson}, {Bezanson}, {Muzzin}, {Franx},
  {MacKenty}, {Leja}, {Kriek}, \& {Marchesini}}]{Mowla:2019aa}
{Mowla}, L.~A., {van Dokkum}, P., {Brammer}, G.~B., {et~al.} 2019, \apj, 880,
  57, \dodoi{10.3847/1538-4357/ab290a}

\bibitem[{{Muller} {et~al.}(2020){Muller}, {Jaswanth}, {Horellou}, \&
  {Mart{\'\i}-Vidal}}]{Muller:2020aa}
{Muller}, S., {Jaswanth}, S., {Horellou}, C., \& {Mart{\'\i}-Vidal}, I. 2020,
  \aap, 641, L2, \dodoi{10.1051/0004-6361/202038978}

\bibitem[{{Narayan} \& {Bartelmann}(1996)}]{Narayan:1996aa}
{Narayan}, R., \& {Bartelmann}, M. 1996, ArXiv Astrophysics e-prints

\bibitem[{{Narayanan} {et~al.}(2018){Narayanan}, {Dav{\'e}}, {Johnson},
  {Thompson}, {Conroy}, \& {Geach}}]{Narayanan:2018ab}
{Narayanan}, D., {Dav{\'e}}, R., {Johnson}, B.~D., {et~al.} 2018, \mnras, 474,
  1718, \dodoi{10.1093/mnras/stx2860}

\bibitem[{{Nardiello} {et~al.}(2022){Nardiello}, {Bedin}, {Burgasser},
  {Salaris}, {Cassisi}, {Griggio}, \& {Scalco}}]{Nardiello:2022aa}
{Nardiello}, D., {Bedin}, L.~R., {Burgasser}, A., {et~al.} 2022, \mnras, 517,
  484, \dodoi{10.1093/mnras/stac2659}

\bibitem[{{Nelson} {et~al.}(2012){Nelson}, {van Dokkum}, {Brammer},
  {F{\"o}rster Schreiber}, {Franx}, {Fumagalli}, {Patel}, {Rix}, {Skelton},
  {Bezanson}, {Da Cunha}, {Kriek}, {Labbe}, {Lundgren}, {Quadri}, \&
  {Schmidt}}]{Nelson:2012aa}
{Nelson}, E.~J., {van Dokkum}, P.~G., {Brammer}, G., {et~al.} 2012, \apjl, 747,
  L28, \dodoi{10.1088/2041-8205/747/2/L28}

\bibitem[{{Nelson} {et~al.}(2016{\natexlab{a}}){Nelson}, {van Dokkum},
  {Momcheva}, {Brammer}, {Wuyts}, {Franx}, {F{\"o}rster Schreiber}, {Whitaker},
  \& {Skelton}}]{Nelson:2016aa}
{Nelson}, E.~J., {van Dokkum}, P.~G., {Momcheva}, I.~G., {et~al.}
  2016{\natexlab{a}}, \apjl, 817, L9, \dodoi{10.3847/2041-8205/817/1/L9}

\bibitem[{{Nelson} {et~al.}(2016{\natexlab{b}}){Nelson}, {van Dokkum},
  {F{\"o}rster Schreiber}, {Franx}, {Brammer}, {Momcheva}, {Wuyts}, {Whitaker},
  {Skelton}, {Fumagalli}, {Hayward}, {Kriek}, {Labb{\'e}}, {Leja}, {Rix},
  {Tacconi}, {van der Wel}, {van den Bosch}, {Oesch}, {Dickey}, \& {Ulf
  Lange}}]{Nelson:2016ab}
{Nelson}, E.~J., {van Dokkum}, P.~G., {F{\"o}rster Schreiber}, N.~M., {et~al.}
  2016{\natexlab{b}}, \apj, 828, 27, \dodoi{10.3847/0004-637X/828/1/27}

\bibitem[{{Oser} {et~al.}(2010){Oser}, {Ostriker}, {Naab}, {Johansson}, \&
  {Burkert}}]{Oser:2010aa}
{Oser}, L., {Ostriker}, J.~P., {Naab}, T., {Johansson}, P.~H., \& {Burkert}, A.
  2010, \apj, 725, 2312, \dodoi{10.1088/0004-637X/725/2/2312}

\bibitem[{{Ostrovski} {et~al.}(2018){Ostrovski}, {Lemon}, {Auger}, {McMahon},
  {Fassnacht}, {Chen}, {Connolly}, {Koposov}, {Pons}, {Reed}, \&
  {Rusu}}]{Ostrovski:2018aa}
{Ostrovski}, F., {Lemon}, C.~A., {Auger}, M.~W., {et~al.} 2018, \mnras, 473,
  L116, \dodoi{10.1093/mnrasl/slx173}

\bibitem[{{Pantoni} {et~al.}(2021){Pantoni}, {Massardi}, {Lapi}, {Donevski},
  {D'Amato}, {Giulietti}, {Pozzi}, {Talia}, {Vignali}, {Cimatti}, {Silva},
  {Bressan}, \& {Ronconi}}]{Pantoni:2021ab}
{Pantoni}, L., {Massardi}, M., {Lapi}, A., {et~al.} 2021, \mnras, 507, 3998,
  \dodoi{10.1093/mnras/stab2346}

\bibitem[{{Pascale} {et~al.}(2022){Pascale}, {Frye}, {Dai}, {Foo}, {Qin},
  {Leimbach}, {Bauer}, {Merlin}, {Coe}, {Diego}, {Yan}, {Zitrin}, {Cohen},
  {Conselice}, {Dole}, {Harrington}, {Jansen}, {Kamieneski}, {Windhorst}, \&
  {Yun}}]{Pascale:2022aa}
{Pascale}, M., {Frye}, B.~L., {Dai}, L., {et~al.} 2022, \apj, 932, 85,
  \dodoi{10.3847/1538-4357/ac6ce9}

\bibitem[{{Pastrav} {et~al.}(2013){Pastrav}, {Popescu}, {Tuffs}, \&
  {Sansom}}]{Pastrav:2013aa}
{Pastrav}, B.~A., {Popescu}, C.~C., {Tuffs}, R.~J., \& {Sansom}, A.~E. 2013,
  \aap, 553, A80, \dodoi{10.1051/0004-6361/201220962}

\bibitem[{{Patel} {et~al.}(2012){Patel}, {Holden}, {Kelson}, {Franx}, {van der
  Wel}, \& {Illingworth}}]{Patel:2012aa}
{Patel}, S.~G., {Holden}, B.~P., {Kelson}, D.~D., {et~al.} 2012, \apjl, 748,
  L27, \dodoi{10.1088/2041-8205/748/2/L27}

\bibitem[{{Peletier} \& {Balcells}(1996)}]{Peletier:1996aa}
{Peletier}, R.~F., \& {Balcells}, M. 1996, \aj, 111, 2238,
  \dodoi{10.1086/117958}

\bibitem[{{Peng} {et~al.}(2002){Peng}, {Ho}, {Impey}, \& {Rix}}]{Peng:2002aa}
{Peng}, C.~Y., {Ho}, L.~C., {Impey}, C.~D., \& {Rix}, H.-W. 2002, \aj, 124,
  266, \dodoi{10.1086/340952}

\bibitem[{{Penney} {et~al.}(2020){Penney}, {Blain}, {Assef}, {Diaz-Santos},
  {Gonz{\'a}lez-L{\'o}pez}, {Tsai}, {Aravena}, {Eisenhardt}, {Jones}, {Jun},
  {Kim}, {Stern}, \& {Wu}}]{Penney:2020aa}
{Penney}, J.~I., {Blain}, A.~W., {Assef}, R.~J., {et~al.} 2020, \mnras, 496,
  1565, \dodoi{10.1093/mnras/staa1582}

\bibitem[{{Perrin} {et~al.}(2014){Perrin}, {Sivaramakrishnan}, {Lajoie},
  {Elliott}, {Pueyo}, {Ravindranath}, \& {Albert}}]{Perrin:2014aa}
{Perrin}, M.~D., {Sivaramakrishnan}, A., {Lajoie}, C.-P., {et~al.} 2014, in
  Society of Photo-Optical Instrumentation Engineers (SPIE) Conference Series,
  Vol. 9143, Space Telescopes and Instrumentation 2014: Optical, Infrared, and
  Millimeter Wave, ed. J.~{Oschmann}, Jacobus~M., M.~{Clampin}, G.~G. {Fazio},
  \& H.~A. {MacEwen}, 91433X, \dodoi{10.1117/12.2056689}

\bibitem[{{Perrin} {et~al.}(2012){Perrin}, {Soummer}, {Elliott}, {Lallo}, \&
  {Sivaramakrishnan}}]{Perrin:2012aa}
{Perrin}, M.~D., {Soummer}, R., {Elliott}, E.~M., {Lallo}, M.~D., \&
  {Sivaramakrishnan}, A. 2012, in Society of Photo-Optical Instrumentation
  Engineers (SPIE) Conference Series, Vol. 8442, Space Telescopes and
  Instrumentation 2012: Optical, Infrared, and Millimeter Wave, ed. M.~C.
  {Clampin}, G.~G. {Fazio}, H.~A. {MacEwen}, \& J.~{Oschmann}, Jacobus~M.,
  84423D, \dodoi{10.1117/12.925230}

\bibitem[{{Petric} {et~al.}(2011){Petric}, {Armus}, {Howell}, {Chan},
  {Mazzarella}, {Evans}, {Surace}, {Sanders}, {Appleton}, {Charmandaris},
  {D{\'\i}az-Santos}, {Frayer}, {Haan}, {Inami}, {Iwasawa}, {Kim}, {Madore},
  {Marshall}, {Spoon}, {Stierwalt}, {Sturm}, {U}, {Vavilkin}, \&
  {Veilleux}}]{Petric:2011aa}
{Petric}, A.~O., {Armus}, L., {Howell}, J., {et~al.} 2011, \apj, 730, 28,
  \dodoi{10.1088/0004-637X/730/1/28}

\bibitem[{{Pilkington} {et~al.}(2012){Pilkington}, {Few}, {Gibson}, {Calura},
  {Michel-Dansac}, {Thacker}, {Moll{\'a}}, {Matteucci}, {Rahimi}, {Kawata},
  {Kobayashi}, {Brook}, {Stinson}, {Couchman}, {Bailin}, \&
  {Wadsley}}]{Pilkington:2012aa}
{Pilkington}, K., {Few}, C.~G., {Gibson}, B.~K., {et~al.} 2012, \aap, 540, A56,
  \dodoi{10.1051/0004-6361/201117466}

\bibitem[{{Pope} {et~al.}(2008){Pope}, {Bussmann}, {Dey}, {Meger}, {Alexander},
  {Brodwin}, {Chary}, {Dickinson}, {Frayer}, {Greve}, {Huynh}, {Lin},
  {Morrison}, {Scott}, \& {Yan}}]{Pope:2008aa}
{Pope}, A., {Bussmann}, R.~S., {Dey}, A., {et~al.} 2008, \apj, 689, 127,
  \dodoi{10.1086/592739}

\bibitem[{{Popping} {et~al.}(2017){Popping}, {Puglisi}, \&
  {Norman}}]{Popping:2017ab}
{Popping}, G., {Puglisi}, A., \& {Norman}, C.~A. 2017, \mnras, 472, 2315,
  \dodoi{10.1093/mnras/stx2202}

\bibitem[{{Popping} {et~al.}(2022){Popping}, {Pillepich}, {Calistro Rivera},
  {Schulz}, {Hernquist}, {Kaasinen}, {Marinacci}, {Nelson}, \&
  {Vogelsberger}}]{Popping:2022ac}
{Popping}, G., {Pillepich}, A., {Calistro Rivera}, G., {et~al.} 2022, \mnras,
  510, 3321, \dodoi{10.1093/mnras/stab3312}

\bibitem[{{Rigby} {et~al.}(2023){Rigby}, {Perrin}, {McElwain}, {Kimble},
  {Friedman}, {Lallo}, {Doyon}, {Feinberg}, {Ferruit}, {Glasse}, \&
  et~al.}]{Rigby:2023ab}
{Rigby}, J., {Perrin}, M., {McElwain}, M., {et~al.} 2023, \pasp, 135, 048001,
  \dodoi{10.1088/1538-3873/acb293}

\bibitem[{{Robertson} {et~al.}(2020){Robertson}, {Smith}, {Massey}, {Eke},
  {Jauzac}, {Bianconi}, \& {Ryczanowski}}]{Robertson:2020ac}
{Robertson}, A., {Smith}, G.~P., {Massey}, R., {et~al.} 2020, \mnras, 495,
  3727, \dodoi{10.1093/mnras/staa1429}

\bibitem[{{Robitaille}(2019)}]{Robitaille:2019aa}
{Robitaille}, T. 2019, {APLpy v2.0: The Astronomical Plotting Library in
  Python}, 2.0, Zenodo,  Zenodo, \dodoi{10.5281/zenodo.2567476}

\bibitem[{{Robitaille} \& {Bressert}(2012)}]{Robitaille:2012aa}
{Robitaille}, T., \& {Bressert}, E. 2012, {APLpy: Astronomical Plotting Library
  in Python}, Astrophysics Source Code Library.
\newblock \doeprint{1208.017}

\bibitem[{{Roussel} {et~al.}(2010){Roussel}, {Wilson}, {Vigroux}, {Isaak},
  {Sauvage}, {Madden}, {Auld}, {Baes}, {Barlow}, {Bendo}, {Bock}, {Boselli},
  {Bradford}, {Buat}, {Castro-Rodriguez}, {Chanial}, {Charlot}, {Ciesla},
  {Clements}, {Cooray}, {Cormier}, {Cortese}, {Davies}, {Dwek}, {Eales},
  {Elbaz}, {Galametz}, {Galliano}, {Gear}, {Glenn}, {Gomez}, {Griffin}, {Hony},
  {Levenson}, {Lu}, {O'Halloran}, {Okumura}, {Oliver}, {Page}, {Panuzzo},
  {Papageorgiou}, {Parkin}, {Perez-Fournon}, {Pohlen}, {Rangwala}, {Rigby},
  {Rykala}, {Sacchi}, {Schulz}, {Schirm}, {Smith}, {Spinoglio}, {Stevens},
  {Srinivasan}, {Symeonidis}, {Trichas}, {Vaccari}, {Wozniak}, {Wright}, \&
  {Zeilinger}}]{Roussel:2010aa}
{Roussel}, H., {Wilson}, C.~D., {Vigroux}, L., {et~al.} 2010, \aap, 518, L66,
  \dodoi{10.1051/0004-6361/201014567}

\bibitem[{{Rowlands} {et~al.}(2018){Rowlands}, {Heckman}, {Wild}, {Zakamska},
  {Rodriguez-Gomez}, {Barrera-Ballesteros}, {Lotz}, {Thilker}, {Andrews},
  {Boquien}, {Brinkmann}, {Brownstein}, {Hwang}, \&
  {Smethurst}}]{Rowlands:2018aa}
{Rowlands}, K., {Heckman}, T., {Wild}, V., {et~al.} 2018, \mnras, 480, 2544,
  \dodoi{10.1093/mnras/sty1916}

\bibitem[{{Rujopakarn} {et~al.}(2011){Rujopakarn}, {Rieke}, {Eisenstein}, \&
  {Juneau}}]{Rujopakarn:2011aa}
{Rujopakarn}, W., {Rieke}, G.~H., {Eisenstein}, D.~J., \& {Juneau}, S. 2011,
  \apj, 726, 93, \dodoi{10.1088/0004-637X/726/2/93}

\bibitem[{{Rusin} \& {Ma}(2001)}]{Rusin:2001ab}
{Rusin}, D., \& {Ma}, C.-P. 2001, \apjl, 549, L33, \dodoi{10.1086/319129}

\bibitem[{{Ryan} {et~al.}(2012){Ryan}, {McCarthy}, {Cohen}, {Yan}, {Hathi},
  {Koekemoer}, {Rutkowski}, {Mechtley}, {Windhorst}, {O'Connell}, {Balick},
  {Bond}, {Bushouse}, {Calzetti}, {Crockett}, {Disney}, {Dopita}, {Frogel},
  {Hall}, {Holtzman}, {Kaviraj}, {Kimble}, {MacKenty}, {Mutchler}, {Paresce},
  {Saha}, {Silk}, {Trauger}, {Walker}, {Whitmore}, \& {Young}}]{Ryan:2012aa}
{Ryan}, R.~E., J., {McCarthy}, P.~J., {Cohen}, S.~H., {et~al.} 2012, \apj, 749,
  53, \dodoi{10.1088/0004-637X/749/1/53}

\bibitem[{{Saito} {et~al.}(2015){Saito}, {Iono}, {Yun}, {Ueda}, {Nakanishi},
  {Sugai}, {Espada}, {Imanishi}, {Motohara}, {Hagiwara}, {Tateuchi}, {Lee}, \&
  {Kawabe}}]{Saito:2015aa}
{Saito}, T., {Iono}, D., {Yun}, M.~S., {et~al.} 2015, \apj, 803, 60,
  \dodoi{10.1088/0004-637X/803/2/60}

\bibitem[{{Sanders} \& {Mirabel}(1996)}]{Sanders:1996aa}
{Sanders}, D.~B., \& {Mirabel}, I.~F. 1996, \araa, 34, 749,
  \dodoi{10.1146/annurev.astro.34.1.749}

\bibitem[{{Sanders} \& {Balamore}(1971)}]{Sanders:1971aa}
{Sanders}, R.~H., \& {Balamore}, D.~S. 1971, \apj, 166, 7,
  \dodoi{10.1086/150938}

\bibitem[{{Schneider} {et~al.}(1992){Schneider}, {Ehlers}, \&
  {Falco}}]{Schneider:1992aa}
{Schneider}, P., {Ehlers}, J., \& {Falco}, E.~E. 1992, {Gravitational Lenses}
  (Springer-Verlag), 112, \dodoi{10.1007/978-3-662-03758-4}

\bibitem[{{Schneider} \& {Weiss}(1986)}]{Schneider:1986aa}
{Schneider}, P., \& {Weiss}, A. 1986, \aap, 164, 237

\bibitem[{{Serjeant}(2012)}]{Serjeant:2012aa}
{Serjeant}, S. 2012, \mnras, 424, 2429,
  \dodoi{10.1111/j.1365-2966.2012.20761.x}

\bibitem[{{Sharma} {et~al.}(2016){Sharma}, {Theuns}, {Frenk}, {Bower}, {Crain},
  {Schaller}, \& {Schaye}}]{Sharma:2016aa}
{Sharma}, M., {Theuns}, T., {Frenk}, C., {et~al.} 2016, \mnras, 458, L94,
  \dodoi{10.1093/mnrasl/slw021}

\bibitem[{{Sharon} {et~al.}(2012){Sharon}, {Gladders}, {Rigby}, {Wuyts},
  {Koester}, {Bayliss}, \& {Barrientos}}]{Sharon:2012aa}
{Sharon}, K., {Gladders}, M.~D., {Rigby}, J.~R., {et~al.} 2012, \apj, 746, 161,
  \dodoi{10.1088/0004-637X/746/2/161}

\bibitem[{{Shen} {et~al.}(2003){Shen}, {Mo}, {White}, {Blanton}, {Kauffmann},
  {Voges}, {Brinkmann}, \& {Csabai}}]{Shen:2003aa}
{Shen}, S., {Mo}, H.~J., {White}, S. D.~M., {et~al.} 2003, \mnras, 343, 978,
  \dodoi{10.1046/j.1365-8711.2003.06740.x}

\bibitem[{{Shibuya} {et~al.}(2015){Shibuya}, {Ouchi}, \&
  {Harikane}}]{Shibuya:2015aa}
{Shibuya}, T., {Ouchi}, M., \& {Harikane}, Y. 2015, \apjs, 219, 15,
  \dodoi{10.1088/0067-0049/219/2/15}

\bibitem[{{Shin} \& {Evans}(2008)}]{Shin:2008aa}
{Shin}, E.~M., \& {Evans}, N.~W. 2008, \mnras, 390, 505,
  \dodoi{10.1111/j.1365-2966.2008.13738.x}

\bibitem[{{Simpson} {et~al.}(2014){Simpson}, {Swinbank}, {Smail}, {Alexander},
  {Brandt}, {Bertoldi}, {de Breuck}, {Chapman}, {Coppin}, {da Cunha},
  {Danielson}, {Dannerbauer}, {Greve}, {Hodge}, {Ivison}, {Karim}, {Knudsen},
  {Poggianti}, {Schinnerer}, {Thomson}, {Walter}, {Wardlow}, {Wei{\ss}}, \&
  {van der Werf}}]{Simpson:2014aa}
{Simpson}, J.~M., {Swinbank}, A.~M., {Smail}, I., {et~al.} 2014, \apj, 788,
  125, \dodoi{10.1088/0004-637X/788/2/125}

\bibitem[{{Simpson} {et~al.}(2015){Simpson}, {Smail}, {Swinbank}, {Almaini},
  {Blain}, {Bremer}, {Chapman}, {Chen}, {Conselice}, {Coppin}, {Danielson},
  {Dunlop}, {Edge}, {Farrah}, {Geach}, {Hartley}, {Ivison}, {Karim}, {Lani},
  {Ma}, {Meijerink}, {Micha{\l}owski}, {Mortlock}, {Scott}, {Simpson},
  {Spaans}, {Thomson}, {van Kampen}, \& {van der Werf}}]{Simpson:2015ab}
{Simpson}, J.~M., {Smail}, I., {Swinbank}, A.~M., {et~al.} 2015, \apj, 799, 81,
  \dodoi{10.1088/0004-637X/799/1/81}

\bibitem[{{Smethurst} {et~al.}(2017){Smethurst}, {Lintott}, {Bamford}, {Hart},
  {Kruk}, {Masters}, {Nichol}, \& {Simmons}}]{Smethurst:2017aa}
{Smethurst}, R.~J., {Lintott}, C.~J., {Bamford}, S.~P., {et~al.} 2017, \mnras,
  469, 3670, \dodoi{10.1093/mnras/stx973}

\bibitem[{{Smethurst} {et~al.}(2016){Smethurst}, {Lintott}, {Simmons},
  {Schawinski}, {Bamford}, {Cardamone}, {Kruk}, {Masters}, {Urry}, {Willett},
  \& {Wong}}]{Smethurst:2016aa}
{Smethurst}, R.~J., {Lintott}, C.~J., {Simmons}, B.~D., {et~al.} 2016, \mnras,
  463, 2986, \dodoi{10.1093/mnras/stw2204}

\bibitem[{{Speagle} {et~al.}(2014){Speagle}, {Steinhardt}, {Capak}, \&
  {Silverman}}]{Speagle:2014aa}
{Speagle}, J.~S., {Steinhardt}, C.~L., {Capak}, P.~L., \& {Silverman}, J.~D.
  2014, \apjs, 214, 15, \dodoi{10.1088/0067-0049/214/2/15}

\bibitem[{{Suess} {et~al.}(2019){Suess}, {Kriek}, {Price}, \&
  {Barro}}]{Suess:2019aa}
{Suess}, K.~A., {Kriek}, M., {Price}, S.~H., \& {Barro}, G. 2019, \apj, 877,
  103, \dodoi{10.3847/1538-4357/ab1bda}

\bibitem[{{Suess} {et~al.}(2022){Suess}, {Bezanson}, {Nelson}, {Setton},
  {Price}, {Dokkum}, {Brammer}, {Labb{\'e}}, {Leja}, {Miller}, {Robertson},
  {Wel}, {Weaver}, \& {Whitaker}}]{Suess:2022aa}
{Suess}, K.~A., {Bezanson}, R., {Nelson}, E.~J., {et~al.} 2022, \apjl, 937,
  L33, \dodoi{10.3847/2041-8213/ac8e06}

\bibitem[{{Tacchella} {et~al.}(2016{\natexlab{a}}){Tacchella}, {Dekel},
  {Carollo}, {Ceverino}, {DeGraf}, {Lapiner}, {Mand elker}, \&
  {Primack}}]{Tacchella:2016aa}
{Tacchella}, S., {Dekel}, A., {Carollo}, C.~M., {et~al.} 2016{\natexlab{a}},
  \mnras, 458, 242, \dodoi{10.1093/mnras/stw303}

\bibitem[{{Tacchella} {et~al.}(2016{\natexlab{b}}){Tacchella}, {Dekel},
  {Carollo}, {Ceverino}, {DeGraf}, {Lapiner}, {Mandelker}, \& {Primack
  Joel}}]{Tacchella:2016ab}
---. 2016{\natexlab{b}}, \mnras, 457, 2790, \dodoi{10.1093/mnras/stw131}

\bibitem[{{Tacchella} {et~al.}(2018){Tacchella}, {Carollo}, {F{\"o}rster
  Schreiber}, {Renzini}, {Dekel}, {Genzel}, {Lang}, {Lilly}, {Mancini},
  {Onodera}, {Tacconi}, {Wuyts}, \& {Zamorani}}]{Tacchella:2018aa}
{Tacchella}, S., {Carollo}, C.~M., {F{\"o}rster Schreiber}, N.~M., {et~al.}
  2018, \apj, 859, 56, \dodoi{10.3847/1538-4357/aabf8b}

\bibitem[{{Tadaki} {et~al.}(2015){Tadaki}, {Kohno}, {Kodama}, {Ikarashi},
  {Aretxaga}, {Berta}, {Caputi}, {Dunlop}, {Hatsukade}, {Hayashi}, {Hughes},
  {Ivison}, {Izumi}, {Koyama}, {Lutz}, {Makiya}, {Matsuda}, {Nakanishi},
  {Rujopakarn}, {Tamura}, {Umehata}, {Wang}, {Wilson}, {Wuyts}, {Yamaguchi}, \&
  {Yun}}]{Tadaki:2015aa}
{Tadaki}, K.-i., {Kohno}, K., {Kodama}, T., {et~al.} 2015, \apjl, 811, L3,
  \dodoi{10.1088/2041-8205/811/1/L3}

\bibitem[{{Tadaki} {et~al.}(2020){Tadaki}, {Iono}, {Yun}, {Aretxaga},
  {Hatsukade}, {Lee}, {Michiyama}, {Nakanishi}, {Saito}, {Ueda}, \&
  {Umehata}}]{Tadaki:2020aa}
{Tadaki}, K.-i., {Iono}, D., {Yun}, M.~S., {et~al.} 2020, \apj, 889, 141,
  \dodoi{10.3847/1538-4357/ab64f4}

\bibitem[{{Talia} {et~al.}(2018){Talia}, {Pozzi}, {Vallini}, {Cimatti},
  {Cassata}, {Fraternali}, {Brusa}, {Daddi}, {Delvecchio}, {Ibar}, {Liuzzo},
  {Vignali}, {Massardi}, {Zamorani}, {Gruppioni}, {Renzini}, {Mignoli},
  {Pozzetti}, \& {Rodighiero}}]{Talia:2018aa}
{Talia}, M., {Pozzi}, F., {Vallini}, L., {et~al.} 2018, \mnras, 476, 3956,
  \dodoi{10.1093/mnras/sty481}

\bibitem[{{Tamura} {et~al.}(2015){Tamura}, {Oguri}, {Iono}, {Hatsukade},
  {Matsuda}, \& {Hayashi}}]{Tamura:2015aa}
{Tamura}, Y., {Oguri}, M., {Iono}, D., {et~al.} 2015, \pasj, 67, 72,
  \dodoi{10.1093/pasj/psv040}

\bibitem[{{Treu} \& {Koopmans}(2004)}]{Treu:2004aa}
{Treu}, T., \& {Koopmans}, L. V.~E. 2004, \apj, 611, 739,
  \dodoi{10.1086/422245}

\bibitem[{{Trujillo} {et~al.}(2004){Trujillo}, {Rudnick}, {Rix}, {Labb{\'e}},
  {Franx}, {Daddi}, {van Dokkum}, {F{\"o}rster Schreiber}, {Kuijken},
  {Moorwood}, {R{\"o}ttgering}, {van der Wel}, {van der Werf}, \& {van
  Starkenburg}}]{Trujillo:2004aa}
{Trujillo}, I., {Rudnick}, G., {Rix}, H.-W., {et~al.} 2004, \apj, 604, 521,
  \dodoi{10.1086/382060}

\bibitem[{{Tsai} {et~al.}(2015){Tsai}, {Eisenhardt}, {Wu}, {Stern}, {Assef},
  {Blain}, {Bridge}, {Benford}, {Cutri}, {Griffith}, {Jarrett}, {Lonsdale},
  {Masci}, {Moustakas}, {Petty}, {Sayers}, {Stanford}, {Wright}, {Yan},
  {Leisawitz}, {Liu}, {Mainzer}, {McLean}, {Padgett}, {Skrutskie}, {Gelino},
  {Beichman}, \& {Juneau}}]{Tsai:2015aa}
{Tsai}, C.-W., {Eisenhardt}, P.~R.~M., {Wu}, J., {et~al.} 2015, \apj, 805, 90,
  \dodoi{10.1088/0004-637X/805/2/90}

\bibitem[{{Unterborn} \& {Ryden}(2008)}]{Unterborn:2008aa}
{Unterborn}, C.~T., \& {Ryden}, B.~S. 2008, \apj, 687, 976,
  \dodoi{10.1086/591898}

\bibitem[{{van der Wel} {et~al.}(2014{\natexlab{a}}){van der Wel}, {Franx},
  {van Dokkum}, {Skelton}, {Momcheva}, {Whitaker}, {Brammer}, {Bell}, {Rix},
  {Wuyts}, {Ferguson}, {Holden}, {Barro}, {Koekemoer}, {Chang}, {McGrath},
  {H{\"a}ussler}, {Dekel}, {Behroozi}, {Fumagalli}, {Leja}, {Lundgren},
  {Maseda}, {Nelson}, {Wake}, {Patel}, {Labb{\'e}}, {Faber}, {Grogin}, \&
  {Kocevski}}]{van-der-Wel:2014aa}
{van der Wel}, A., {Franx}, M., {van Dokkum}, P.~G., {et~al.}
  2014{\natexlab{a}}, \apj, 788, 28, \dodoi{10.1088/0004-637X/788/1/28}

\bibitem[{{van der Wel} {et~al.}(2014{\natexlab{b}}){van der Wel}, {Chang},
  {Bell}, {Holden}, {Ferguson}, {Giavalisco}, {Rix}, {Skelton}, {Whitaker},
  {Momcheva}, {Brammer}, {Kassin}, {Martig}, {Dekel}, {Ceverino}, {Koo},
  {Mozena}, {van Dokkum}, {Franx}, {Faber}, \& {Primack}}]{van-der-Wel:2014ab}
{van der Wel}, A., {Chang}, Y.-Y., {Bell}, E.~F., {et~al.} 2014{\natexlab{b}},
  \apjl, 792, L6, \dodoi{10.1088/2041-8205/792/1/L6}

\bibitem[{{van Dokkum} {et~al.}(2010){van Dokkum}, {Whitaker}, {Brammer},
  {Franx}, {Kriek}, {Labb{\'e}}, {Marchesini}, {Quadri}, {Bezanson},
  {Illingworth}, {Muzzin}, {Rudnick}, {Tal}, \& {Wake}}]{van-Dokkum:2010aa}
{van Dokkum}, P.~G., {Whitaker}, K.~E., {Brammer}, G., {et~al.} 2010, \apj,
  709, 1018, \dodoi{10.1088/0004-637X/709/2/1018}

\bibitem[{{van Dokkum} {et~al.}(2013){van Dokkum}, {Leja}, {Nelson}, {Patel},
  {Skelton}, {Momcheva}, {Brammer}, {Whitaker}, {Lundgren}, {Fumagalli},
  {Conroy}, {F{\"o}rster Schreiber}, {Franx}, {Kriek}, {Labb{\'e}},
  {Marchesini}, {Rix}, {van der Wel}, \& {Wuyts}}]{van-Dokkum:2013aa}
{van Dokkum}, P.~G., {Leja}, J., {Nelson}, E.~J., {et~al.} 2013, \apjl, 771,
  L35, \dodoi{10.1088/2041-8205/771/2/L35}

\bibitem[{{van Dokkum} {et~al.}(2014){van Dokkum}, {Bezanson}, {van der Wel},
  {Nelson}, {Momcheva}, {Skelton}, {Whitaker}, {Brammer}, {Conroy},
  {F{\"o}rster Schreiber}, {Fumagalli}, {Kriek}, {Labb{\'e}}, {Leja},
  {Marchesini}, {Muzzin}, {Oesch}, \& {Wuyts}}]{van-Dokkum:2014aa}
{van Dokkum}, P.~G., {Bezanson}, R., {van der Wel}, A., {et~al.} 2014, \apj,
  791, 45, \dodoi{10.1088/0004-637X/791/1/45}

\bibitem[{{Veilleux} {et~al.}(2009){Veilleux}, {Rupke}, {Kim}, {Genzel},
  {Sturm}, {Lutz}, {Contursi}, {Schweitzer}, {Tacconi}, {Netzer}, {Sternberg},
  {Mihos}, {Baker}, {Mazzarella}, {Lord}, {Sanders}, {Stockton}, {Joseph}, \&
  {Barnes}}]{Veilleux:2009aa}
{Veilleux}, S., {Rupke}, D.~S.~N., {Kim}, D.~C., {et~al.} 2009, \apjs, 182,
  628, \dodoi{10.1088/0067-0049/182/2/628}

\bibitem[{{Vulcani} {et~al.}(2014){Vulcani}, {Bamford}, {H{\"a}u{\ss}ler},
  {Vika}, {Rojas}, {Agius}, {Baldry}, {Bauer}, {Brown}, {Driver}, {Graham},
  {Kelvin}, {Liske}, {Loveday}, {Popescu}, {Robotham}, \&
  {Tuffs}}]{Vulcani:2014aa}
{Vulcani}, B., {Bamford}, S.~P., {H{\"a}u{\ss}ler}, B., {et~al.} 2014, \mnras,
  441, 1340, \dodoi{10.1093/mnras/stu632}

\bibitem[{{Wallington} \& {Narayan}(1993)}]{Wallington:1993aa}
{Wallington}, S., \& {Narayan}, R. 1993, \apj, 403, 517, \dodoi{10.1086/172222}

\bibitem[{{Wang} {et~al.}(2019{\natexlab{a}}){Wang}, {Lilly}, {Pezzulli}, \&
  {Matthee}}]{Wang:2019aa}
{Wang}, E., {Lilly}, S.~J., {Pezzulli}, G., \& {Matthee}, J.
  2019{\natexlab{a}}, \apj, 877, 132, \dodoi{10.3847/1538-4357/ab1c5b}

\bibitem[{{Wang} {et~al.}(2019{\natexlab{b}}){Wang}, {Schreiber}, {Elbaz},
  {Yoshimura}, {Kohno}, {Shu}, {Yamaguchi}, {Pannella}, {Franco}, {Huang},
  {Lim}, \& {Wang}}]{Wang:2019ab}
{Wang}, T., {Schreiber}, C., {Elbaz}, D., {et~al.} 2019{\natexlab{b}}, \nat,
  572, 211, \dodoi{10.1038/s41586-019-1452-4}

\bibitem[{{Wang} {et~al.}(2017){Wang}, {Faber}, {Liu}, {Guo}, {Pacifici},
  {Koo}, {Kassin}, {Mao}, {Fang}, {Chen}, {Koekemoer}, {Kocevski}, \&
  {Ashby}}]{Wang:2017ab}
{Wang}, W., {Faber}, S.~M., {Liu}, F.~S., {et~al.} 2017, \mnras, 469, 4063,
  \dodoi{10.1093/mnras/stx1148}

\bibitem[{{Werner} {et~al.}(2004){Werner}, {Roellig}, {Low}, {Rieke}, {Rieke},
  {Hoffmann}, {Young}, {Houck}, {Brandl}, {Fazio}, {Hora}, {Gehrz}, {Helou},
  {Soifer}, {Stauffer}, {Keene}, {Eisenhardt}, {Gallagher}, {Gautier}, {Irace},
  {Lawrence}, {Simmons}, {Van Cleve}, {Jura}, {Wright}, \&
  {Cruikshank}}]{Werner:2004aa}
{Werner}, M.~W., {Roellig}, T.~L., {Low}, F.~J., {et~al.} 2004, \apjs, 154, 1,
  \dodoi{10.1086/422992}

\bibitem[{{Windhorst} {et~al.}(2023){Windhorst}, {Cohen}, {Jansen}, {Summers},
  {Tompkins}, {Conselice}, {Driver}, {Yan}, {Coe}, {Frye}, {Grogin},
  {Koekemoer}, {Marshall}, {O'Brien}, {Pirzkal}, {Robotham}, {Ryan}, {Willmer},
  {Carleton}, {Diego}, {Keel}, {Porto}, {Redshaw}, {Scheller}, {Wilkins},
  {Willner}, {Zitrin}, {Adams}, {Austin}, {Arendt}, {Beacom}, {Bhatawdekar},
  {Bradley}, {Broadhurst}, {Cheng}, {Civano}, {Dai}, {Dole}, {D'Silva},
  {Duncan}, {Fazio}, {Ferrami}, {Ferreira}, {Finkelstein}, {Furtak}, {Gim},
  {Griffiths}, {Hammel}, {Harrington}, {Hathi}, {Holwerda}, {Honor}, {Huang},
  {Hyun}, {Im}, {Joshi}, {Kamieneski}, {Kelly}, {Larson}, {Li}, {Lim}, {Ma},
  {Maksym}, {Manzoni}, {Meena}, {Milam}, {Nonino}, {Pascale}, {Petric},
  {Pierel}, {del Carmen Polletta}, {R{\"o}ttgering}, {Rutkowski}, {Smail},
  {Straughn}, {Strolger}, {Swirbul}, {Trussler}, {Wang}, {Welch}, {B. Wyithe},
  {Yun}, {Zackrisson}, {Zhang}, \& {Zhao}}]{Windhorst:2023aa}
{Windhorst}, R.~A., {Cohen}, S.~H., {Jansen}, R.~A., {et~al.} 2023, \aj, 165,
  13, \dodoi{10.3847/1538-3881/aca163}

\bibitem[{{Winn} {et~al.}(2004){Winn}, {Rusin}, \& {Kochanek}}]{Winn:2004aa}
{Winn}, J.~N., {Rusin}, D., \& {Kochanek}, C.~S. 2004, \nat, 427, 613,
  \dodoi{10.1038/nature02279}

\bibitem[{{Wong} {et~al.}(2015){Wong}, {Suyu}, \& {Matsushita}}]{Wong:2015aa}
{Wong}, K.~C., {Suyu}, S.~H., \& {Matsushita}, S. 2015, \apj, 811, 115,
  \dodoi{10.1088/0004-637X/811/2/115}

\bibitem[{{Wright}(2006)}]{Wright:2006aa}
{Wright}, E.~L. 2006, \pasp, 118, 1711, \dodoi{10.1086/510102}

\bibitem[{{Wright} {et~al.}(2010){Wright}, {Eisenhardt}, {Mainzer}, {Ressler},
  {Cutri}, {Jarrett}, {Kirkpatrick}, {Padgett}, {McMillan}, {Skrutskie},
  {Stanford}, {Cohen}, {Walker}, {Mather}, {Leisawitz}, {Gautier}, {McLean},
  {Benford}, {Lonsdale}, {Blain}, {Mendez}, {Irace}, {Duval}, {Liu}, {Royer},
  {Heinrichsen}, {Howard}, {Shannon}, {Kendall}, {Walsh}, {Larsen}, {Cardon},
  {Schick}, {Schwalm}, {Abid}, {Fabinsky}, {Naes}, \& {Tsai}}]{Wright:2010aa}
{Wright}, E.~L., {Eisenhardt}, P.~R.~M., {Mainzer}, A.~K., {et~al.} 2010, \aj,
  140, 1868, \dodoi{10.1088/0004-6256/140/6/1868}

\bibitem[{{Wu} {et~al.}(2012){Wu}, {Tsai}, {Sayers}, {Benford}, {Bridge},
  {Blain}, {Eisenhardt}, {Stern}, {Petty}, {Assef}, {Bussmann}, {Comerford},
  {Cutri}, {Evans}, {Griffith}, {Jarrett}, {Lake}, {Lonsdale}, {Rho},
  {Stanford}, {Weiner}, {Wright}, \& {Yan}}]{Wu:2012aa}
{Wu}, J., {Tsai}, C.-W., {Sayers}, J., {et~al.} 2012, \apj, 756, 96,
  \dodoi{10.1088/0004-637X/756/1/96}

\bibitem[{{Wu} {et~al.}(2014){Wu}, {Bussmann}, {Tsai}, {Petric}, {Blain},
  {Eisenhardt}, {Bridge}, {Benford}, {Stern}, {Assef}, {Gelino}, {Moustakas},
  \& {Wright}}]{Wu:2014aa}
{Wu}, J., {Bussmann}, R.~S., {Tsai}, C.-W., {et~al.} 2014, \apj, 793, 8,
  \dodoi{10.1088/0004-637X/793/1/8}

\bibitem[{{Wuyts} {et~al.}(2007){Wuyts}, {Labb{\'e}}, {Franx}, {Rudnick}, {van
  Dokkum}, {Fazio}, {F{\"o}rster Schreiber}, {Huang}, {Moorwood}, {Rix},
  {R{\"o}ttgering}, \& {van der Werf}}]{Wuyts:2007aa}
{Wuyts}, S., {Labb{\'e}}, I., {Franx}, M., {et~al.} 2007, \apj, 655, 51,
  \dodoi{10.1086/509708}

\bibitem[{{Wuyts} {et~al.}(2012){Wuyts}, {F{\"o}rster Schreiber}, {Genzel},
  {Guo}, {Barro}, {Bell}, {Dekel}, {Faber}, {Ferguson}, {Giavalisco}, {Grogin},
  {Hathi}, {Huang}, {Kocevski}, {Koekemoer}, {Koo}, {Lotz}, {Lutz}, {McGrath},
  {Newman}, {Rosario}, {Saintonge}, {Tacconi}, {Weiner}, \& {van der
  Wel}}]{Wuyts:2012aa}
{Wuyts}, S., {F{\"o}rster Schreiber}, N.~M., {Genzel}, R., {et~al.} 2012, \apj,
  753, 114, \dodoi{10.1088/0004-637X/753/2/114}

\bibitem[{{Wuyts} {et~al.}(2013){Wuyts}, {F{\"o}rster Schreiber}, {Nelson},
  {van Dokkum}, {Brammer}, {Chang}, {Faber}, {Ferguson}, {Franx}, {Fumagalli},
  {Genzel}, {Grogin}, {Kocevski}, {Koekemoer}, {Lundgren}, {Lutz}, {McGrath},
  {Momcheva}, {Rosario}, {Skelton}, {Tacconi}, {van der Wel}, \&
  {Whitaker}}]{Wuyts:2013aa}
{Wuyts}, S., {F{\"o}rster Schreiber}, N.~M., {Nelson}, E.~J., {et~al.} 2013,
  \apj, 779, 135, \dodoi{10.1088/0004-637X/779/2/135}

\bibitem[{{Yang} {et~al.}(2022){Yang}, {Morishita}, {Leethochawalit},
  {Castellano}, {Calabr{\`o}}, {Treu}, {Bonchi}, {Fontana}, {Mason}, {Merlin},
  {Paris}, {Trenti}, {Roberts-Borsani}, {Bradac}, {Vanzella}, {Vulcani},
  {Marchesini}, {Ding}, {Nanayakkara}, {Birrer}, {Glazebrook}, {Jones},
  {Boyett}, {Santini}, {Strait}, \& {Wang}}]{Yang:2022ac}
{Yang}, L., {Morishita}, T., {Leethochawalit}, N., {et~al.} 2022, \apjl, 938,
  L17, \dodoi{10.3847/2041-8213/ac8803}

\bibitem[{{Yuan} {et~al.}(2011){Yuan}, {Kewley}, {Swinbank}, {Richard}, \&
  {Livermore}}]{Yuan:2011aa}
{Yuan}, T.-T., {Kewley}, L.~J., {Swinbank}, A.~M., {Richard}, J., \&
  {Livermore}, R.~C. 2011, \apjl, 732, L14, \dodoi{10.1088/2041-8205/732/1/L14}

\bibitem[{{Yun} {et~al.}(1994){Yun}, {Scoville}, \& {Knop}}]{Yun:1994aa}
{Yun}, M.~S., {Scoville}, N.~Z., \& {Knop}, R.~A. 1994, \apjl, 430, L109,
  \dodoi{10.1086/187450}

\bibitem[{{Zhuang} \& {Shen}(2023)}]{Zhuang:2023aa}
{Zhuang}, M.-Y., \& {Shen}, Y. 2023, arXiv e-prints, arXiv:2304.13776,
  \dodoi{10.48550/arXiv.2304.13776}

\bibitem[{{Zolotov} {et~al.}(2015){Zolotov}, {Dekel}, {Mandelker}, {Tweed},
  {Inoue}, {DeGraf}, {Ceverino}, {Primack}, {Barro}, \&
  {Faber}}]{Zolotov:2015aa}
{Zolotov}, A., {Dekel}, A., {Mandelker}, N., {et~al.} 2015, \mnras, 450, 2327,
  \dodoi{10.1093/mnras/stv740}

\end{thebibliography}






\end{document}